\documentclass[aps,prx,twocolumn,notitlepage,letterpaper,longbibliography]{revtex4-2}

\usepackage{amsmath}
\usepackage{amssymb}
\usepackage{amsfonts}
\usepackage{bm}
\usepackage{mathrsfs}
\usepackage{tensor}
\usepackage{dsfont}
\usepackage{graphicx}
\usepackage[caption=false]{subfig}
\usepackage[usenames,dvipsnames]{xcolor}
\usepackage[hyperindex,pdftex, breaklinks,colorlinks = true,linkcolor = blue,urlcolor=blue,citecolor=blue]{hyperref}
\usepackage{physics}

\begin{document}
	
	%\title{Reconciling magnetoelectric response and time-reversal symmetry \\
	%	in non-magnetic $\mathbb{Z}_2$ topological insulators}
        %\title{Aspects of symmetry, topology, and geometry in the \\ 
        %magnetoelectric response of non-magnetic $\mathbb{Z}_2$ topological insulators}
	\title{Symmetry, topology, and geometry: The many faces of the \\
        topological magnetoelectric effect}
	\author{Perry T. Mahon}
	\email{perry.mahon@austin.utexas.edu}
	\affiliation{Department of Physics, University of Texas at Austin, Austin, Texas 78712, USA}
	\author{Chao Lei}
	\affiliation{Department of Physics, University of Texas at Austin, Austin, Texas 78712, USA}
	\author{Allan H. MacDonald}
	\affiliation{Department of Physics, University of Texas at Austin, Austin, Texas 78712, USA}
	
	\date{\today}
	
\begin{abstract}
A delicate tension complicates the relationship between the topological magnetoelectric effect (TME) in three-dimensional (3D) $\mathbb{Z}_2$ topological insulators (TIs) and time-reversal symmetry (TRS). TRS underlies a particular $\mathbb{Z}_2$ topological classification of the electronic ground state of crystalline band insulators and the associated quantization of the magnetoelectric response  
coefficient calculated using bulk linear response theory but, according to standard symmetry arguments, simultaneously forbids a nonzero magnetoelectric coefficient in any physical finite-size system. 
This contrast between theories of magnetoelectric response in formal bulk models and 
in real finite-sized materials originates from the distinct approaches required to 
introduce notions of (electronic) polarization and orbital magnetization in these 
fundamentally different environments. 
In this work we argue for a modified interpretation of the bulk linear response calculations in non-magnetic $\mathbb{Z}_2$ TIs that is more plainly
consistent with TRS, and use this interpretation to discuss the effect's observation -- 
still absent over a decade after its prediction.
Our analysis is reinforced by microscopic bulk and thin film calculations carried out using 
a simplified but still realistic effective model for the well established V$_2$VI$_3$ ($\text{V}=\text{(Sb,Bi)}$ and $\text{VI}=\text{(Se,Te)}$) family of non-magnetic $\mathbb{Z}_2$ TIs. 
When a uniform dc magnetic field is included in this model, the anomalous $n=0$ Landau levels (LLs) play the central role, both in thin films and in bulk. 
%Within that model we show that the central role is played by the 
%anomalous $n=0$ Landau levels (LLs), both in thin films and in bulk. 
In the former case, only the $n=0$ LL eigenfunctions can support a dipole moment, which vanishes if there are no magnetic surface dopants and is quantized in the thick-film limit if magnetic dopants at the top and bottom surfaces have opposite orientation. 
In the latter case, the Hamiltonian projected into the $n=0$ LL subspace is a one-dimensional Su-Schrieffer-Heeger model with ground state polarization that is quantized in accordance with 
%the topological properties of 
the bulk linear response coefficient calculated for (a lattice regularization of) the starting 3D model.
Motivated by analytical results, we conjecture a type of microscopic bulk-boundary correspondence:
a bulk insulator with (generalized) TRS supports a magnetoelectric coefficient that is purely itinerant (which is generically related to the geometry of the ground state) if and only if magnetic surface dopants are 
required for the TME to manifest in finite samples thereof.
%We argue that this bulk partitioning also distinguishes the physical mechanism leading to the TME in non-magnetic and anti-ferromagnetic TIs.
We conclude that in non-magnetic $\mathbb{Z}_2$
TIs the TME is activated by magnetic surface dopants, that the charge density response to a uniform dc magnetic field is localized at the surface and specified by the configuration of those dopants, and that the TME is qualitatively less robust against disorder than the integer quantum Hall effect.
%\textcolor{red}{In particular, in a toy model of thin films with magnetic dopants at the surface, the magnetic field dependent electric dipole moment arises entirely from the $\mathcal{N}=0$ Landau level subspace and vanishes when there is no dopants. 
%In a more realistic model of thick films, the electric dipole moment reduces to that of a Su-Schrieffer-Heeger model and is quantized in accordance with 
%the topological properties of the starting 3D model, which we explicitly demonstrate via lattice regularization.}
%Supported by semi-analytic bulk calculations, we speculate that this magnetic-dopant qualification in non-magnetic TIs can manifest in %the electric-field-induced bulk orbital magnetization. 
\end{abstract}

\maketitle

\section{Introduction}
One of Peierls' {\it Surprises in Theoretical Physics} \cite{peierls1979surprises} is that the orbital magnetization of metals can be correctly calculated using infinite lattice models that neglect the surfaces of realistic material samples, even though that magnetization can be understood to arise
from bound currents at those surfaces.
This mysterious success is now often taken for granted. 
In recent times, strong interest in the multitude of topological insulators (TIs) and the linear response thereof
has highlighted similar issues, in particular when disentangling the roles played by the topology ascribed to the bulk electronic ground state and by topologically protected electronic surface states.
The integer quantum Hall effect (IQHE) provides a familiar example. 
In a two-dimensional (2D) crystalline band insulator the IQHE occurs when the vector bundle of occupied electronic Bloch states over the Brillouin zone (BZ) torus is characterized by a nonzero Chern invariant \cite{TKNN1982,Simon1983,KOHMOTO1985}. 
Meanwhile, experiments are 
often interpreted in terms of the gapless chiral edge states \cite{halperin1982quantized,MacDonald1984,buttiker1988absence,thouless1993edge,tong2016lectures} whose presence is more immediately related to charge conduction. 
In the IQHE, it is generally accepted that this bulk topology of a band insulator
implies a quantized bulk Hall conductivity but also implies 
the existence of chiral edge states that yield a consistent conductance in 
finite-size samples thereof, unifying the two interpretations. 
Notably, breaking time-reversal symmetry (TRS) is necessary for a nonzero Chern invariant and, by the usual symmetry arguments, for a nonzero Hall conductivity in finite-sized systems.
The topological magnetoelectric effect (TME) presents an even more stark puzzle.
%The topological magnetoelectric effect (TME) refers to the linear response of the electronic orbital magnetization (polarization) to a uniform dc electric (magnetic) field in three-dimensional (3D) band insulators 
%that exhibit time-reversal symmetry (TRS) and presents an even more stark puzzle.
In three-dimensional (3D) crystalline band insulators, TRS leads to a $\mathbb{Z}_{2}$ topological classification of the 
electronic ground state and in $\mathbb{Z}_{2}$-odd phases
to a magnetoelectric linear response coefficient that is argued \cite{Zhang2008,Vanderbilt2009} to equal $(n+1/2) e^2/hc$ for $n\in\mathbb{Z}$. 
On the other hand, in any finite-sized system with TRS, the usual symmetry arguments dictate that the magnetoelectric coefficient must vanish.
Indeed, the requirement of magnetic surface dopants for realization
of the TME has been noted previously \cite{Zhang2008,Vanderbilt2009,Rosenow2013,Burnell2013,Witten2016}. 
There is no physical bulk response directly related to the 3D $\mathbb{Z}_{2}$ invariant, but a non-trivial $\mathbb{Z}_{2}$
invariant implies the existence of surface states that can then be gapped to activate the TME.
This more convoluted connection has implications for the experimental 
robustness of the effect, making it qualitatively less robust against disorder compared with the 
IQHE.

\begin{figure}[t!]
    \centering
    \includegraphics[width=0.4\textwidth]{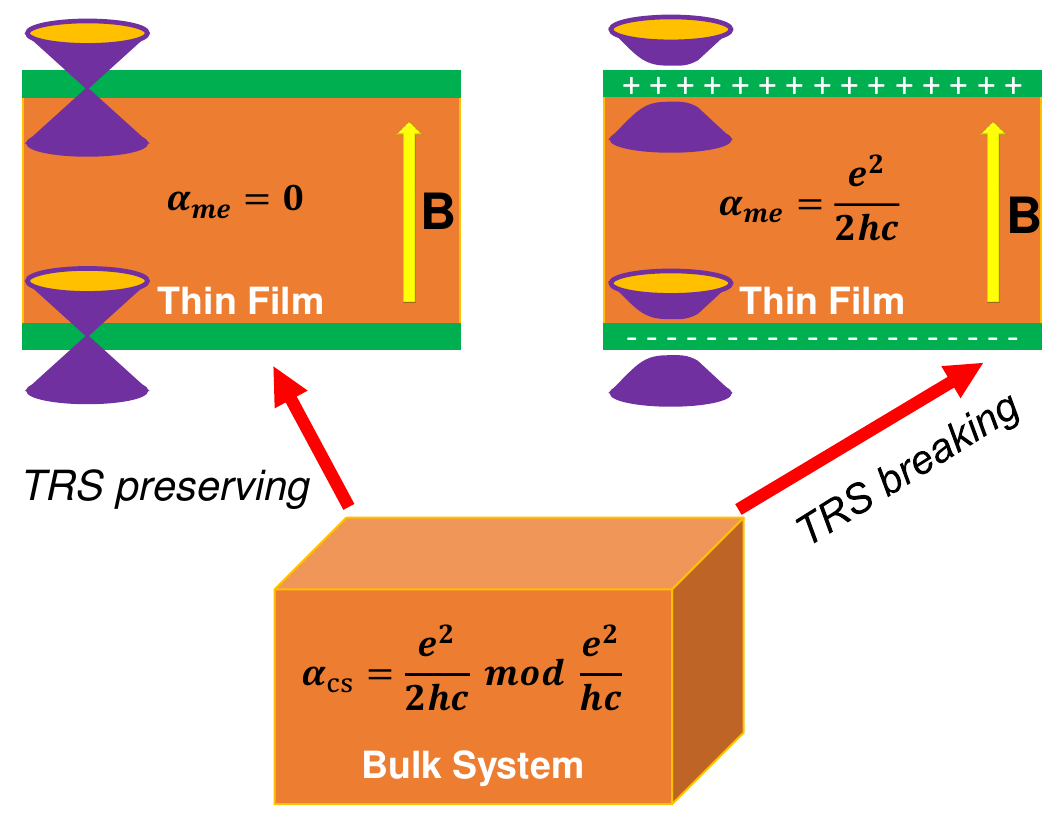}
    \caption{Relationship between $\alpha_{\text{CS}}$ in bulk insulators that exhibit time-reversal symmetry (TRS) and $\alpha_{\text{me}}$ in thin films thereof. In the bulk, TRS leads to a $\mathbb{Z}_2$ topological classification of the electronic ground state, to which $\alpha_{\text{CS}}$ is sensitive. In three-dimensional $\mathbb{Z}_2$-odd phases, $\alpha_{\text{CS}}$ is quantized at $e^2/2 h c$ up to an integer multiple of $e^2/h c$. In non-magnetic thin films, however, the physically realized magnetoelectric response $\alpha_{\text{me}}$ is nonzero only when the energy dispersion of the topologically protected surface states is everywhere gapped
    by magnetic surface dopants which break TRS locally.}
    \label{fig:lattice_model}
\end{figure}

It seems therefore that the magnetoelectric response of non-magnetic $\mathbb{Z}_{2}$ TIs is a rare violation of the Peierls principle referred to above, 
\textit{i.e.}~that evaluating a quantity using a band theoretic description
does not always correctly produce the value of that quantity in a large finite-sized sample thereof (see Fig.~\ref{fig:lattice_model}). 
In this paper we explicitly address the relationship between the magnetoelectric
linear response coefficient of a bulk insulator that exhibits TRS
and that of a finite size sample counterpart. 
As always, the bulk crystal is merely a convenient theoretical construct. Our interest is in understanding when it yields the physically correct magnetoelectric response.
In Sec.~\ref{Sec:GeneralLinearResponse} we describe this conundrum in more detail and highlight the main results of the
calculations that follow.
We argue for a slightly different interpretation of the 
%finite size implications of the 
bulk magnetoelectric coefficient derived in 
previous work \cite{Essin2010}, which is informed by the relationship (see Fig.~\ref{fig:lattice_model})
between particular linear response coefficients in finite-sized samples and bulk that we explicitly establish in Sec.~\ref{Sec:II} and \ref{Sec:III}. 
Since the magnetoelectric coefficient lacks physical significance in macroscopically uniform bulk insulators that exhibit TRS, the bulk interpretation that we propose implicitly involves the consideration of surfaces, a seemingly unavoidable feature.
In this interpretation the magnetoelectric coefficient adheres to the Peierls principle and we reach a conclusion that is 
manifestly consistent with TRS; in non-magnetic $\mathbb{Z}_{2}$ TIs, magnetoelectric response occurs only when the surface states are gapped by magnetic dopants, the magnetoelectric coefficient is quantized only
when the surface magnetization configuration satisfies stringent conditions, 
and currents proportional to the magnetic field
can flow through the material when the
surface magnetization profile is changed.

In Sec.~\ref{Sec:II} we extract the magnetoelectric coefficient for thin film and bulk samples of 
non-magnetic $\mathbb{Z}_2$ TIs in the V$_2$VI$_3$ family of materials
by directly including a uniform dc magnetic field in a coupled-Dirac cone model of the low-energy electronic states,
then computing the polarization of the occupied energy eigenfunctions.
In the case of thin films, we find that when the Dirac cones associated with the top and bottom surfaces 
are gapped by magnetic dopants of opposite magnetization, the magnetoelectric coefficient $\alpha_{\text{me}}$ is nonzero and approaches a quantized value as the film thickness is increased.
The induced polarization is sensitive to the magnetic dopant configuration and $\alpha_{\text{me}}$ moves continuously through zero to $-\alpha_{\text{me}}$ as the component of magnetization perpendicular to the surface is continuously taken to its opposite value.
When there is no surface magnetization, there is no magnetoelectric response.
We show that the quantization of $\alpha_{\text{me}}$ in the thick film limit can be understood in terms of the topological properties of a Su-Schrieffer-Heeger model that arises in the bulk description.

In Sec.~\ref{Sec:III} we employ the 3D tight-binding model that we obtain from a lattice regularization 
of the coupled-Dirac cone model used in Sec.~\ref{Sec:II} 
and calculate $\alpha_{\text{CS}}$, the purported magnetoelectric coefficient in crystalline band insulators that exhibit TRS, semi-analytically using the well-known bulk linear response expression; we reproduce the expected quantization.
Calculating $\alpha_{\text{CS}}$ is technically challenging because it must be evaluated with respect to
a smooth global gauge of the vector bundle of occupied Bloch states over the 3D BZ. 
%We leverage the time-reversal and inversion symmetries of the model to perform
%a semi-analytical calculation of $\alpha_{\text{CS}}$ in a smooth Hamiltonian gauge and reproduce the expected quantization.
%Although it is typically convenient to choose a Hamiltonian gauge for susceptibility calculations, in a $\mathbb{Z}_2$ TI there exists no such gauge that is smooth over the $\text{BZ}_{\text{3D}}$ unless there is some combination of symmetries that results in globally degenerate energy bands. 
Generically, Bloch energy eigenvectors that are smooth over the entire $\text{BZ}$ do not exist in a $\mathbb{Z}_2$ TI \cite{Monaco2017} and a Wannierization-like procedure \cite{Marzari2012} is required to obtain an adequate gauge choice \footnote{See, e.g., Essin \textit{et al.}~\cite{Vanderbilt2009}.}. 
Fortunately, the model we employ exhibits a fermionic time-reversal symmetry and an inversion symmetry, thus each eneregy band is at least two-fold degenerate over the $\text{BZ}$. 
In fact, the energy bands are everywhere pairwise isolated. Thus there exists a smooth global Hamiltonian gauge, and with respect to such a gauge we make our calculations.
In Sec.~\ref{Sec:IV} we analytically demonstrate that, in a particular smooth global Hamiltonian gauge, $\alpha_{\text{CS}}$ is entirely attributed to the itinerant portion of the electric-field-induced orbital magnetization, and speculate that this is a universal feature unique to non-magnetic $\mathbb{Z}_{2}$ TIs. 
Our calculations explicitly demonstrate that for $\mathbb{Z}_{2}$ TIs,
the bulk theory of magnetoelectric response does not always capture the properties of large finite size samples.

 \section{Linear Response Theory}
\label{Sec:GeneralLinearResponse}
The interaction between macroscopic electromagnetic fields and the charged constituents of
material media is often described at the level of linear response by introducing phenomenological 
susceptibility tensors that are nonlocal in space and time, and relate changes in the (macroscopic) charge $\varrho (\boldsymbol{r},t)$ and current $\boldsymbol{J} (\boldsymbol{r},t)$ densities of the material to the applied electric $\boldsymbol{E}(\boldsymbol{r},t)$ and magnetic $\boldsymbol{B}(\boldsymbol{r},t)$ Maxwell fields that induce those changes. The charge and current densities and the Maxwell fields are understood to be coarse grained \footnote{See, e.g., Chapter 2 of Swiecicki \cite{SwiecickiThesis} and references therein.} over a length scale 
intermediate between the atomic spacing and the wavelength of light; indeed local field corrections, which can be important, are not of interest here and are neglected.
It can often be assumed that a bulk material is spatially uniform at that intermediate length scale, in which case the susceptibility tensors are translationally invariant.
Then, for example, the linear response of $\boldsymbol{J} (\boldsymbol{r},t)$ is of the form
    \begin{align}
	J^{i(1)}(\boldsymbol{q},\omega)=\sigma^{il}(\boldsymbol{q},\omega)E^{l}(\boldsymbol{q},\omega),
	\label{current}
    \end{align}
where $\sigma^{il}(\boldsymbol{q},\omega)$ is the effective conductivity tensor, $\boldsymbol{q}$ and $\omega$ are respectively the wave vectors and frequencies that arise in the Fourier transforms of $\boldsymbol{E}(\boldsymbol{r},t)$ and $\boldsymbol{B}(\boldsymbol{r},t)$, and 
superscript indices here and below identify Cartesian components that are summed over when repeated. If $\boldsymbol{E}(\boldsymbol{r},t)$ and $\boldsymbol{B}(\boldsymbol{r},t)$ only involve wavelengths in the optical regime, then $\sigma^{il}(\boldsymbol{q},\omega)$ can be expanded in powers of $\boldsymbol{q}$,
    \begin{align}
        \sigma^{il}(\boldsymbol{q},\omega)=\sigma^{il}(\omega)+\sigma^{ilj}(\omega)\,q^{j}+\ldots,
	\label{conductivity}
    \end{align} 
where $\sigma^{il}(\omega)\equiv\sigma^{il}(\boldsymbol{0},\omega)$ and $\sigma^{ilj}(\omega)\equiv(\partial\sigma^{il}(\boldsymbol{q},\omega)/\partial q^{j})|_{\boldsymbol{q}=\boldsymbol{0}}$. 
Using Eq.~(\ref{conductivity}) in (\ref{current}) and performing a Fourier transform to coordinate space yields
    \begin{align}
        J^{i(1)} (\boldsymbol{r},\omega)=\sigma^{il}(\omega)E^{l} (\boldsymbol{r},\omega)-i\sigma^{ilj}(\omega)\,\frac{\partial E^{l} (\boldsymbol{r},\omega)}{\partial r^{j}}+\ldots.
        \label{eq:conductivity_qexpansion}
    \end{align}
The first term in
Eq.~(\ref{eq:conductivity_qexpansion}) is the familiar long-wavelength frequency-dependent conductivity tensor.
Using Faraday's law, the second term can be shown to include contributions that involve
$\boldsymbol{B} (\boldsymbol{r},\omega)$ and symmeterized spatial derivatives of $\boldsymbol{E} (\boldsymbol{r},\omega)$.
Any susceptibility in a material that is uniform at the course grained length scale can be analyzed in this way.
	
Physical insight into the distribution and dynamics of the charged (quasi-)particles that constitute a material medium can be gleaned by identifying bound and free contributions to $\varrho(\boldsymbol{r},t)$ and $\boldsymbol{J}(\boldsymbol{r},t)$, and associating the former with 
macroscopic polarization $\boldsymbol{\mathscr{P}} (\boldsymbol{r},t)$ and (orbital) magnetization $\boldsymbol{\mathscr{M}} (\boldsymbol{r},t)$ fields \footnote{See, e.g., Chapter 6.7 of Jackson \cite{Jackson} or Chapter 2 of Swiecicki \cite{SwiecickiThesis}.} such that
    \begin{align}
	\varrho (\boldsymbol{r},t)&=-\boldsymbol{\nabla}\cdot\mathscr{\boldsymbol{P}} (\boldsymbol{r},t)+\varrho_{F}   (\boldsymbol{r},t),\nonumber\\
	\boldsymbol{J} (\boldsymbol{r},t)&=\frac{\partial\boldsymbol{\mathscr{P}} (\boldsymbol{r},t)}{\partial t}+c\boldsymbol{\nabla}\cross\boldsymbol{\mathscr{M}} (\boldsymbol{r},t)+\boldsymbol{J}_{F} (\boldsymbol{r},t).\label{PM}
    \end{align}
For a given $\varrho(\boldsymbol{r},t)$ and $\boldsymbol{J}(\boldsymbol{r},t)$ there is always ambiguity in defining
$\boldsymbol{\mathscr{P}} (\boldsymbol{r},t)$, $\boldsymbol{\mathscr{M}} (\boldsymbol{r},t)$, $\varrho_{F}(\boldsymbol{r},t)$, $\boldsymbol{J}_{F} (\boldsymbol{r},t)$ that satisfy Eq.~(\ref{PM}).
Nevertheless, let us assume that definitions for $\boldsymbol{\mathscr{P}}(\boldsymbol{r},t)$ and $\boldsymbol{\mathscr{M}}(\boldsymbol{r},t)$ in a bulk material have been made and that the applied Maxwell fields are in the linear response regime so that expansions of the electric and magnetic dipole moments 
\footnote{In writing Eq.~(\ref{PM}), $\boldsymbol{\mathscr{P}} (\boldsymbol{r},t)$ and $\boldsymbol{\mathscr{M}} (\boldsymbol{r},t)$ can be understood as an infinite sum of electric and magnetic multipole moments (see, e.g., Chapter 6.7 of Jackson \cite{Jackson} or Chapter 2 of Swiecicki \cite{SwiecickiThesis}). It is often taken implicitly that it is the spontaneous electric and magnetic dipole moments and their linear response that is of primary interest. Of course, this is not always true. In fact, the electric quadrupole response contributes to the linear response of charge and current densities at first-order in $\boldsymbol{q}$, just like the magnetoelectric response. Nevertheless, in this work we will not explicitly consider response beyond that of the dipole moments.}
as sums of spontaneous and field-dependent contributions,
	\begin{align}
		P^{i} (\boldsymbol{r},\omega)&=P^{i(0)}+\chi_{E}^{il}(\omega)E^{l} (\boldsymbol{r},\omega)+\alpha_{P}^{il}(\omega)B^{l} (\boldsymbol{r},\omega)+\ldots,\nonumber\\
		M^{i} (\boldsymbol{r},\omega)&=M^{i(0)}+\alpha^{il}_{M}(\omega)E^{l} (\boldsymbol{r},\omega)+\ldots,
		\label{PMpert}
	\end{align}
are justified \footnote{Here ``$\ldots$'' denote other contributions to the linear response of the electric and magnetic dipole moments. For example, the linear response of $\boldsymbol{P}$ to the symmeterized spatial derivative of $\boldsymbol{E}$ and of $\boldsymbol{M}$ to $\boldsymbol{B}$ are contained in ``$\ldots$''.}.
	
At low frequencies $\omega$, $\alpha^{il}_{P}(\omega)$ and $\alpha^{il}_{M}(\omega)$ are well approximated by their static $\omega=0$ values, which are in fact related by a thermodynamic Maxwell relation 
	\begin{align}
		\alpha^{il}_{P}(\omega=0)=\alpha^{li}_{M}(\omega=0)\equiv\alpha^{il}.
		\label{alpha}
	\end{align}
In this case, using Eq.~(\ref{PMpert}) in (\ref{PM}) and comparing with Eq.~(\ref{conductivity}) yields
	\begin{align}
		\sigma^{ilj}&=-ic\alpha^{ik}\epsilon^{kjl}+ic\epsilon^{ijb}\alpha^{bl}+\ldots,
		\label{conductivity1}
	\end{align}
where ``$\ldots$'' denotes purely electric contributions to $\sigma^{ilj}$.
%Under these conditions, (\ref{conductivity1}) is only sensitive to the traceless part of $\alpha^{il}$.

Explicit forms for phenomenological susceptibilities can be obtained from quantum linear response theory. 
The semi-classical Hamiltonian of the coupled light-matter system
consists of electron and Maxwell contributions, and an interaction term that results from minimal coupling. 
The latter involves the electromagnetic scalar and vector potentials multiplied by charge and current density operators that are 
obtained from components of the Noether 4-current of the electron theory. 
In a finite sized material those charge and current densities are nonzero only in a localized region of space, in which case there exists well-motivated definitions for polarization and magnetization fields developed by Power, Zinau and Wooley (PZW) \footnote{See, e.g., Ref.~\cite{PZW} and references therein.}, Healy \footnote{See, e.g., Ref.~\cite{Healybook} and references therein.}, and others.
This approach begins with a unitary transformation of the minimal-coupling Hamiltonian (differential operator) to yield the physically equivalent PZW Hamiltonian. 
After defining the polarization and magnetization fields, 
the interaction of the electronic degrees of freedom with the electromagnetic field as described in the PZW Hamiltonian involves 
terms in which the polarization and magnetization fields are multiplied by $\boldsymbol{E} (\boldsymbol{r},t)$ and $\boldsymbol{B} (\boldsymbol{r},t)$, respectively \footnote{For an overview of this procedure, see, e.g., Chapter 2 of Swiecicki \cite{SwiecickiThesis}.}; notably this formulation does not require an explicit choice of the electromagnetic gauge. 
When a multipole expansion of those polarization and magnetization fields is made, the PZW Hamiltonian takes the classically anticipated form of a multipole Hamiltonian \footnote{Compare, e.g., Eqs.~(4.24), (5.72) of Jackson \cite{Jackson} with Eq.~(2.86) of Swiecicki \cite{SwiecickiThesis}.}.
Thus, the PZW definitions for polarization and magnetization fields are physically reasonable, and from them the (minimally-coupled) electronic charge and current density expectation values, and thus the finite size sample analog of the susceptibility tensors mentioned above (which are position dependent), can be rigorously obtained.
Moreover, that multipole expansion results in Hermitian operators in the electronic Hilbert space; that is, the electric and magnetic multipole moments are genuine physical observables.
For example, the ground state expectation value of the electronic contribution to the PZW electric dipole moment per unit volume (which is equal to the interior polarization of the material) is
	\begin{equation}
		\boldsymbol{P}^{(0)}_{\text{el}}=-\frac{e}{\Omega} \sum_{E < E_F} \int\boldsymbol{r}|\Psi_{E}(\boldsymbol{r})|^2d\boldsymbol{r},
		\label{Pfinite}
	\end{equation}
where $e>0$ is the elementary charge, $\Omega$ is the finite volume of the sample, $E_F$ is the Fermi energy, and $\Psi_{E}(\boldsymbol{r})$ are the electronic energy eigenfunctions of the unperturbed Hamiltonian \footnote{In Sec.~\ref{Sec:II} we do not consider the static magnetic field as a perturbation and therefore $\boldsymbol{P}^{(0)}_{\text{el}}$ will there involve $\boldsymbol{B}$.}.
	
Unfortunately the PZW approach cannot be directly applied to crystalline solids since, even in the absence of electromagnetic fields, the electronic charge and current density expectation values are expressed in terms of Bloch energy eigenfunctions, the support of which is all of space \footnote{In deriving the PZW Hamiltonian, it is assumed that the charge and current densities do not extend to infinity so integrals involving these quantities and spatial derivatives may be integrated by parts and the surface terms can be taken to vanish.}.
Indeed, this is related to the property that the usual position operator is not well-defined in the Hilbert space of Bloch functions \cite{Resta1998}.
There are a variety of different approaches \cite{Vanderbilt1993,Resta1998,Resta2005,Resta2006,Hughes2018,Mahon2019} that can be used to extend the notions of polarization and magnetization to bulk crystals. 
Inspired by the PZW approach, Sipe \textit{et al.}~have developed a formalism \cite{Swiecicki2013,Swiecicki2014,Mahon2019} applicable to general extended systems, crystalline or otherwise, which has been employed to account for spatial variation of electromagnetic fields in crystal insulators \cite{Mahon2020,Mahon2020a}, and more \cite{Mahon2023_CI,Mahon2023_Metals,Duff2022,Duff2023,Kattan2023}. 
However, none of these bulk crystal approaches are able to define electric and magnetic multipole moments as 
genuine physical observables in the PZW sense.

The most commonly used approach to sidestep difficulties in defining the
electric and magnetic dipole moments in crystalline insulators is the so-called modern theories of polarization \cite{Vanderbilt1993,Resta1994} and (orbital) magnetization \cite{Resta2005,Resta2006}. 
Focusing on the former, one aims to deduce an electric dipole moment from calculation of the current density, rather than to propose a general definition for it.
To do so, the macroscopic free current density (appearing in the second of Eq.~(\ref{PM})) is assumed to vanish at linear response in typical insulators, which seems 
to be a sensible assignment, thus the current density is related only to the polarization and magnetization fields.
It is less obvious that this assignment is sensible in Chern insulators, but materials of that type are not considered here.
Next, focus is restricted to describing the influence of electric and magnetic fields 
that are spatially uniform, in which case all of the macroscopic densities appearing in Eq.~(\ref{PM}) are also expected to be uniform. 
Thus the polarization and magnetization fields are entirely described by their dipole moments, and $\boldsymbol{J}(t)=\partial\boldsymbol{P}(t)/\partial t$.

The magnetoelectric susceptibility $\alpha^{il}$ can generally be written as the sum of an isotropic $\delta^{il}\alpha_{\theta}$ 
and a traceless contribution \cite{Essin2010}. 
%The TME is associated with the isotropic contribution.
Examining Eq.~(\ref{conductivity1}) makes it evident that in a static and macroscopically uniform bulk material, 
the $\sigma^{ilj}$ tensor is insensitive to the former. 
That is, $\alpha_{\theta}$ cannot be determined by calculating the wave vector dependent optical conductivity.
In other words, the physical implications of $\alpha_{\theta}$ are not equivalent to those of a wave-vector dependent conductivity.

If the isotropic magnetoelectric response cannot be calculated from the bulk current response to non-uniform 
electric fields, how can it be found? Two independent strategies that can be used to obtain $\alpha_{\theta}$ have
been identified in the literature: (i) Essin \textit{et al.}~\cite{Essin2010} consider an insulating state of a Bloch Hamiltonian that is adiabatically varied in time, such that $\alpha^{il}$ becomes time-dependent and there is an additional contribution to Eq.~$(\ref{conductivity1})$ (originating from $\partial\boldsymbol{P}/\partial t$) proportional to $(\partial\alpha^{il}/\partial t)B^{l}$; 
(ii) Malashevich \textit{et al.}~\cite{Malashevich2010} consider a finite sized material, in which case $\alpha^{il}$ becomes position-dependent and there is an additional contribution to Eq.~$(\ref{conductivity1})$ (originating from $\boldsymbol{\nabla}\cross\boldsymbol{M}$) that 
is proportional to $(\epsilon^{iab}\partial\alpha^{bl}/\partial r^{a})E^{l}$. Both approaches reach the same conclusion, which is that in a crystalline insulator initially occupying its zero-tempetature electronic ground state
    \begin{align}
	\alpha^{il}=\alpha^{il}_{\text{G}}+\delta^{il}\alpha_{\text{CS}},
    \label{alpha+cs}
    \end{align}
where the explicit form of $\alpha^{il}_{\text{G}}$ resembles that of a conventional linear response tensor, involving cross-gap matrix elements between 
Bloch states that are initially occupied and unoccupied,
and $\alpha_{\text{CS}}$ (the Chern-Simons contribution) can be expressed in terms of 
occupied Bloch states alone. If the bulk insulator exhibits TRS then $\alpha^{il}=\delta^{il}\alpha_{\text{CS}}$ and is quantized since $\alpha_{\text{CS}}$ evaluates to an 
element of a discrete lattice of values \cite{Zhang2008}, which does not include zero in the case of $\mathbb{Z}_{2}$ TIs.
It is at the step that the conundrum mentioned in the Introduction is introduced, since 
this is where the conclusion is reached that $\alpha^{ii}\text{ mod } e^{2}/hc$ \footnote{For a discussion on the physical basis of the discrete ambiguity that is inherent to $\alpha^{ii}$, see, e.g., Chapter 6.2 of Vanderbilt \cite{VanderbiltBook}.} can be nonzero in insulators with TRS.
%If TRS is absent, then $\alpha_{\theta}$ can have a contribution from the conventional cross-gap term.
 
The strategy (i) of Essin \textit{et al.}~\cite{Essin2010} considers
a 3D crystalline insulator described by a Bloch Hamiltonian that varies in time via 
adiabatic variation of the crystal Hamiltonian. 
They show that in the presence of a static and uniform magnetic field $\boldsymbol{B}$, 
adiabatic linear response to slow changes in the crystal Hamiltonian 
induces a spatially uniform macroscopic (minimal-coupling) electronic current density of the form $\boldsymbol{J}^{(B)}(t)=\boldsymbol{J}_{\text{G}}^{(B)}(t)+\boldsymbol{J}_{\text{CS}}^{(B)}(t)$ 
\footnote{Importantly, $\boldsymbol{J}^{(B)}(t)$ is not the magnetic-field-induced electronic current in a material with unperturbed electrons governed by the Bloch Hamiltonian evaluated at $t$; this would be given by $\boldsymbol{J}_{\text{G}}^{(B)}(t)$. Instead, $\boldsymbol{J}^{(B)}(t)$ also involves Streda-type transport, which manifests through the time-dependence of the ground state density matrix and leads to $\boldsymbol{J}_{\text{CS}}^{(B)}(t)$ (see Eq.~(35b,35c) of Ref.~\cite{Essin2010}).}.
Unlike above, where the time dependence of $\boldsymbol{J}^{(1)}(t)$ resulted from that of the dynamical electromagnetic field, $\bm{B}$ is here taken to be static and the $t$-dependence results from that of the Bloch Hamiltonian itself; $t$-dependence manifests implicitly in the Bloch energy eigenvectors and eigenvalues.
Although their strategy applies more generally, we focus on the special case in which the Bloch Hamiltonian exhibits TRS at every instant $t$. Then $\boldsymbol{J}_{\text{G}}^{(B)}(t)=\boldsymbol{0}$ and \cite{Niu2009,Essin2010}
\begin{align}
    \boldsymbol{J}^{(B)}(t) = \boldsymbol{J}_{\text{CS}}^{(B)}(t) = -\frac{e^{2}\boldsymbol{B}}{2hc}\int_{\text{BZ}} \left(\frac{\partial}{\partial t} Q_{\text{3}}({\bm k};t)\right)d^{3}k .
    \label{eq:linearresponseA}
\end{align} 
The explicit expression for $Q_{\text{3}}(\bm{k};t)$, the Chern-Simons 3-form \footnote{See, e.g., Chapter 11 of Nakahara \cite{NakaharaBook}.} at instant $t$,
is discussed in Sec.~\ref{Sec:III}. 
It is important to note that $Q_{\text{3}}(\bm{k};t)$ explicitly involves only the Berry connection and curvature that are defined on the vector bundle of occupied Bloch states over the Brillouin zone torus at instant $t$.
That is, at every $t$, $Q_{\text{3}}(\bm{k};t)$ is a purely geometrical quantity.
In Ref.~\cite{Essin2010} the derivative with respect to time in Eq.~(\ref{eq:linearresponseA}) is taken outside of the Brillouin zone integral to obtain
\begin{align}
    \boldsymbol{J}^{(B)}(t) = \frac{\partial \boldsymbol{P}^{(B)}(t)}{\partial t} = \boldsymbol{B} \frac{\partial}{\partial t} \alpha_{\text{CS}}(t),
    \label{eq:linearresponseB}
\end{align}
%and $\boldsymbol{P}^{(1)}$ is identified with $\alpha_{\text{CS}}\boldsymbol{B}$, 
where 
\begin{equation}
 \alpha_{\text{CS}}(t) \equiv -\frac{e^{2}}{2hc}\int_{\text{BZ}} Q_{\text{3}}(\boldsymbol{k};t)d^{3}k.
 \label{alphaCSgeneral}
\end{equation}
At each instant $t$,
\begin{align}
\int_{\text{BZ}} Q_{\text{3}}(\boldsymbol{k};t)d^{3}k \in \mathbb{Z}
\end{align}
is determined (modulo 2) by the TRS-induced $\mathbb{Z}_{2}$ topological classification \cite{Zhang2008,Freed,Kaufmann_2016} of the electronic ground state \footnote{More precisely, TRS induces a topological classification of the vector bundle (for reference, see, e.g., Chapter 10 of Lee \cite{LeeBook} or Chapter 9.3 of Nakahara \cite{NakaharaBook}) that is naturally constructed using the Bloch states that are occupied in the ground state of a band insulator \cite{Panati2007,Brouder}. 
We adopt the common terminology that ascribes the topology of that vector bundle to the ground state itself. 
In this context, the $\mathbb{Z}_{2}$ topological classification of the ground state is most intuitively understood in terms of the types of gauge choices (frames) of that bundle. If there exists a smooth global TRS frame of that bundle then the ground state is called $\mathbb{Z}_{2}$-even, otherwise it is called $\mathbb{Z}_{2}$-odd.}
and the value of $\alpha_{\text{CS}}(t)$ is therefore
fixed (modulo $e^2/hc$) by that topology;
if the ground state is $\mathbb{Z}_2$-even then $\alpha_{\text{CS}}$ evaluates to $0\text{ mod }e^2/hc$ and if it is $\mathbb{Z}_2$-odd then $\alpha_{\text{CS}}$ evaluates to $e^2/2hc \text{ mod }e^2/hc$ \cite{Zhang2008,Vanderbilt2009}.
Importantly, $ Q_{\text{3}}(\boldsymbol{k};t)$ and indeed $\alpha_{\text{CS}}(t)$ are gauge dependent -- they are both sensitive to the smooth global frame of the vector bundle of occupied Bloch states over the Brillouin zone torus that is used for their evaluation -- which leads to the above described discrete ambiguity.
Therefore, even when considering instants $t_1$ and $t_2$ between which a band insulating ground state remains in the same topological class, $\alpha_{\text{CS}}(t_1)$ need not equal $\alpha_{\text{CS}}(t_2)$ since at each $t$ one has the freedom to choose one of many gauges and with respect to each of those gauge choices a different value of $\alpha_{\text{CS}}(t)\in(e^2/2hc)\mathbb{Z}$ can result; in this scenario it is only guaranteed that $\alpha_{\text{CS}}(t_1)\text{ mod }e^2/hc$ equals $\alpha_{\text{CS}}(t_2)\text{ mod }e^2/hc$.

To obtain Eq.~(\ref{eq:linearresponseA}) it must be assumed that the gauge choice used to express $Q_{3}(\bm{k};t)$ is smooth in $\boldsymbol{k}\in\text{BZ}$ for each $t$. 
And in order to move from Eq.~(\ref{eq:linearresponseA}) to (\ref{eq:linearresponseB}), that gauge choice must also be continuous in $t$ \footnote{For there to exist a smooth global gauge of the vector bundle of occupied Bloch states over $\text{BZ}$ at instant $t$ implies that the bulk band gap does not vanish at that $t$. Therefore, this assumption can only be satisfied if the bulk band gap is nonzero throughout the duration of time during which the crystalline parameters are varied.}.
The latter condition implies that $\alpha_{\text{CS}}(t)$ is continuous in $t$. 
If there is TRS at each $t$ then $\alpha_{\text{CS}}(t)\in(e^{2}/2hc)\mathbb{Z}$ and thus $\alpha_{\text{CS}}(t)$ must be constant in $t$.
%This is crucial in order for Eq.~(\ref{eq:linearresponseB}) to be physically reasonable,
%for if it were not so then $\boldsymbol{J}^{(B)}(t)$ and therefore $\int_{t_{i}}^{t_{f}}\boldsymbol{J}^{(B)}(t)dt$ could be gauge dependent (\textit{i.e.}~multivalued).
%Although this was previously claimed \cite{Essin2010} to be true, the assumption of a gauge choice that is continuous in $t$ was not recognized.
Indeed our explicit calculations in Sec.~\ref{Sec:III} demonstrate that once a gauge choice is made at any $t$, the value of $\alpha_{\text{CS}}(t)$ is fixed for all $t$ if that gauge is continuous in $t$. 
Then under any TRS preserving adiabatic variation of the Bloch Hamiltonian in time, the right hand side of Eq.~(\ref{eq:linearresponseB}) evaluates to zero; no bulk currents flow when the parameters of the Hamiltonian change within the space of a given topological phase and the adiabatic approximation is valid.
Indeed this result should be expected since Eq.~(\ref{eq:linearresponseB}) is consistent with TRS
only if nonzero values for $\partial\alpha_{\text{CS}}(t)/\partial t$ are activated by some time-reversal symmetry breaking perturbation.

If a Bloch Hamiltonian is adiabatically varied in $t$ without necessarily exhibiting TRS at each $t$, then $\alpha_{\text{CS}}(t)$ need not evaluate to an element of $(e^{2}/2hc)\mathbb{Z}$ for each $t$ \cite{Zhang2008} and therefore $\alpha_{\text{CS}}(t)$ is not necessarily constant in $t$. For example, if there is TRS at some initial $t_{i}$ and final $t_{f}$ times, then $\alpha_{\text{CS}}(t_{i}),\, \alpha_{\text{CS}}(t_{f})\in(e^{2}/2hc)\mathbb{Z}$ and if the electronic ground state at $t_{i}$ ($t_{f}$) is $\mathbb{Z}_{2}$-even ($\mathbb{Z}_{2}$-odd), then $\int _{t_{i}}^{t_{f}}\boldsymbol{J}^{(B)}(t)dt \text{ mod } e^{2}\bm{B}/hc$ equals $e^{2}\bm{B}/2hc$. 
Crucially, this ambiguity in $\int _{t_{i}}^{t_{f}}\boldsymbol{J}^{(B)}(t)dt$ is not a consequence of the gauge dependence of $\alpha_{\text{CS}}$, but rather a manifestation of the fact that the current pumped through the material depends on how the model parameters are changed in time. 
Indeed if the Bloch Hamiltonian evaluated at $t_{i}$ and at $t_{f}$ equal, then one can define a vector bundle of occupied electronic states over $\text{BZ}\times S^{1}$ and the topology of this structure is characterized by a second Chern invariant which determines $\int _{t_{i}}^{t_{f}}\boldsymbol{J}^{(B)}(t)dt$ \cite{Zhang2008,Souza2017_tmeSurface}.
A similar situation arises in the case of the unperturbed bulk 1D polarization, where a certain first Chern invariant determines the bound charge \cite{Vanderbilt1993_Bulk-Surface,VanderbiltBook}.

The linear response result (\ref{eq:linearresponseB}) for crystalline insulators that exhibit TRS has been interpreted \cite{Essin2010} by identifying the
magnetic-field-induced bulk polarization $\boldsymbol{P}^{(B)}(t)$ with $\alpha_{\text{CS}}(t)\boldsymbol{B}$. 
For a static insulator of interest, its bulk $\boldsymbol{P}^{(B)}$ is then 
obtained by evaluating $\alpha_{\text{CS}}(t)\boldsymbol{B}$ 
at any time $t$ at which the time-dependent Bloch Hamiltonian describes that insulator. 
If we accept this identification, then
$\boldsymbol{P}^{(B)}$ can differ from the physically realized electric dipole 
response in a large but finite sample \cite{Souza2011}.
For example, consider a finite crystallite of a non-magnetic $\mathbb{Z}_{2}$ TI that has TRS everywhere. 
TRS implies that there is no electric dipole induced by $\bm{B}$ in linear response.
The bulk expression for the polarization is evidently misleading in the absence of 
magnetic dopants somewhere.
Because this identification of $\boldsymbol{P}^{(B)}$ does not imply induced bulk charge or current 
densities it is still technically acceptable as a bulk quantity, 
but associating physical significance with it requires additional insight.
Indeed the necessity of magnetic surface dopants for the topological magnetoelectric effect to manifest in non-magnetic TIs has been noted \cite{Zhang2008,Vanderbilt2009,Souza2011,Zhang2015} previously. 
A ubiquitous aspect of bulk theories of polarization and orbital magnetization is the importance of interpretation \cite{VanderbiltBook}.
We believe it is useful to formulate an alternative, physically equivalent, perspective to that of Ref.~\cite{Souza2011}.
By combining the known topological constraints in bulk with the known symmetry constraints in finite sized systems, we produce below a
physically meaningful notion of $\boldsymbol{P}^{(B)}$ in bulk
insulators that exhibit TRS, one that adheres to the Peierls principle and implicitly accounts for additional surface-related criteria. 
To that end, the V$_{2}$VI$_{3}$ family of TIs are an ideal test bed on which to develop that notion.

We have argued above that even in the strategy (i) of Ref.~\cite{Essin2010}, wherein the Bloch Hamiltonian for a bulk insulator involves crystal parameters that depend on time, 
$\alpha_{\text{CS}}(t)\bm{B}$ lacks physical consequences if that Hamiltonian exhibits TRS at each $t$. 
Thus, in this setting any physically significant identification of $\boldsymbol{P}^{(B)}(t)$ must involve the consideration of surfaces.
As a result, any such identification will be \textit{ad hoc} in that one has to artificially include the role 
played by the surface in a proposed bulk quantity. 
This is an aesthetically unappealing, but a seemingly unavoidable, feature of \textit{any} 
physically meaningful identification of $\boldsymbol{P}^{(B)}(t)$ in bulk insulators with TRS.

\iffalse
A physically meaningful identification of $\boldsymbol{P}^{(B)}(t)$ could be obtained by 
integrating $\bm{J}^{(B)}(t)$ of Essin \textit{et al.}~in time, replacing the time-dependent Bloch Hamiltonian with one that describes a macroscopically large
but finite sized material with time-dependence only at its surfaces, where 
magnetic moments order to break 
TRS and open gaps in the Dirac-like energy dispersion of the surface states.
We will assume that magnetic dopants are present only at the surface of the material and that the effect of interactions that break TRS is local \cite{Pournaghavi2021}, such that the current within 
the bulk-like interior region (where TRS is maintained) is given by Eq.~(\ref{eq:linearresponseB}),
and near the boundary is controlled by the dopant configuration on the nearby surface.
\fi

\begin{figure}[t!]
    \centering
    \includegraphics[width=0.47\textwidth]{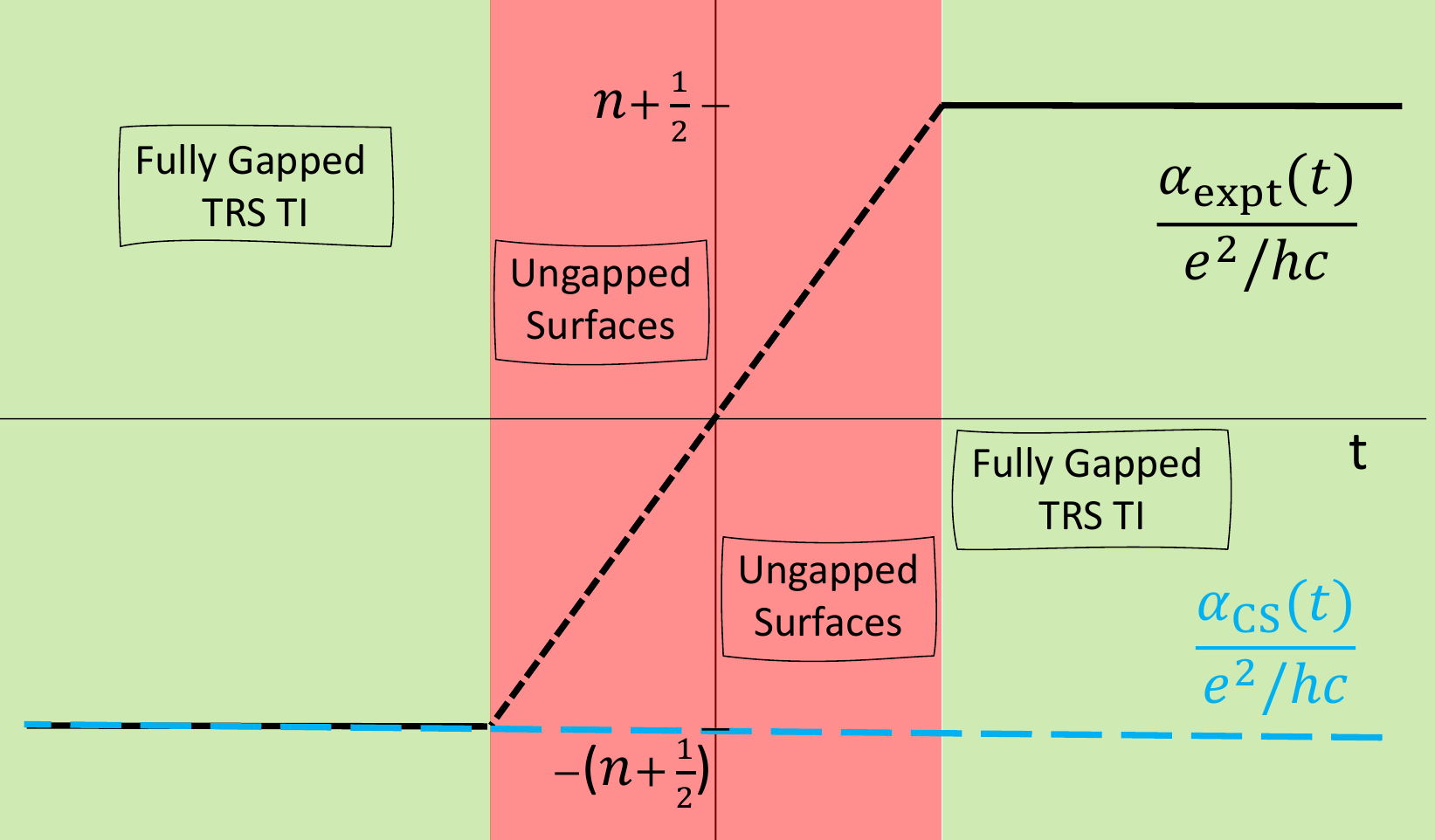}
    \caption{Relationship between the bulk magnetoelectric coefficient $\alpha_{\text{expt}}(t)$ (black line) that we propose
    and $\alpha_{\text{CS}}(t)$ (blue line) as obtained from a simple interpretation of the bulk current that is adiabatically induced by slow changes of a Bloch Hamiltonian describing a non-magnetic bulk $\mathbb{Z}_{2}$ TI.
    The illustrated time interval is imagined to be one over which the magnetization describing a configuration of surface magnetic dopants that initially gaps the energy dispersion of the surface states is changed to its opposite. 
    The red region, where this magnetization vanishes, is given finite width for illustrative purposes.
    We imagine making a particular gauge choice at some initial time such that $\alpha_{\text{CS}}(t)$ exactly agrees with $\alpha_{\text{expt}}(t)$ until a surface gap closes.
    A purely bulk linear response calculation \cite{Essin2010} uses the adiabatic approximation to show that $\alpha_{\text{CS}}(t)$ is time-independent and the (magnetic field dependent) bulk current 
    $\bm{J}^{(B)}(t)$ vanishes (no currents flow) when an instantaneous band insulating state is varied in a manner that maintains TRS.
    We argue that in finite samples $\int \bm{J}^{(B)}(t) dt$ is insensitive to 
    gapped surfaces and is therefore given correctly by the bulk linear response calculation 
    at early and late times within the illustrated time interval.
    The adiabatic approximation fails to describe finite samples when the surface gap vanishes, 
    in which case the physically realized $\bm{J}^{(B)}(t)$ is nonzero and the polarization can change. In practice, the sub-interval within which
    $\bm{J}^{(B)}(t) \ne \bm{0}$ will always be finite because of disorder at the surface. 
    We argue that the interior polarization at a particular point along the surface is a local quantity \cite{Zhang2008} that changes sign with the local orientation of magnetic dopants, 
    in agreement with TRS.}
    \label{fig:linearresponse}
\end{figure}

One such identification of $\boldsymbol{P}^{(B)}(t)$ could be obtained in a manner similar to that of Essin \textit{et al.}~\cite{Essin2010} --
by considering the magnetic field dependent current $\bm{J}^{(B)}(t)$ that flows in a material as its properties are changed -- but, rather than considering a time-dependent Bloch Hamiltonian, consider a Hamiltonian that describes a macroscopically large
but finite sized material with time-dependence only at its surfaces, where 
magnetic moments order to break TRS and open gaps in the Dirac-like energy dispersion of the surface states.
(TRS is maintained in the interior region.) 
If the effect of interactions that break TRS is local \cite{Pournaghavi2021}, then charge density response to $\bm{B}$ can only occur near the surfaces.
Our interpretation for non-magnetic $\mathbb{Z}_{2}$ TIs \footnote{In other bulk insulators that exhibit an effective TRS, such as anti-ferromagnetic $\mathbb{Z}_{2}$ TIs \cite{Moore2010}, that symmetry is typically broken merely by the existence of a surface. In that case the following is unnecessary since there is no conundrum related to TRS and magnetoelectric response, and the consideration of magnetic surface dopants seems unnecessary.},
which is supported by later calculations in this paper,
imagines replacing the bulk quantity $\alpha_{\text{CS}}(t)$ 
by a new function $\alpha_{\text{expt}}(t)$ that 
compares more directly to experiment \footnote{In principle, experiments only probe polarization differences (see, e.g., Chapter 4.4 of Vanderbilt \cite{VanderbiltBook}).} by implicitly accounting for the role of surface magnetic dopants.  The (physically meaningful) bulk polarization in these materials is  
\begin{equation}
    \boldsymbol{P}^{(B)}(t) = \alpha_{\text{expt}}(t) \bm{B},
    \label{eq:Pexpt}
\end{equation}
where $\alpha_{\text{expt}}(t)$ must have the following properties:
\begin{enumerate}
\item In the absence of magnetic surface dopants $\boldsymbol{J}^{(B)}(t)$ vanishes in finite samples, which is consistent with the bulk prediction (\ref{eq:linearresponseB}). In this case, TRS dictates that $\alpha_{\text{expt}}(t)\equiv 0$ and the identification of $\boldsymbol{P}^{(B)}(t)$ with $\alpha_{\text{CS}}(t)\boldsymbol{B}$ for $\mathbb{Z}_{2}$ TI materials is incorrect. 
\item (green regions in Fig.~\ref{fig:linearresponse}) 
During time intervals over which the magnetic dopant configuration on top and bottom surfaces is opposite and such that the energy dispersion of the surface states remains gapped, 
the physically realized response is that of the bulk; the prediction (\ref{eq:linearresponseB}) that $\boldsymbol{J}^{(B)}(t)$ vanishes and the identification of $\boldsymbol{P}^{(B)}(t)$ with $\alpha_{\text{CS}}(t)\boldsymbol{B}$ are correct (the latter modulo $e^{2}\bm{B}/hc$), {\it i.e.}~$\alpha_{\text{expt}}(t)=\alpha_{\text{CS}}(t)$.
$\boldsymbol{P}^{(B)}(t)$ is constant in $t$ as long as the surface states remain gapped. 
\item (red region in Fig.~\ref{fig:linearresponse}) During time intervals over which the magnetic dopant configuration is such that there is a finite surface
density-of-states, a current proportional to $\bm{B}$ flows through the material. %even though the interior always exhibits TRS. 
Indeed, the bulk prediction (\ref{eq:linearresponseB}) using the adiabatic approximation fails to describe these
finite size samples and $\alpha_{\text{expt}}(t)\neq \alpha_{\text{CS}}(t)$.
That current, and the pattern of polarization that develops, is sensitive to the magnetic configuration of the surface dopants.
For realistic 
samples with potential and magnetic disorder, the surface gaps will remain closed over a finite range of 
dopant configurations. Under surface magnetization reversal, currents proportional to $\bm{B}$ will flow through the material during the time interval
over which there is no surface energy gap. $\alpha_{\text{expt}}(t)$ now has a contribution that is sensitive to the surface magnetization, 
which allows the polarization to interpolate between the topologically allowed values of gapped states.
\iffalse
\item (red region in Fig.~\ref{fig:linearresponse}) During time intervals over which the magnetic dopant configuration is such that there is a finite surface
density-of-states, the form of Eq.~(\ref{eq:linearresponseB}) remains valid but the adiabatic approximation fails to describe 
finite size samples and Eq.~(\ref{alphaCSgeneral}) is no longer correct, 
\textit{i.e.}~$\alpha_{\text{expt}}(t)\neq \alpha_{\text{CS}}(t)$. The current that flows in the presence of a magnetic field and the 
pattern of polarization that develops is sensitive to the magnetic configuration of the surface dopants. For realistic 
samples with potential and magnetic disorder, \textcolor{red}{the surface gaps will remain closed} over a finite range of 
dopant configurations. Under surface magnetization reversal, interior currents will flow during the time interval
over which there is no surface energy gap. $\alpha_{\text{expt}}(t)$ in Eq.~(\ref{eq:Pexpt}) now has a contribution that is sensitive to the surface magnetization, 
which allows the polarization to interpolate between the topologically allowed values of gapped states.
\fi
\end{enumerate}

Our interpretation is that during time intervals $t\in[t_{i},t_{f}]$ over which the surface 
magnetization is reversed, the value of $\boldsymbol{P}^{(B)}(t)$ changes and these changes 
are associated with the flow of currents through the bulk of the material.
Varying the configuration of magnetic surface dopants in such a way that the total magnetization always vanishes but that surface gaps close then reopen 
induces bulk currents in much the same way that varying crystal parameters such that at intermediary times TRS is broken in the bulk 
\footnote{See, e.g., the discussion at the beginning of Sec.~3 of Ref.~\cite{Essin2010} or Chapter 6.4.2 of Ref.~\cite{VanderbiltBook}. If the crystalline parameters are adiabatically varied such that the infinite material remains macroscopically uniform, then $\partial \alpha^{ii}/\partial t$ could be nonzero throughout the material. Indeed, as described above, if there is TRS at $t_{i}$ and $t_{f}$, and if the adiabatic approximation applies, then $\partial \alpha^{ii}/\partial t$ equals $\partial \alpha_{\text{CS}}/\partial t$, which can only be nonzero if TRS is broken at intermediary times.} 
does; notably, neither mechanism yields a bulk charge density.
The change in polarization may be obtained by integrating $(\partial \alpha_{\text{expt}}(t)/\partial t)\bm{B}$ over time, which, due to TRS in the bulk, will integrate to the difference 
\footnote{This assumes non-pathological variation of the magnetization profile in time such that the jumps of $|\boldsymbol{P}^{(B)}(t)|$ do not resemble, for example, a Cantor ternary function. If it did, then even though $\boldsymbol{P}^{(B)}(t)$ is continuous almost everywhere (except possibly on a set of measure zero) the integral of the derivative of $\boldsymbol{P}^{(B)}(t)$ is not equal to the difference of $\boldsymbol{P}^{(B)}(t)$ evaluated at the endpoints.}
between two quantized values of $\alpha_{\text{CS}}(t)$ that differ in sign.
Since in our interpretation we assume that the TRS-induced bulk topology is unaffected by the manipulation of the magnetic surface dopants, 
$\alpha_{\text{expt}}(t_{f})-\alpha_{\text{expt}}(t_{i})\in(e^{2}/hc)\mathbb{Z}$.
If the magnetization of the dopants at $t_{f}$ is opposite to that at $t_{i}$, we require $\alpha_{\text{expt}}(t_{f})=-\alpha_{\text{expt}}(t_{i})$.

Under this identification, the magnetoelectric response in non-magnetic $\mathbb{Z}_{2}$ TIs is nonzero only when magnetic surface dopants are present, solving the TRS conundrum. 
If there is no surface magnetization, then there will never be a surface energy gap everywhere and no bulk current flows 
in linear response to a magnetic field. In the following sections of this paper we describe 
electronic polarization (orbital magnetization) calculations in model non-magnetic $\mathbb{Z}_{2}$ TIs subject to a uniform dc magnetic (electric) field, 
which support this interpretation. In particular, in sufficiently thick films the polarization changes and interior currents flow
only when the surface-normal magnetization component is varied through zero.

\section{Magnetoelectric response in \\ $\mathbb{Z}_2$ topological insulator thin films}
\label{Sec:II}
In this section we employ a simplified but realistic model of the electronic states near the Fermi energy in thin films of V$_2$VI$_3$-type non-magnetic $\mathbb{Z}_2$ TIs in which it is possible to 
account for the presence or absence of magnetic surface dopants. We calculate
the magnetoelectric coefficient $\alpha_{\text{me}}$ by introducing a magnetic field that is perpendicular to the thin film and accounting for the Landau quantization it produces. We begin
in Sec.~\ref{Sec:IIa} by considering a toy version of this model in which
the only electronic states retained are the topologically protected surface states modeled by two coupled 2D Dirac cones, one associated with each surface. The surface magnetic dopants couple to these Dirac-like states separately via an exchange mass $m$, and for finite thickness films those states are hybridized by an inter-surface tunneling parameter $\Delta$. This model is amenable to analytic analysis, from which it is found that in the limit $m\rightarrow0$ $\alpha_{\text{me}}\rightarrow0$, as expected for finite-sized material samples with TRS, and in the limit $\Delta\rightarrow0$ $\alpha_{\text{me}}\rightarrow e^{2}/2hc$, as expected for bulk $\mathbb{Z}_2$ TIs. In Sec.~\ref{Sec:IIb} we consider a more realistic model, which involves many coupled Dirac cones, and reach the same conclusions.
We show that the quantization of $\alpha_{\text{me}}$ in the thick film limit can be understood in terms of the topological properties of a Su-Schrieffer-Heeger model that arises in the bulk description.
	
\subsection{Simplified toy model of the surface states}
\label{Sec:IIa}
The analytic calculations in this section illustrate the essential microscopic physics of 
magnetic-field-induced charge redistribution in finite thickness thin-films of non-magnetic $\mathbb{Z}_2$ TIs. 
In materials of this type, the electronic states that are nearest to the Fermi energy occur about the $\boldsymbol{k}_{\Gamma}$ point $(k_{\Gamma,x},k_{\Gamma,y})=(0,0)$ and are localized at the surface of the sample \cite{Mele2007,Zhang2009}.
Indeed the bulk topology of a $\mathbb{Z}_2$ TI implies the existence of an odd number of Dirac points at each surface of a finite sample thereof \cite{Kane2010,Qi2011}.
A model of a thin film can therefore be constructed
using two copies of a 2D $\boldsymbol{k}\cdot\boldsymbol{p}$ Dirac Hamiltonian, one associated with each surface.
We account for the (finite) film thickness in the surface-normal direction (taken to be $\boldsymbol{z}$) by hand by endowing the basis states of each copy of the $\boldsymbol{k}\cdot\boldsymbol{p}$ Hamiltonian with $z$-dependence to encode its association with a particular surface. 

We therefore consider two Dirac Hamiltonians written in a basis of states $\ket{\bar{u}_{(\alpha,\sigma),\boldsymbol{k}=\boldsymbol{k}_{\Gamma}}}$ that are orthogonal linear combinations of the $\boldsymbol{k}\cdot\boldsymbol{p}$ basis states and are associated with the bottom $(\alpha=0)$ and top $(\alpha=1)$ surfaces of the film. At $\boldsymbol{k}_{\Gamma}$, the $\boldsymbol{k}\cdot\boldsymbol{p}$ Hamiltonian will commute with $s_{z}$, thus the corresponding spin-up and spin-down eigenvalues $\sigma=\pm\hbar/2$ ($\equiv$ $\uparrow$, $\downarrow$) can be used to identify Hilbert space basis states at $\boldsymbol{k}_{\Gamma}$ (and therefore label basis states of any $\boldsymbol{k}\cdot\boldsymbol{p}$ Hamiltonian about $\boldsymbol{k}_{\Gamma}$). We include an exchange coupling $m_{\alpha}$ associated with each surface, here describing the effect of magnetic dopants at that surface, that couples to the $s_z$ spin components of the states at the surface $\alpha$; this interaction breaks TRS.
We take the exchange mass to be of opposite 
\footnote{When the two exchange masses have the same sign, the model describes a quantum anomalous Hall insulator \cite{Lei2020}.} value at the top and bottom surfaces, $m_{0}=-m_{1}\equiv m$ \cite{Zhang2015,Nagaosa2015}.
We also include an inter-surface hybridization parameter $\Delta$. 
In the basis $(\ket{\bar{u}_{(0,\uparrow),\boldsymbol{k}_{\Gamma}}},\ket{\bar{u}_{(0,\downarrow),\boldsymbol{k}_{\Gamma}}},\ket{\bar{u}_{(1,\uparrow),\boldsymbol{k}_{\Gamma}}},\ket{\bar{u}_{(1,\downarrow),\boldsymbol{k}_{\Gamma}}})$, the Hamiltonian is specified by \cite{Lei2021}
\begin{align}
	\mathcal{H}_{\text{eff}}(\boldsymbol{k})=\left(
	\begin{array}{cccc}
		m & i\hbar v_{\text{D}}k_{-} & \Delta& 0 \\
		-i\hbar v_{\text{D}}k_{+} & -m & 0 & \Delta \\
		\Delta & 0 & -m & -i\hbar v_{\text{D}}k_{-} \\
		0 & \Delta & i\hbar v_{\text{D}}k_{+} & m \\
	\end{array}
	\right),
	\label{Heff1}
\end{align}
where $k_{\pm}\equiv k_{x}\pm ik_{y}$. 

To account for the presence of a uniform dc magnetic field that is parallel to the surface-normal of the thin film, we first implement the usual prescription \footnote{See, e.g., Chapter 2 of Winkler \cite{WinklerBook}.} to obtain an envelope function approximated (EFA) Hamiltonian that is related to a given $\boldsymbol{k}\cdot\boldsymbol{p}$ Hamiltonian. Following this, we minimally couple the electronic degrees of freedom to the corresponding vector potential. That is, first take $\hbar\boldsymbol{k}\rightarrow \boldsymbol{\mathfrak{p}}(\boldsymbol{r})\equiv-i\hbar\boldsymbol{\nabla}$ followed by $\boldsymbol{\mathfrak{p}}(\boldsymbol{r})\rightarrow \boldsymbol{\mathfrak{p}}_{\text{mc}}(\boldsymbol{r})=-i\hbar\boldsymbol{\nabla}+\frac{e}{c}\boldsymbol{A}(\boldsymbol{r})$, where $-e$ is the electronic charge. We take $\boldsymbol{B}=(0,0,-B)$ and in the Landau gauge $\boldsymbol{A}(\boldsymbol{r})=(0,-Bx,0)$.
The 2D EFA Hamiltonian that is obtained by employing this prescription to Eq.~(\ref{Heff1}) is invariant under translations along $\boldsymbol{y}$ within the film. 
Thus, we seek energy eigenfunctions of the form 
\begin{align}
	\Psi_{E,q_{y}}(x,y)=\frac{e^{iq_{y}y}\Phi_{E,q_{y}}(x)}{\sqrt{L_{y}}}.
	\label{LLeigen}
\end{align}
Introducing the usual differential ladder operators $a\equiv\frac{1}{\sqrt{2}}(\tilde{x}+\frac{\partial}{\partial\tilde{x}})$ and $a^{\dagger}\equiv\frac{1}{\sqrt{2}}(\tilde{x}-\frac{\partial}{\partial\tilde{x}})$ \footnote{See, e.g., pg.~89-94 of Sakurai \cite{Sakurai}.}, where $\tilde{x}\equiv l_{B}q_{y}-x/l_{B}$, $l_{B}^2\equiv\hbar c/eB$, and $[a,a^{\dagger}]=1$, the EFA Hamiltonian that is related to (\ref{Heff1}) acts on the $\Phi_{E,q_{y}}(x)$ via
\begin{align}
	H_{\text{EFA}}\Big(x,\frac{\partial}{\partial x};q_{y}\Big) &= \hbar \omega_c \tau_z\otimes \begin{pmatrix} 0 & a^{\dagger} \\ a & 0 \end{pmatrix} \nonumber\\
	&+ m \tau_z\otimes\sigma_z+ \Delta \tau_x\otimes\sigma_0,
	\label{Hamiltonian_eff}
\end{align}
\iffalse
\begin{align}
	H_{\text{EFA}}\Big(x,\frac{\partial}{\partial x};q_{y}\Big) &= \frac{1}{2}\hbar \omega_c \tau_z\otimes( \sigma_{+} a^{\dagger} + \sigma_{-} a ) \nonumber\\
	&+ m \tau_z\otimes\sigma_z+ \Delta \tau_x\otimes\sigma_0,
	%\label{Hamiltonian_eff}
\end{align}
\fi
where $\omega_c\equiv\sqrt{2}v_{\text{D}}/l_{B}$, $\tau_{a}$ are the Pauli matrices acting on the surface label (\textit{i.e.}~orbital type) degree of freedom, and $\sigma_{a}$ are the Pauli matrices on spin degree of freedom.
	
A general eigenfunction of Eq.~(\ref{Hamiltonian_eff}) has the
form $\Phi_{E_{n},q_{y}}(x)=\Phi_{n}(\tilde{x})$, where
\begin{align}
	\Phi_{n}(\tilde{x})&= \Big( C_{n,(\text{b},\uparrow)}\chi_{n}(\tilde{x}), \, C_{n,(\text{b},\downarrow)}\chi_{n-1}(\tilde{x}),\nonumber\\
	&\qquad C_{n,(\text{t},\uparrow)}\chi_{n}(\tilde{x}), \, C_{n,(\text{t},\downarrow)}\chi_{n-1}(\tilde{x})\Big)
	\label{wavefunction_eff}
\end{align}
for integers $n>0$ and
\begin{align}
	\Phi_{0}(\tilde{x})&=\big( C_{0,(\text{b},\uparrow)}\chi_{0}(\tilde{x}), \, 0, \, C_{0,(\text{t},\uparrow)}\chi_{0}(\tilde{x}), \, 0\big)
	\label{wavefunction_eff2}
\end{align}
for $n=0$, where $\chi_{n,q_{y}}(x)=\chi_{n}(\tilde{x})$ are normalized eigenfunctions of $a^{\dagger}a$.
The corresponding eigenvalues are 
\begin{align}
	E^{\pm}_{n}=\pm\sqrt{\hbar^2\omega_c^2 n +m^2+\Delta^2},
\end{align} 
with the $n=0$ eigenvalues non-degenerate and the $n>0$ eigenvalues each two-fold degenerate.
We adopt the notation $\Phi^{\pm}_{0}(\tilde{x})$ and $\Phi^{\pm,1}_{n}(\tilde{x})$, $\Phi^{\pm,2}_{n}(\tilde{x})$ for the corresponding eigenfunctions. 

We endow the $\Psi^{\pm}_{n,q_{y}}(x,y)$ with $z$-dependence by multiplying the components of (\ref{wavefunction_eff},\ref{wavefunction_eff2}) that are associated with the top (bottom) surface of the thin film by $\delta(z-z_{\text{t}(\text{b})})$, where $z_\text{t} = a/2$ and $z_\text{b} = -a/2$. Then, taking $E_{F}=0$ such that the ($\Psi^{+}_{n,q_{y}}(\boldsymbol{r})$) $\Psi^{-}_{n,q_{y}}(\boldsymbol{r})$ are (un)occupied, and using $\sum_{q_{y}}(1)=\sum_{q_{y}}\big(\int|\chi_{n}(\tilde{x})|^2dx\big)=eBL_{x}L_{y}/hc$, the number of $q_{y}$ per Landau level $n$, employing Eq.~(\ref{Pfinite}) yields
\begin{equation}
	P^{z}_{\text{el}} = -\frac{e^2B}{2hc}\sum_{\substack{n\ge0 \\ \sigma\in\{\uparrow,\downarrow\}}} \Big(|C^{-}_{n,(\text{t},\sigma)}|^2 - |C^{-}_{n,(\text{b},\sigma)}|^2\Big),
	\label{Pfinite2}
\end{equation}
from which we can identify $\alpha_{\text{me}}\equiv\alpha^{zz}$ via $P^{z}_{\text{el}}=\alpha^{zz}B^{z}$.
	
The $n=0$ anomalous Landau levels $\Phi^{\pm}_{0}(\tilde{x})$ are spin polarized, and the action of (\ref{Hamiltonian_eff}) on (\ref{wavefunction_eff2}) simplifies such that $\big(C^{\pm}_{0,(\text{b},\uparrow)},C^{\pm}_{0,(\text{t},\uparrow)}\big)$ is obtained by diagonalizing $m\tau_z+\Delta\tau_{x}$. At half filling, the occupied $n=0$ eigenfunction is
\begin{align*}
	\Phi^{-}_{0}(\tilde{x}) = \mathcal{N} \left(-\Delta\chi_{0}(\tilde{x}), \, 0, \, \left(m+\sqrt{m^2+\Delta^2}\right)\chi_{0}(\tilde{x}), \, 0\right),
\end{align*}
where $\mathcal{N} =1/\sqrt{ 2\big(m^2 + \Delta^2+ m\sqrt{m^2 + \Delta^2}\big)}$. Using (\ref{Pfinite2}), the 
contribution to the magnetoelectric coefficient $\alpha_{\text{me}}$ from $\Phi^{-}_{0}(\tilde{x})$ is then
\begin{equation}
    \alpha^{n=0}_{\text{me}}=\frac{e^2}{2hc}\frac{m}{\sqrt{m^2+\Delta^2}}.
    \label{alphaFinite}
\end{equation}

In the bulk limit of this toy model, the top and bottom surface states decouple and 
$\Delta\rightarrow0$, in which case Eq.~(\ref{alphaFinite}) yields the quantized magnetoelectric coefficient $e^{2}/2hc$ independent of $m$. Meanwhile in thin films with TRS, which implies $m=0$, Eq.~(\ref{alphaFinite}) gives that $\alpha^{n=0}_{\text{me}}$ vanishes.
	
Of course, the occupied $n>0$ Landau levels may also contribute to (\ref{Pfinite2}).
Unfortunately, the eigenvector-eigenvalue equations are not as simple as the $n=0$ case. 
Moreover, due to the two-fold degeneracy of each eigenvalue, the energy eigenvectors are non-unique. 
One convenient non-orthogonal pair is \footnote{Here we denote $\big(C_{n,(\text{b},\uparrow)},C_{n,(\text{b},\downarrow)},C_{n,(\text{t},\uparrow)},C_{n,(\text{t},\downarrow)}\big)$ by $\big(C_{n,(\alpha,\sigma)}\big)_{\alpha,\sigma}$.}
\begin{align*}
	\big(C^{\pm,1}_{n,(\alpha,\sigma)}\big)_{\alpha,\sigma}&=\mathcal{N}_{n}^{\pm,1}\left(\frac{\Delta\pm\hbar\omega_c 
		\sqrt{n}}{E^{\pm}_{n}-m},\pm1,1,\frac{\pm\Delta-\hbar\omega_c 
		\sqrt{n}}{E^{\pm}_{n}-m}\right),\\
	\big(C^{\pm,2}_{n,(\alpha,\sigma)}\big)_{\alpha,\sigma}&=\mathcal{N}_{n}^{\pm,2}\left(\pm1,\frac{\Delta\pm\hbar\omega_c 
		\sqrt{n}}{E^{\pm}_{n}+m},\frac{\pm\Delta-\hbar\omega_c 
		\sqrt{n}}{E^{\pm}_{n}+m},1\right),
\end{align*}
where $\mathcal{N}_{n}^{\pm,1}$ and $\mathcal{N}_{n}^{\pm,2}$ are normalization factors. These eigenvectors can be orthogonalized and their contributions to the induced polarization calculated. In the limit $\Delta\rightarrow0$, which is of primary interest, these eigenvectors are indeed orthogonal and are moreover equally weighted combinations of states that are associated with the top and bottom surfaces. Thus, at half filling and in the limit $\Delta\rightarrow0$, it is only the $\Phi^{-}_{0}(\tilde{x})$ anomalous Landau level that contributes to the magnetic-field-induced polarization.
	
\subsection{Coupled Dirac cone model of layered thin films}
\label{Sec:IIb}
The results for $\alpha_{\text{me}}$ obtained using the simple model described in Sec.~\ref{Sec:IIa} can be 
confirmed by considering a more realistic model for finite thickness films of V$_2$VI$_3$-type TIs. Materials in this family
consist of many stacked 2D layers. They are more accurately modelled 
by introducing a pair of Dirac cones for each layer, one associated with its top and bottom surface, then coupling the Dirac-like states within the same layer via $\Delta_{\text{S}}$ and coupling those associated with the closest surfaces of the nearest-neighbor layers via $\Delta_{\text{D}}$ \cite{Burkov2011,Lei2020}.
This model has the merits that it is readily solved numerically for films that contain many van der Waals layers and that 
it is readily solved in the presence of a perpendicular external magnetic field, allowing the magnetic field dependence of the electronic polarization to be evaluated
explicitly. 

The influence of a static and uniform magnetic field that is parallel to the surface-normal of the thin film is accounted for layer-by-layer;
a 2D EFA Hamiltonian that is minimally coupled to $\bm{B}$ is assumed to describe the electronic dynamics within an isolated layer, and is taken as Eq.~(\ref{Hamiltonian_eff}) with $\Delta\rightarrow\Delta_{\text{S}}$ and $m\rightarrow M_{l_{z}}$ for $l_{z}\in\{1,\ldots,N\}$ a layer index.
We then couple $N$ copies (each indexed by an $l_{z}$) of this 2D isolated layer Hamiltonian via $\Delta_{\text{D}}$ as described above.
We take $M_{l_{z}}\neq 0$ only if $l_{z}=1$ or $N$ to account for the presence of magnetic dopants only in the outermost top and bottom layer.
As in Sec.~\ref{Sec:IIa}, we take the orientation of the magnetization describing the configuration of dopants at the top and bottom surface to be opposite ($M_{1}=-M_{N}\equiv m$) \footnote{
In future studies of anti-ferromagnetic materials we consider $M_{l_{z}}=M_{l_{z}+1}$ for $l_{z}\in\{1,\ldots,N-1\}$.}.

\begin{figure}[t!]
\centering
\includegraphics[width=0.485\textwidth]{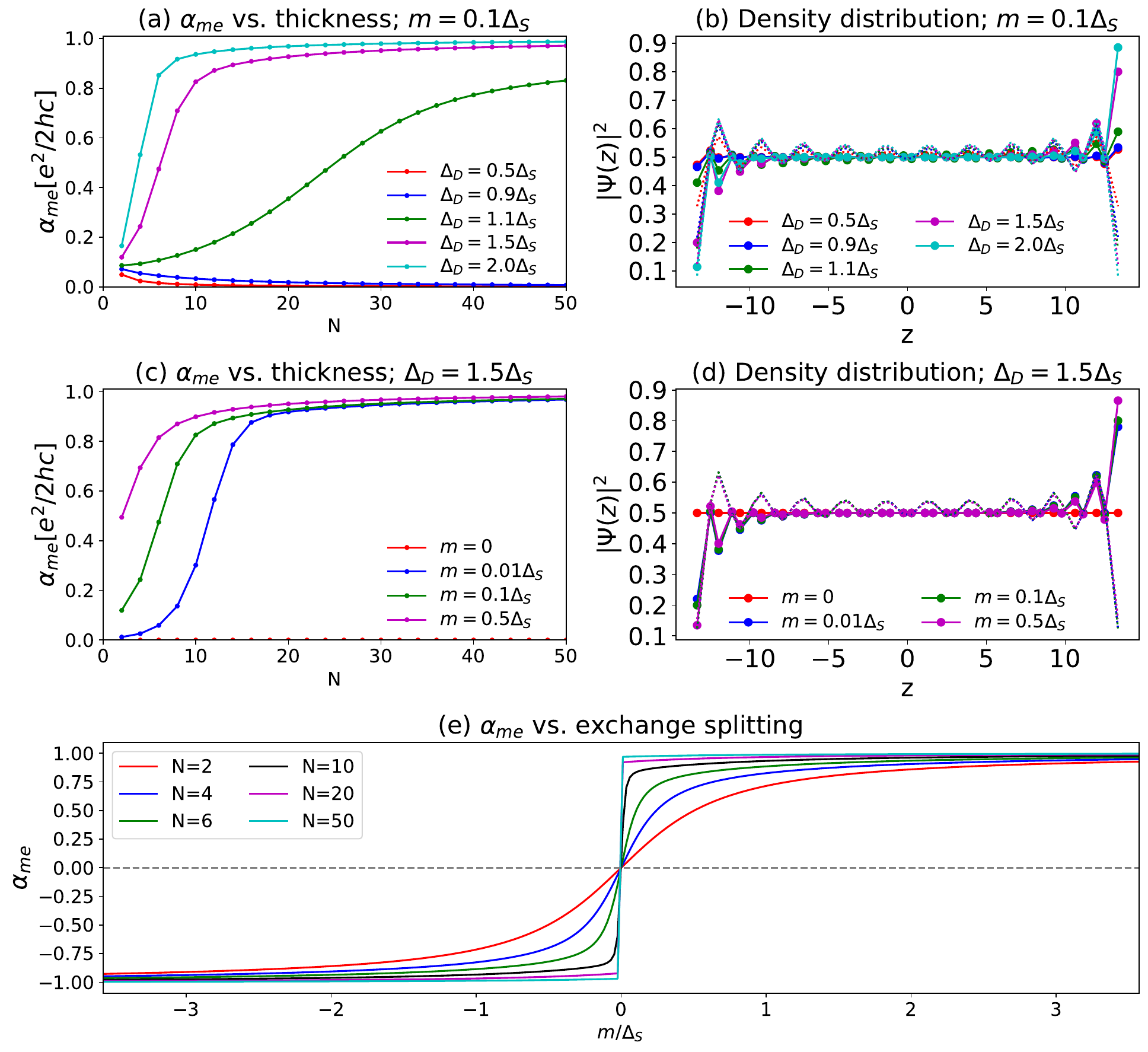}
\caption{Magnetoelectric coefficient $\alpha_{\text{me}}$ vs.~film thickness and the distribution of the electronic charge density in the stacking direction ($\boldsymbol{z}$). 
    (a) $\alpha_{\text{me}}$ vs.~film thickness for various $\Delta_{\text{D}}$. Here the exchange splitting $m$ of the Dirac cones associated with the top and bottom layers (resulting from the presence of surface magnetic dopants) is taken to be $0.1\Delta_{\text{S}}$ and taken to vanish for all other layers.
    (b) Distribution of the electronic charge density along the stacking direction for the same values of $\Delta_{\text{D}}$ used in (a), with the film thickness taken to be 20 layers and $z=0$ taken to coincide with the middle of the center layer. The dotted (solid) lines identify the contribution of the $n=1$ ($n=0$) Landau levels to this distribution, which is (anti-)symmetric in $z$.
    (c) $\alpha_{\text{me}}$ vs.~film thickness for various exchange splittings $m$, with $\Delta_{\text{D}}$ taken as $1.5 \Delta_{\text{S}}$;
    (d) Distribution of the electronic charge density in the stacking direction for the same values of $m$ used in (c) for a 20-layer thin film;
    (e) $\alpha_{\text{me}}$ vs.~exchange splitting $m$ due to magnetic surface dopants for films of varying thickness.}
\label{fig:alphathickness}
\end{figure}

This model can be solved numerically for a varying number $N\ge2$ of layers, which we always take to be even, and a general energy eigenfunction is again of the form \cite{lei2023afm} $\Psi^{r}_{n,q_{y}}(x,y)=e^{iq_{y}y}\Phi^{r}_{n}(\tilde{x})/\sqrt{L_{y}}$, where
\begin{equation}
	\Phi^{r}_{n}(\tilde{x}) = \prod_{j\in\{0,\ldots,2N-1\}}\left( C^{r}_{n,(j,\uparrow)}\chi_{n}(\tilde{x}) , \, C^{r}_{n,(j,\downarrow)}\chi_{n-1}(\tilde{x}) \right)
\end{equation}
for $n>0$ and 
\begin{equation}
	\Phi^{r}_{0}(\tilde{x}) = \prod_{j\in\{0,\ldots,2N-1\}}\left( C^{r}_{0,(j,\uparrow)}\chi_{0}(\tilde{x}) , \, 0 \right)
\end{equation}
for $n=0$, and even (odd) values of $j$ identify components that are associated with the bottom (top) surface of layer number $j/2$ ($(j-1)/2$) enumerated in ascending order along the stacking axis $\bm{z}$. We again endow the $\Psi^{r}_{n,q_{y}}(x,y)$ with $z$-dependence by taking $C^{r}_{n,(j,\uparrow)}\rightarrow C^{r}_{n,(j,\uparrow)}\delta(z-z_{j})$; choosing the origin to be at the center of the material, for even values of $j$ we have $z_{j}=a(j-N)/2$ and $z_{j+1}=z_{j}+a$ (see Fig.~1 of Lei \textit{et al.}~\cite{Lei2020}). The electronic polarization (\ref{Pfinite}) can then be written
\begin{align}
	P^{z}_{\text{el}} = -\frac{e^2B}{aNhc}\sum_{j\in\{0,\ldots,2N-1\} }\sum_{\substack{E_{n,r}<0 \\ \sigma\in\{\uparrow,\downarrow\}}}z_{j}|C^{r}_{n,(j,\sigma)}|^2.
\end{align}
The results are shown in Fig.~\ref{fig:alphathickness}. In particular, as found in Sec.~\ref{Sec:IIa}, if $m=0$ then $\alpha_{\text{me}}=0$ any $N$, whereas 
if $m\ne0$ then $\alpha_{\text{me}}$ approaches the expected quantized bulk value for $N \to \infty$.
That is, $\alpha_{\text{me}} \rightarrow e^{2}/2hc$ for $|\Delta_{\text{D}}| > |\Delta_{\text{S}}|$ (in the 
$\mathbb{Z}_{2}$-odd regime of the model) and 
$\alpha_{\text{me}} \rightarrow 0 $ for $|\Delta_{\text{D}}| < |\Delta_{\text{S}}|$ (in the $\mathbb{Z}_{2}$-even regime).
Moreover, numerical sums of the $n>0$ ($n=0$) energy eigenfunction distributions below the Fermi energy at half filling show that the charge density is (anti-)symmetric across the thin film, 
leading to a (non)vanishing contribution to the ground state dipole moment and thus to $\alpha_{\text{me}}$. 
This is shown for $n=1$ ($n=0$) by the dotted (solid) curves in Fig.~\ref{fig:alphathickness} (b) and (d).

The quantization of $\alpha_{\text{me}}$ in thick films can be understood by formally examining the model for an infinite number layers. In this case the analog of Eq.~(\ref{Hamiltonian_eff}) is
\begin{align}
	H_{\text{EFA}}\Big(x,\frac{\partial}{\partial x};q_{y},k_{z}\Big) &= \left(
	\begin{array}{cccc}
		0 & \hbar\omega_c a^{\dagger} & \Delta_{k_{z}}^{*}& 0 \\
		\hbar\omega_c a & 0 & 0 & \Delta_{k_{z}}^{*} \\
		\Delta_{k_{z}} & 0 & 0 & -\hbar\omega_c a^{\dagger} \\
		0 & \Delta_{k_{z}} & -\hbar\omega_c a & 0 \\
	\end{array}
	\right),
	\label{Hamiltonian_eff_bulk}
\end{align}
where $\Delta_{k_{z}}\equiv\Delta_{\text{S}}+e^{ik_{z}a}\Delta_{\text{D}}$. 
To obtain Eq.~(\ref{Hamiltonian_eff_bulk}) we have taken the basis states $\ket{\bar{u}_{(\alpha,\uparrow),(k_{x},k_{y})}}\rightarrow\ket{\bar{u}_{(\alpha,\uparrow),(k_{x},k_{y},R)}}$ for $R\in a\mathbb{Z}$ as hybrid WFs \cite{Marzari2012} (in $\mathbb{R}^{3}$) that are spatially localized in $\bm{z}$ about the surface $\alpha$ in the unit cell at $R\bm{z}$.
In the non-magnetic bulk there is one layer per unit cell and therefore $\alpha\in\{0,1\}$.
%; they therefore satisfy $\bar{u}_{(\alpha,\uparrow),(k_{x},k_{y},R)}(z-al_{z})=\bar{u}_{(\alpha,\uparrow),(k_{x},k_{y},R-al_{z})}(z)$.
Again projecting into the $n=0$ subspace, Eq.~(\ref{Hamiltonian_eff_bulk}) acts on $\big( C^{r}_{0,(\text{b},\uparrow)}, \, C^{r}_{0,(\text{t},\uparrow)}\big)$ via
\begin{align}
	H_{n=0}(k_{z}) &= \begin{pmatrix}
		0 & \Delta_{k_{z}}^{*} \\
		\Delta_{k_{z}} & 0
	\end{pmatrix},
	\label{Hamiltonian_eff_SSH}
\end{align}
which is a Su-Schrieffer-Heeger (SSH) model with $\Delta_{\text{S}}$ and $\Delta_{\text{D}}$ playing the role of the hopping parameters.
This model has particle-hole symmetry: $U^{\dagger}_{\mathcal{C}}H_{n=0}(-k_{z})^{*}U_{\mathcal{C}}=-H_{n=0}(k_{z})$ for $U_{\mathcal{C}}=\tau_{z}$ and therefore
%admits quantized polarizations is related \cite{ryu2010topological} to the presence of a $\mathcal{T}^{2}=+1$ time-reversal symmetry: $H_{\text{AFM}}(-k_{z})^{*}=H_{\text{AFM}}(k_{z})$ and a \textcolor{red}{$\mathcal{C}^{2}=+1$} chiral symmetry: $U^{\dagger}_{\mathcal{C}}H_{\text{AFM}}(-k_{z})^{*}U_{\mathcal{C}}=-H_{\text{AFM}}(k_{z})$ for $U_{\mathcal{C}}=(\lambda_{x}\text{cos}dk_{z}-\lambda_{y}\text{sin}dk_{z})\otimes\tau_{z}$. 
at half-filling supports topologically distinct ground states
characterized by a $\mathbb{Z}_{2}$-valued invariant \footnote{See Sec.~2D of Ref.~\cite{Zhang2008} and Ref.~\cite{Ryu2010}.}, 
which happens to be obtained by performing the BZ$_{\text{1D}}$-integral in the expression for the 1D bulk electronic ground state polarization \cite{Vanderbilt1993,Vanderbilt1993_Bulk-Surface}.
Indeed, the $n=0$ contribution to the bulk polarization $P^{z}$ of Eq.~(\ref{Hamiltonian_eff_bulk}) (which is related to adiabatically induced electronic currents in $\boldsymbol{z}$ \footnote{See, e.g., Chapter 4 of Ref.~\cite{VanderbiltBook} and Ref.~\cite{Resta2007}.}) is proportional to that 1D polarization and,
in-line with the well-known results of the usual SSH model, at half-filling we find
\begin{align}
    P_{n=0}^{z}\text{ mod } \frac{e^2}{hc}B^{z}=\begin{cases}
    0, &\text{if $|\Delta_{\text{S}}|>|\Delta_{\text{D}}|$}\\
    \frac{e^2}{2hc}B^{z}, &\text{if $|\Delta_{\text{S}}|<|\Delta_{\text{D}}|$}
    \end{cases}.
    \label{SSH_polarization}
\end{align}
In contrast, the projected Hamiltonian in the $n>0$ subspace is symmetric under 1D center-of-inversion (about the layer center) times spin-flip transformation and therefore does not support an electronic polarization \cite{lei2023afm}.

\section{Semi-analytic calculation of $\alpha_{\text{CS}}$ in a 3D tight-binding model}
\label{Sec:III}
	
In this section our primary aim is to calculate the Chern-Simons coefficient $\alpha_{\text{CS}}$ in bulk 3D non-magnetic V$_2$VI$_3$-type insulators. Effective models for the electronic states near the Fermi energy in these materials have been developed \cite{Zhang2009,Burkov2011,Lei2020}, but since effective models are not generally defined over the entire BZ, they are inadequate for this purpose. The reason for this can be understood by noting that the topology of the vector bundle of occupied Bloch states over the Brillouin zone torus constructed for the true Hamiltonian $H(\boldsymbol{x},\boldsymbol{\mathfrak{p}}(\boldsymbol{x}))$ of a crystalline insulator manifests as an obstruction to the existence of a smooth \textit{global} gauge thereof \cite{Panati2007,Monaco2017}, but smooth local gauges always exist \footnote{This follows immediately from the definition of the vector bundle of occupied states as a vector bundle \cite{Panati2007,Monaco2014}.}. In particular, as explained previously, the linear response expression for $\alpha_{\text{CS}}$ as a $\text{BZ}$-integral of the Chern-Simons 3-form is valid \textit{only} in a smooth global gauge of that bundle. We emphasize this issue in Appendix \ref{Appendix:kdotp}, where we explicitly show that in a low-energy effective model an incorrect half quantization of $\alpha_{\text{CS}}$ can occur, which is reminiscent of the situation that arises when attempting to calculate the first Chern invariant of a 2D $\boldsymbol{k}\cdot\boldsymbol{p}$ model. Thus, lattice regularization of the previously developed effective models is required, and that is where we begin.
	
\iffalse
A proper lattice regularization consists of modifying the Hamiltonian matrix elements of the $\boldsymbol{k}\cdot\boldsymbol{p}$ theory such that the result is $\Gamma^*$-periodic and yields the original $\boldsymbol{k}\cdot\boldsymbol{p}$ theory as a low-energy approximation about a point in BZ. In addition, one needs to specify the set of WFs (or the Bloch-type functions to which there exists a bijection) that the tight-binding model is written with respect to. The $\boldsymbol{k}\cdot\boldsymbol{p}$ theory is of little use in this regard because it is written with respect to a constant basis in $\boldsymbol{k}$, but the crystal field considerations leading to it provide some physical insight.
\fi
	
\subsection{Construction of a 3D tight-binding model}
\label{Sec:IIIa}
In this subsection we construct a lattice regularization of a previously presented \cite{Burkov2011} effective model for the low-energy electronic states in bulk 3D V$_2$VI$_3$-type insulators. Since a formally similar effective model describes anti-ferromagnetic $\mathbb{Z}_{2}$ TIs including $\text{Mn(Sb}_x\text{Bi}_{1-x})_2\text{X}_4$ (where X = Se, Te) \cite{Lei2020}, which will be the focus of future work, we in fact construct a sufficiently general lattice regularization scheme that can be applied to both families of materials. Notably, both families of materials can be viewed as layered compounds, with each layer having 3-fold rotational symmetry about the stacking ($\boldsymbol{z}$) axis and 2-fold rotational symmetry about an axis perpendicular to the stacking axis \cite{Zhang2009,Wang2019}. The bulk crystal structure of the non-magnetic (magnetic) family of materials consists of stacked 5-atom (7-atom) layers, called quintuple (septuple) layers, and has a center-of-inversion symmetry within each such layer \cite{Zhang2009,Wang2019}. That layered structure motivated the development of low-energy effective models \cite{Burkov2011,Lei2020} in which discrete 2D $\boldsymbol{k}\cdot\boldsymbol{p}$ continuum models in the planes perpendicular to the stacking axis are coupled to one another with interlayer hopping parameters $\Delta_{\text{D}}$ to yield a 1D tight-binding model along the stacking axis for each 2D momentum (see Appendix \ref{Appendix:kdotp}); these models can be thought of as the 3D bulk limit of the quasi-2D Hamiltonian employed in Sec.~\ref{Sec:IIb}. Since the electronic states within each layer are Dirac-like, effective Hamiltonians of this type are called coupled-Dirac cone models. We construct 3D tight-binding models to describe both families of materials by first developing a 2D lattice regularization of the $\boldsymbol{k}\cdot\boldsymbol{p}$ model of an isolated layer (obtained by taking $\Delta_{\text{D}}=0$), then re-introduce $\Delta_{\text{D}}$ as in those past works.
	
\subsubsection{A lattice regularized description of an isolated 2D layer}
Two proposed low-energy effective models of 3D V$_2$VI$_3$-type band insulators include the above described coupled-Dirac cone model \cite{Burkov2011,Lei2020} and a traditional 3D $\boldsymbol{k}\cdot\boldsymbol{p}$ model \cite{Zhang2009}. When restricted to a single layer (\textit{i.e.}~the $k_x$--$k_y$ plane), the 2D $\boldsymbol{k}\cdot\boldsymbol{p}$ models that result from these descriptions are unitarily equivalent to first order in $k_{x}$ and $k_{y}$ (when certain parameter values are taken to coincide), as one would expect.
Furthermore, it has been observed \cite{Zhang2010} that the restriction of the 3D $\boldsymbol{k}\cdot\boldsymbol{p}$ model to the $k_x$--$k_y$ plane is unitarily equivalent to the Bernevig-Hughes-Zhang (BHZ) model \cite{BHZ2006} for certain parameter values. 
Thus, when the coupled-Dirac cone model is employed for an isolated layer of a non-magnetic material (\textit{i.e.}~$\Delta_{\text{D}}=0$ and $M_{l_{z}}=0$), it is unitarily equivalent to the BHZ model to first order in $k_{x}$ and $k_{y}$. In fact, in this limit of the coupled-Dirac cone model, exact unitary equivalence with the BHZ model occurs if we generalize $\Delta_{\text{S}}\rightarrow\Delta_{\text{S}}-B(k_{x}^2+k_{y}^2)$ and if $\Delta_{\text{S}}$ equals $M$ of BHZ. Rather than generalizing $\Delta_{\text{S}}\rightarrow\Delta_{\text{S}}-B(k_{x}^2+k_{y}^2)$, we could instead obtain exact unitary equivalence by restricting the BHZ model to first order in $k_{x}$ and $k_{y}$ by taking $B=0$. However, doing so would result in the lattice regularization that we employ below to admit band gap minima at points in the $\text{BZ}_{\text{2D}}$ in addition to that at $(k_x,k_y)=(0,0)$, in contrast to the known monolayer band structure \cite{Kanatzidis2004}. Indeed, it will turn out that the parameter regime of immediate interest is $B/\Delta_{\text{S}}<0$.
	
In their seminal work on 2D $\mathbb{Z}_2$ TIs \cite{BHZ2006}, BHZ present a square lattice regularization of their $\boldsymbol{k}\cdot\boldsymbol{p}$ Hamiltonian. Thus, we seek a square lattice regularization of Eq.~(1) of Lei \textit{et al.}~\cite{Lei2020} (generalized by taking $\Delta_{\text{S}}\rightarrow\Delta_{\text{S}}-B(k_{x}^2+k_{y}^2)$) that, when restricted to the non-magnetic case and applied to a single layer, is unitarily equivalent to Eq.~(5) of BHZ \cite{BHZ2006}. In doing so, we generally consider Hamiltonian operators of the form
\begin{align}
	\hat{H}^{(d)}&=\int_{\text{BZ}_{d}}\hat{c}_{\boldsymbol{k}}^{\dagger}\mathcal{H}^{(d)}(\boldsymbol{k})\hat{c}_{\boldsymbol{k}}d^{d}k,
	\label{Hoperator_general}
\end{align}
where $d$ is the spatial dimension of the crystal, $\boldsymbol{k}\in\text{BZ}_{d}$ for $\text{BZ}_{d}$ a $d$-dimensional first Brillouin zone of $\Gamma^{*}_{H}$ -- $\Gamma_{H}\subset\mathbb{R}^{d}$ is the Bravais lattice of the spatially periodic Bloch Hamiltonian under consideration and $\Gamma^{*}_{H}\subset\mathbb{R}^{d}$ its dual -- and $\hat{c}_{\boldsymbol{k}}\equiv\big(\hat{c}_{(0,\uparrow),\boldsymbol{k}},\hat{c}_{(0,\downarrow),\boldsymbol{k}},\hat{c}_{(1,\uparrow),\boldsymbol{k}},\hat{c}_{(1,\downarrow),\boldsymbol{k}}\big)^{\text{T}}$ and $\hat{c}^{\dagger}_{\boldsymbol{k}}\equiv\Big(\hat{c}^{\dagger}_{(0,\uparrow),\boldsymbol{k}},\hat{c}^{\dagger}_{(0,\downarrow),\boldsymbol{k}},\hat{c}^{\dagger}_{(1,\uparrow),\boldsymbol{k}},\hat{c}^{\dagger}_{(1,\downarrow),\boldsymbol{k}}\Big)$ are tuples of fermionic operators that act in the electronic Fock space. Products in Eq.~(\ref{Hoperator_general}) are the usual matrix multiplication. 
The fermionic operators involved in Eq.~(\ref{Hoperator_general}) are of the form $\hat{c}_{I,\boldsymbol{k}}$, $\hat{c}^{\dagger}_{I,\boldsymbol{k}}$, for $I$ a general Wannier type label.
We require those operators to be such that the following is satisfied: $\ket{\bar{\psi}_{I,\boldsymbol{k}}}\equiv\hat{c}^{\dagger}_{I,\boldsymbol{k}}\ket{\text{vac}}$ are Bloch-type vectors \footnote{We require that the position space representation $\bar{\psi}_{I,\boldsymbol{k}}(\boldsymbol{r})\equiv\braket{\boldsymbol{r}}{\bar{\psi}_{I,\boldsymbol{k}}}$ of each vector $\ket{\bar{\psi}_{I,\boldsymbol{k}}}$ to be of Bloch's form $\forall\boldsymbol{R}\in\Gamma_{H}:\bar{\psi}_{I,\boldsymbol{k}}(\boldsymbol{r}+\boldsymbol{R})=e^{i\boldsymbol{k}\cdot\boldsymbol{R}}\bar{\psi}_{I,\boldsymbol{k}}(\boldsymbol{r})$. In general, any such $\ket{\bar{\psi}_{I,\boldsymbol{k}}}$ need not be an energy eigenvector.} that are smooth over $\text{BZ}_{d}$, orthonormal such that $\braket{\bar{\psi}_{I,\boldsymbol{k}}}{\bar{\psi}_{J,\boldsymbol{k}'}}=\delta_{I,J}\delta(\boldsymbol{k}-\boldsymbol{k}')$, and constitute a basis of the ``relevant'' electronic Hilbert space \footnote{In general, all tight-binding constructions aim to model the electronic energy eigenvalues and eigenvectors within some ``relevant'' subspace of the full Hilbert space. This space can either be thought of as spanned by a set of ``relevant'' WFs or equivalently a set of ``relevant'' Bloch functions. Typically one aims to model those eigenvalues and vectors near the Fermi energy, usually starting with a representation in terms of the former (e.g., include only ``nearest-neighbor hopping,'' etc.) then mapping to the latter before diagonalizing the model Hamiltonian.}. The penultimate and final criteria imply that $\{\hat{c}_{I,\boldsymbol{k}},\hat{c}^{\dagger}_{J,\boldsymbol{k}}\}=\delta_{I,J}\delta(\boldsymbol{k}-\boldsymbol{k}')$, which follows from the anti-commutation of the electron field operators themselves.
These are the usual criteria that are employed when constructing a tight-binding model and, for a given crystal, there are many sets of vectors $\ket{\bar{\psi}_{I,\boldsymbol{k}}}$ (and therefore operators $\hat{c}_{I,\boldsymbol{k}}$, $\hat{c}^{\dagger}_{I,\boldsymbol{k}}$) that satisfy them. An equivalent 
description can be obtained by noting that a set of Bloch-type vectors $\ket{\bar{\psi}_{I,\boldsymbol{k}}}$ that satisfy the above criteria bijectively maps to a set of exponentially localized Wannier functions (WFs) \cite{Brouder,Vanderbilt2011,Marzari2012},
\begin{align}
	\ket{W_{I,\boldsymbol{R}}}=\sqrt{\frac{\Omega_{uc}}{(2\pi)^d}}\int_{\text{BZ}_{d}}d^{d}ke^{-i\boldsymbol{k}\cdot\boldsymbol{R}}\ket{\bar{\psi}_{I,\boldsymbol{k}}},
	\label{WF}
\end{align}
where $\boldsymbol{R}\in\Gamma_{H}$.
We emphasize that in order for the WFs $W_{I,\boldsymbol{R}}(\boldsymbol{r})\equiv\braket{\boldsymbol{r}}{W_{I,\boldsymbol{R}}}$ to be well-localized, the $\ket{\bar{\psi}_{I,\boldsymbol{k}}}$ must be smooth over $\text{BZ}_{d}$ \cite{Marzari2012}.
Typically it is with respect to the fermionic operators $\hat{c}_{I,\boldsymbol{R}}$, $\hat{c}^{\dagger}_{I,\boldsymbol{R}}$ defined by $\ket{W_{I,\boldsymbol{R}}}\equiv\hat{c}^{\dagger}_{I,\boldsymbol{R}}\ket{\text{vac}}$ that a tight-binding Hamiltonian is specified.

In-line with typical tight-binding approaches, we further \textit{assume} the existence of a Wannier type label $I$ of the form $I=(\alpha,\sigma)$, where $\alpha\in\{0,1\}$ is an orbital index that identifies WFs of distinct spatial distribution and $\sigma\in\{\uparrow,\downarrow\}$ identifies the $s_{z}$ spin component 
\footnote{\label{Footnote:Topology} We simply include the spin degree of freedom as an additional component $\sigma$ of the Wannier index $I$, and associate spin-up and spin-down components with the corresponding eigenvectors of $s_{z}$. 
In particular, we assume that the Hermitian bundle constructed using all relevant states can be decomposed as a product of Hermitian bundles corresponding with spin-up and spin-down sectors \cite{Prodan2009}, and that each of these subbundles is topologically trivial \cite{Panati2007,Monaco2014} such that they each admit smooth global frames (see, e.g, Theorem 4.2.19 of Hamilton \cite{HamiltonBook}). One instance of a smooth global frame for one such subbundle is our $\left(\ket{\bar{u}_{(\alpha,\sigma),\boldsymbol{k}}}\right)_{\alpha\in\{0,1\}}$ for a given $\sigma$. Therefore, the WFs $\{\ket{W_{(\alpha,\sigma),\boldsymbol{R}}}\}_{\alpha\in\{0,1\}}$ exist and can be obtained via Eq.~(\ref{WF}). Since the eigenvectors of $s_{z}$ are an orthonormal basis of the ``spin space,'' the set of such WFs that result is orthonormal and complete.}. 
To coincide with those coupled-Dirac cone effective models \cite{Burkov2011,Lei2020} for which we develop this lattice regularization, we take $\alpha=0$ (1) to label states that are associated with the bottom (top) surface of a layer; this identification will not explicitly enter calculations in this paper, but will provide physical motivation for the terms that appear in the 3D Hamiltonian of the following subsection.
We further discuss (see preceding footnote) the topological implications related to the assumed existence of a Wannier type label of the form $I=(\alpha,\sigma)$ in the following subsections, where we also address additional ubiquitous constraints that are imposed on the WFs (\ref{WF}).
	
A 2D low-energy effective model for the electronic states in a single layer of a MnV$_2$VI$_4$ magnetic (V$_2$VI$_3$ non-magnetic) van der Waals semiconductor is obtained by taking $\Delta_{\text{D}}=0$ (and $M_{l_z}=0$) in Eq.~(1) of Lei \textit{et al.}~\cite{Lei2020}. 
A lattice regularization thereof is motivated by BHZ \cite{BHZ2006}, and is specified by taking $d=2$ and the general $\mathcal{H}^{(d)}(\boldsymbol{k})$ of (\ref{Hoperator_general}) to be
\begin{widetext}
\begin{align}
    \mathcal{H}^{(\text{2D})}_{\text{reg}}(\boldsymbol{k};l_{z})&=\left(\begin{array}{cccc}
	J_{\text{S}}M_{l_{z}} & iA(\text{s}k_{x}-i\text{s}k_{y}) & \Delta_{\text{S}}(k_x,k_y) & 0	\\
	-iA(\text{s}k_{x}+i\text{s}k_{y}) & -J_{\text{S}}M_{l_{z}} & 0 & \Delta_{\text{S}}(k_x,k_y)	\\
	\Delta_{\text{S}}(k_x,k_y) & 0 & J_{\text{S}}M_{l_{z}} & -iA(\text{s}k_{x}-i\text{s}k_{y})	\\
	0 & \Delta_{\text{S}}(k_x,k_y) & iA(\text{s}k_{x}+i\text{s}k_{y}) & -J_{\text{S}}M_{l_{z}}
    \end{array}\right),
    \label{HlayerTB}
\end{align}
\end{widetext}
where $A$ is related to $\hbar v_{\text{D}}$ and fixed later by fitting the 2D band structure to the $\boldsymbol{k}\cdot\boldsymbol{p}$ dispersion near $\Gamma$ and
\begin{align}
	\Delta_{\text{S}}(k_x,k_y)\equiv\Delta_{\text{S}}-2B(2-\text{c}k_{x}-\text{c}k_{y}),
\end{align}
where $\text{s}k_{x}\equiv\sin(k_{x})$, $\text{c}k_{x}\equiv\cos(k_{x})$, etc. We have also introduced a layer index $l_{z}\in\mathbb{Z}$ in anticipation of the generalization to 3D in the following subsection. 
This is a square lattice regularization since $\mathcal{H}^{(\text{2D})}_{\text{reg}}(\boldsymbol{k}+\boldsymbol{G};l_{z})=\mathcal{H}^{(\text{2D})}_{\text{reg}}(\boldsymbol{k};l_{z})$ for any $\boldsymbol{G}\in\Gamma^*_{\text{2D}}\equiv\text{span}_{\mathbb{Z}}(\{2\pi\boldsymbol{x},2\pi\boldsymbol{y}\})$, thus $\Gamma_{\text{2D}}=\text{span}_{\mathbb{Z}}(\{\boldsymbol{x},\boldsymbol{y}\})$; the 2D lattice constant is taken to unity, and $k_x$, $k_y$ are taken to be dimensionless. Of course, in the physical materials under consideration $\Gamma_{\text{2D}}$ is a triangular lattice \cite{Zhang2009,Wang2019}. Thus, there likely exists a lattice regularization that better captures the lattice scale physics, but for our purposes a square lattice
regularization is sufficient. 
	
In Eq.~(\ref{HlayerTB}) the terms involving $M_{l_{z}}$ act to generate an exchange mass within the layer $l_{z}$ in 
the magnetic case, and arise from the exchange interactions of the dynamical electronic degrees of freedom deemed relevant
with those well below the Fermi energy and thus approximated as static.
We assume that this interaction is approximated by a Heisenberg interaction between dynamic spins 
and the quenched magnetic moments. We assume the direction of the static magnetization 
is parallel to the stacking axis (the $\boldsymbol{z}$ direction) \cite{Lei2020}. The terms involving $A$ and $\Delta_{\text{S}}$ can be understood as arising from spin-preserving electronic transitions between some relevant WFs of the layer that are associated with its top and bottom surfaces.
	
The eigenvalues of the single-layer Hamiltonian (\ref{HlayerTB})
demonstrate that it is indeed unitarily equivalent to that of BHZ \cite{BHZ2006} when our 
$M_{l_{z}}=0$, their $C=D=0$, and our $\Delta_{\text{S}}$ equals their $M$. If $M_{l_{z}}=0$, 
then we similarly find that all of the energy bands of (\ref{HlayerTB}) are degenerate at $\Gamma$ when $\Delta_{\text{S}}=0$. Similar degeneracies appear at the other high-symmetry points $(k_{x},k_{y})=(\pi,0)$ and $(0,\pi)$ when $\Delta_{\text{S}}-4B=0$, and at $(\pi,\pi)$ when $\Delta_{\text{S}}-8B=0$.
Indeed it was previously found \cite{Lei2020} that the materials of immediate interest, for which the band gap at half filling is about $\Gamma$, are described by $\Delta_{\text{S}}>0$ and thus $B<0$.
At half filling, BHZ find that the zero-temperature ground state of their model is insulating and in a $\mathbb{Z}_{2}$-even (odd) phase for $M<0$ ($0<M/2B<2$) \cite{BHZ2006}; varying $M/2B$ across $0$ ($2$) closes and re-opens the band gap(s) at $\Gamma$ (at $(\pi,0)$ and $(0,\pi)$), and a topological phase transition occurs. However, this 2D topology is not directly related to that captured by $\alpha_{\text{CS}}$ in 3D, which will vanish in the limit of isolated layers independent of the parameters of a single-layer, even though the symmetries on which the classification is based are the same. 
Our focus in this paper is on the properties of the 3D crystals formed when these 2D layers are stacked and their states hybridized, not on the properties of individual layers.
	
\subsubsection{A non-magnetic 3D multi-layer Hamiltonian}
\label{Sec:IIIa2}
We consider 3D models in which the electronic states associated with adjacent surfaces of nearest-neighbor layers are hybridized 
by a phenomenological interlayer hopping parameter $\Delta_{\text{D}}$: 
\begin{widetext}
\begin{align}
\hat{H}_{\text{reg}}^{(\text{3D})}&=\sum_{l_z\in\mathbb{Z}}\hat{H}_{\text{reg}}^{(\text{2D})}(l_z)+\Delta_{\text{D}}\sum_{\substack{\boldsymbol{R}\in\Gamma_{\text{2D}}\times\{0\}\\l_z\in\mathbb{Z}\\\sigma\in\{\uparrow,\downarrow\}}}\Big(\hat{c}^{\dagger}_{(0,\sigma),\boldsymbol{R}+(l_z+1)a\boldsymbol{z}}\hat{c}_{(1,\sigma),\boldsymbol{R}+l_za\boldsymbol{z}}+h.c.\Big),
\label{H3d}
\end{align}
\end{widetext}
where we have assumed that all layers are identical apart from a layer-dependent $M_{l_{z}}$, that each layer is described by the 2D Hamiltonian (\ref{HlayerTB}), and that each pair of nearest-neighbor layers are separated by the same distance $a$ in the $\boldsymbol{z}$ direction. 
In writing Eq.~(\ref{H3d}) we have assumed a non-magnetic or ferromagnetic configuration of the static magnetic moments, in which case 
$\Delta_{\text{D}}$ is always identified with a hopping parameter between WFs associated with different unit cells. 
For general magnetic configurations there are two issues 
that need be addressed related to the precise definition of the WFs implicit in defining fermionic operators $\hat{c}_{(\alpha,\sigma),\boldsymbol{R}+l_za\boldsymbol{z}}$, $\hat{c}^{\dagger}_{(\alpha,\sigma),\boldsymbol{R}+l_za\boldsymbol{z}}$ 
that involve the layer label $l_z$. First, depending on $\{M_{l_{z}}\}_{l_{z}\in\mathbb{Z}}$, the group of translations $\Gamma_{H}$ under which a single-particle 3D Bloch Hamiltonian $H(\boldsymbol{r},\boldsymbol{\mathfrak{p}}(\boldsymbol{r}))$ -- from which the tight-binding model might be obtained -- is invariant may not be equal to the group of translations $\Gamma_{\text{3D}}\equiv\Gamma_{\text{2D}}\times a\mathbb{Z}$ under which the multi-layer crystal lattice is invariant ($\Gamma_{H}\subseteq\Gamma_{\text{3D}}$). Then, although $\boldsymbol{R}+l_{z}a\boldsymbol{z}\in\Gamma_{\text{3D}}$ for all $l_{z}\in\mathbb{Z}$, generically $\boldsymbol{R}+l_{z}a\boldsymbol{z}\notin\Gamma_{H}$ and therefore \textit{cannot} be used to label WFs or their corresponding operators. In the non-magnetic ($M_{l_z}=0$) and ferromagnetic ($M_{l_z}=M_{l_z'}$) cases, the magnetic and chemical unit cells coincide, $\Gamma_{H}=\Gamma_{\text{3D}}$, and $\boldsymbol{R}+l_za\boldsymbol{z}\in\Gamma_{H}$ for all $l_{z}\in\mathbb{Z}$. However, in the anti-ferromagnetic case ($M_{l_z}=-M_{l_z\pm1}$) this is not so and, strictly speaking,
Eq.~(\ref{H3d}) requires modification. For completeness, we present the tight-binding Hamiltonian for the anti-ferromagnetic case in Appendix \ref{Appendix:AFM} and focus on the non-magnetic case in the remainder of this paper. Second, a set of 3D WFs of a multi-layer crystal (whose corresponding operators appear in Eq.~(\ref{H3d})) is not generally equal to the set of all $l_{z}\boldsymbol{a}_3$-translated 2D WFs of an individual layer (whose corresponding operators appear in Eq.~(\ref{HlayerTB})) thereof. In principle, to construct an accurate 3D tight-binding model one indeed requires WFs of the 3D Bloch Hamiltonian $H(\boldsymbol{r},\boldsymbol{\mathfrak{p}}(\boldsymbol{r}))$. 
However, such details will not be relevant in our calculation of $\alpha_{\text{CS}}$ as we will not employ actual WFs but rather make simplifying approximations regarding their form. Consequently, this second type of imprecision in writing Eq.~(\ref{H3d}) is inconsequential to our calculation. Meanwhile the line of reasoning leading to Eq.~(\ref{H3d}) yields a physically well-motivated form of $\hat{H}^{(\text{3D})}_{\text{reg}}$.

\begin{figure*}[t!]
\subfloat{\includegraphics[width=8cm]{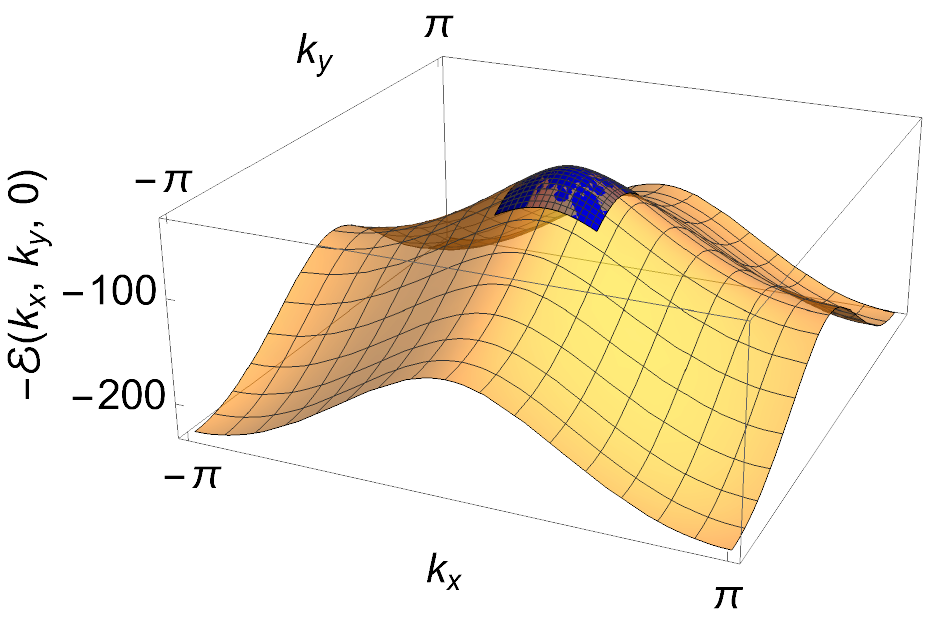}}\quad
\subfloat{\includegraphics[width=8cm]{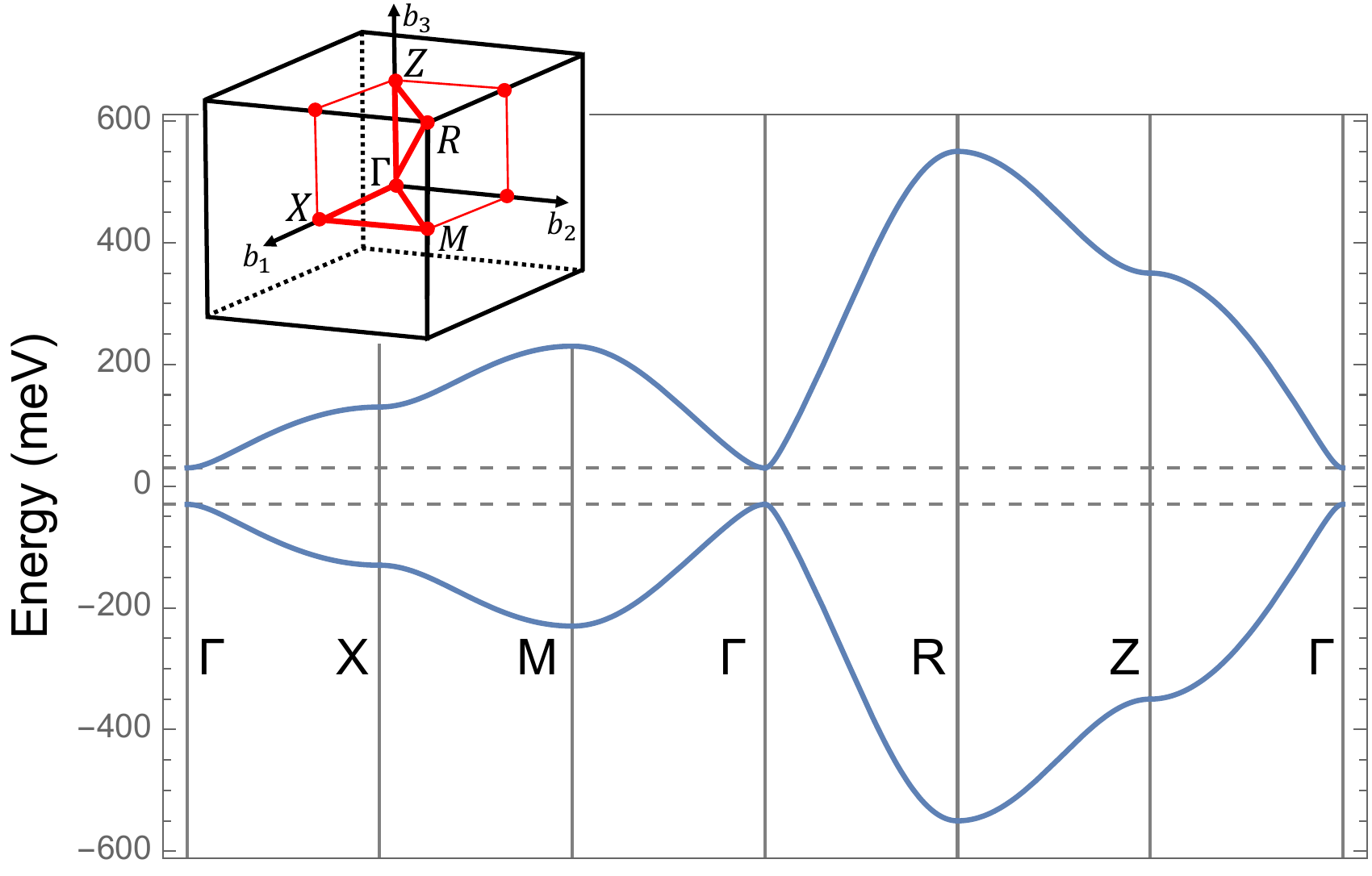}}
\caption{Energy dispersion of the 3D tight-binding model of bulk V$_{2}$VI$_{3}$-type insulators for (somewhat arbitrary) parameters $A=52.7$ meV, $B=-25$ meV, $\Delta_{\text{S}}=190$ meV, $\Delta_{\text{D}}=-160$ meV. (a) The value of $A$ was chosen to fit $-\varepsilon(\boldsymbol{k})$ (orange surface) to the in-plane dispersion of a single layer of Bi$_{2}$Te$_{3}$ near $(k_{\Gamma,x},k_{\Gamma,y})=(0,0)$ (blue surface) when the regularizing lattice has a lattice constant of 1 nm. (b) Bulk 3D energy dispersion along the high symmetry axes.}
\label{Fig:latticeEnergies}
\end{figure*}
	
Focusing on the non-magnetic case ($M_{l_z}=0$) such that Eq.~(\ref{H3d}) is valid, we use Eq.~(\ref{WF},\ref{HlayerTB}) and
\begin{align}
	\frac{\Omega_{uc}}{(2\pi)^d}\sum_{\boldsymbol{R}\in\Gamma_{H}}e^{i(\boldsymbol{k}-\boldsymbol{k}')\cdot\boldsymbol{R}}=\delta(\boldsymbol{k}-\boldsymbol{k}')
	\label{deltaIdentity}
\end{align} 
for $d=3$ and $\boldsymbol{k},\boldsymbol{k}'\in\text{BZ}_{\text{3D}}$ to recast Eq.~(\ref{H3d}) as a $\text{BZ}_{\text{3D}}$-integral of the form (\ref{Hoperator_general}). In this case the general $\mathcal{H}^{(d)}(\boldsymbol{k})$ is taken to be
	\iffalse
	\begin{align}
		\hat{H}^{(\text{3D})}_{\text{reg}}&=\int_{\text{BZ}}\frac{d^3k}{(2\pi)^3}\left(\hat{c}^{\dagger}_{(0,\uparrow),\boldsymbol{k}},\hat{c}^{\dagger}_{(0,\downarrow),\boldsymbol{k}},\hat{c}^{\dagger}_{(1,\uparrow),\boldsymbol{k}},\hat{c}^{\dagger}_{(1,\downarrow),\boldsymbol{k}}\right)\nonumber\\
		&\qquad\times\left(\begin{array}{cccc}
			0 & iA(\text{s}k_x-i\text{s}k_y) & \Delta_{\boldsymbol{k}}^* & 0	\\
			- iA(\text{s}k_x+i\text{s}k_y) & 0 & 0 & \Delta_{\boldsymbol{k}}^*	\\
			\Delta_{\boldsymbol{k}} & 0 & 0 & - iA(\text{s}k_x-i\text{s}k_y)	\\
			0 & \Delta_{\boldsymbol{k}} & iA(\text{s}k_x+i\text{s}k_y) & 0
		\end{array}\right)\left(\begin{array}{c}
			\hat{c}_{(0,\uparrow),\boldsymbol{k}} \\ \hat{c}_{(0,\downarrow),\boldsymbol{k}} \\ \hat{c}_{(1,\uparrow),\boldsymbol{k}} \\ \hat{c}_{(1,\downarrow),\boldsymbol{k}},
		\end{array}\right),
		%\label{H_nonmag}
	\end{align}
	\fi
\begin{widetext}
\begin{align}
	\mathcal{H}^{(\text{3D})}_{\text{reg}}(\boldsymbol{k})=\left(\begin{array}{cccc}
		0 & iA(\text{s}k_x-i\text{s}k_y) & \Delta_{\boldsymbol{k}}^* & 0	\\
		- iA(\text{s}k_x+i\text{s}k_y) & 0 & 0 & \Delta_{\boldsymbol{k}}^*	\\
		\Delta_{\boldsymbol{k}} & 0 & 0 & - iA(\text{s}k_x-i\text{s}k_y)	\\
		0 & \Delta_{\boldsymbol{k}} & iA(\text{s}k_x+i\text{s}k_y) & 0
	\end{array}\right)
\label{H_nonmag}
\end{align}
\end{widetext}
and $\Delta_{\boldsymbol{k}}\equiv\Delta_{\text{S}}(k_x,k_y)+e^{iak_z}\Delta_{\text{D}}$. The eigenvalues of Eq.~(\ref{H_nonmag}) are $E_{1,2}(\boldsymbol{k})=-\varepsilon(\boldsymbol{k})$ and $E_{3,4}(\boldsymbol{k})=\varepsilon(\boldsymbol{k})$ (see Fig.~\ref{Fig:latticeEnergies}), where
\begin{align}
	\varepsilon(\boldsymbol{k})\equiv\sqrt{\frac{1}{2}A^2(2-\text{c}2k_x-\text{c}2k_y)+|\Delta_{\boldsymbol{k}}|^2}.
 \label{energyNonmagnetic}
\end{align}
% Unfortunately in the non-magnetic case considered here, the approach of coupled-Dirac cones to generate a $\boldsymbol{k}\cdot\boldsymbol{p}$ model does not accurately describe the low-energy excitations about $(k_{x},k_{y})=(0,0)$ (see Supplementary Material of Lei \textit{et al.}~\cite{Lei2020}). This feature is inherited by any lattice regularization thereof, including this one. Indeed, the effect of remote bands in addition to the Dirac-like excitations appear physically relevant in these materials. Nevertheless, we continue to consider this model as we will only aim for a description of the topological classification, which may not be sensitive to such details.
Crucially, in this model there is a two-fold degeneracy at each $\boldsymbol{k}\in\text{BZ}_{\text{3D}}$, which is a consequence \footnote{See, e.g., Chapter 2 of Vanderbilt \cite{VanderbiltBook}.} of a center-of-inversion symmetry and a (fermionic) time-reversal symmetry of (\ref{H_nonmag}); we explicitly demonstrate these symmetries in Appendix \ref{Appendix:Symmetries}. Consequently, eigenvectors of (\ref{H_nonmag}) are highly non-unique, in a more general sense than the usual $\boldsymbol{k}$-dependent phase ambiguity associated with Bloch's theorem. However, as described in Appendix \ref{Appendix:kdotp}, the coupled-Dirac cone effective model that motivated (\ref{H_nonmag}) has a natural set of eigenvectors, and under the substitutions $\hbar v_{\text{D}}k_a\rightarrow A\text{s}k_a$ and $\Delta_{\text{S}}\rightarrow\Delta_{\text{S}}(k_x,k_y)$, those eigenvectors are related to a set of orthogonal eigenvectors of (\ref{H_nonmag}),
\begin{widetext}
\begin{align}
    \ket{\psi_{1,\boldsymbol{k}}}&=\frac{1}{\sqrt{2}}\left(-\frac{\Delta_{\boldsymbol{k}}^*}{\varepsilon(\boldsymbol{k})}\ket{\bar{\psi}_{(0,\uparrow),\boldsymbol{k}}}+\ket{\bar{\psi}_{(1,\uparrow),\boldsymbol{k}}}-\frac{iA(\text{s}k_x+i\text{s}k_y)}{\varepsilon(\boldsymbol{k})}\ket{\bar{\psi}_{(1,\downarrow),\boldsymbol{k}}}\right),\nonumber \\
    \ket{\psi_{2,\boldsymbol{k}}}&=\frac{1}{\sqrt{2}}\left(-\frac{iA(\text{s}k_x-i\text{s}k_y)}{\varepsilon(\boldsymbol{k})}\ket{\bar{\psi}_{(0,\uparrow),\boldsymbol{k}}}+\ket{\bar{\psi}_{(0,\downarrow),\boldsymbol{k}}}-\frac{\Delta_{\boldsymbol{k}}}{\varepsilon(\boldsymbol{k})}\ket{\bar{\psi}_{(1,\downarrow),\boldsymbol{k}}}\right), \nonumber\\
    \ket{\psi_{3,\boldsymbol{k}}}&=\frac{1}{\sqrt{2}}\left(\frac{\Delta_{\boldsymbol{k}}^*}{\varepsilon(\boldsymbol{k})}\ket{\bar{\psi}_{(0,\uparrow),\boldsymbol{k}}}+\ket{\bar{\psi}_{(1,\uparrow),\boldsymbol{k}}}+\frac{iA(\text{s}k_x+i\text{s}k_y)}{\varepsilon(\boldsymbol{k})}\ket{\bar{\psi}_{(1,\downarrow),\boldsymbol{k}}}\right), \nonumber\\
    \ket{\psi_{4,\boldsymbol{k}}}&=\frac{1}{\sqrt{2}}\left(\frac{iA(\text{s}k_x-i\text{s}k_y)}{\varepsilon(\boldsymbol{k})}\ket{\bar{\psi}_{(0,\uparrow),\boldsymbol{k}}}+\ket{\bar{\psi}_{(0,\downarrow),\boldsymbol{k}}}+\frac{\Delta_{\boldsymbol{k}}}{\varepsilon(\boldsymbol{k})}\ket{\bar{\psi}_{(1,\downarrow),\boldsymbol{k}}}\right),
\label{eigenvectorsNonMagneticTB}
\end{align}
\end{widetext}
written here in the basis of Bloch-type vectors $\ket{\bar{\psi}_{(\alpha,\sigma),\boldsymbol{k}}}$ that are assumed smooth over $\text{BZ}_{\text{3D}}$ (recall the discussion below Eq.~(\ref{Hoperator_general})). 
The energy eigenvectors in (\ref{eigenvectorsNonMagneticTB}) are of Bloch's form, satisfying
$\ket{\psi_{n,\boldsymbol{k}}}=\ket{\psi_{n,\boldsymbol{k}+\boldsymbol{G}}}$ for any 
$\boldsymbol{G}\in\Gamma_{\text{3D}}^*=\text{span}_{\mathbb{Z}}(\{2\pi\boldsymbol{x},2\pi\boldsymbol{y},\frac{2\pi}{a}\boldsymbol{z}\})$. When $\varepsilon(\boldsymbol{k})>0$ for all $\boldsymbol{k}\in\text{BZ}_{\text{3D}}$, \textit{i.e.}~when the model is a band insulator at half filling (the case of primary interest), these $\ket{\psi_{n,\boldsymbol{k}}}$ are smooth over $\text{BZ}_{\text{3D}}$. We can also identify from Eqs.~(\ref{eigenvectorsNonMagneticTB}) a $\Gamma^{*}_{\text{3D}}$-periodic unitary matrix $T(\boldsymbol{k})$ relating Wannier and energy Bloch-type vectors,
\begin{align}
    \ket{\bar{\psi}_{(\alpha,\sigma),\boldsymbol{k}}}=\sum_{n=1}^{4}\ket{\psi_{n,\boldsymbol{k}}}T_{n,(\alpha,\sigma)}(\boldsymbol{k}).
\label{smoothFrame}
\end{align}
Since both $\bar{\psi}_{(\alpha,\sigma),\boldsymbol{k}}(\boldsymbol{r})\equiv\braket{\boldsymbol{r}}{\bar{\psi}_{(\alpha,\sigma),\boldsymbol{k}}}$ and $\psi_{n,\boldsymbol{k}}(\boldsymbol{r})\equiv\braket{\boldsymbol{r}}{\psi_{n,\boldsymbol{k}}}$ are of Bloch's form, they have associated $\Gamma_{\text{3D}}$-periodic functions $\bar{u}_{(\alpha,\sigma),\boldsymbol{k}}(\boldsymbol{r})\equiv\braket{\boldsymbol{r}}{\bar{u}_{(\alpha,\sigma),\boldsymbol{k}}}=(2\pi)^{3/2}e^{-i\boldsymbol{k}\cdot\boldsymbol{r}}\bar{\psi}_{(\alpha,\sigma),\boldsymbol{k}}(\boldsymbol{r})$ and $u_{n,\boldsymbol{k}}(\boldsymbol{r})\equiv\braket{\boldsymbol{r}}{u_{n,\boldsymbol{k}}}=(2\pi)^{3/2}e^{-i\boldsymbol{k}\cdot\boldsymbol{r}}\psi_{n,\boldsymbol{k}}(\boldsymbol{r})$, respectively, which are similarly related to one another by that $T(\boldsymbol{k})$ and are smooth (modulo $\exp(-i\boldsymbol{G}\cdot\boldsymbol{r})$ for all $\boldsymbol{G}\in\Gamma_{H}^{*}$) over $\text{BZ}_{\text{3D}}$.
	
The Bloch energy eigenvectors could alternately be chosen 
to simultaneously diagonalize the symmetry operators and the Hamiltonian itself.
However, topological considerations forbid one from making that choice if the goal is to 
compute $\alpha_{\text{CS}}$ in a gauge defined by those energy eigenvectors, as we now describe.
	
Many of the surprising single-particle properties of the various types of topological insulators can be understood by studying the structure of a particular (complex) vector bundle \cite{Panati2007,Monaco2014,Monaco2017}, the Hermitian bundle of occupied Bloch states over the Brillouin zone torus denoted $\mathcal{V}\xrightarrow{\pi_{\mathcal{V}}}\text{BZ}_{d}$. 
Band insulators are special in that the Hilbert space of occupied Bloch states at $\bm{k}\in\text{BZ}_{d}$, $\mathcal{V}_{\boldsymbol{k}}\equiv\text{span}_{\mathbb{C}}(\{\ket{\psi_{n,\boldsymbol{k}'}}:\boldsymbol{k}'=\boldsymbol{k} \text{ and }E_{n}(\boldsymbol{k}')<E_{F}\})$,
has a dimension that is constant through $\text{BZ}_{d}$, 
thus $\mathcal{V}_{\boldsymbol{k}}\cong\mathcal{V}_{\boldsymbol{k}'}$ holds for all $\boldsymbol{k},\boldsymbol{k}'\in\text{BZ}_{d}$.
If there are $N$ fully occupied energy bands then $\mathcal{V}_{\boldsymbol{k}}\cong\mathbb{C}^{N}$ for all ${\boldsymbol{k}}\in\text{BZ}_{d}$.
The essential mathematical object is (isomorphic to) $\bigcup_{\boldsymbol{k}\in\text{BZ}_{d}}\{\boldsymbol{k}\}\times\mathcal{V}_{\boldsymbol{k}}$, which is the smooth manifold that is obtained by attaching to each $\boldsymbol{k}\in\text{BZ}_{d}$ the corresponding $\mathcal{V}_{\boldsymbol{k}}$ and equipping this set with a topology and smooth structure such that the natural projection map $(\boldsymbol{k},\ket{\psi_{\boldsymbol{k}}})\mapsto \boldsymbol{k}$ is smooth \footnote{This is a typical approach to construct a vector bundle given a collection of desired fibers, one associated with each point of the base manifold, and is employed in the textbook example of the tangent bundle (see, e.g., Chapter 3 of Lee \cite{LeeBook}). However, in that prescription an additional choice is required, specifically a local frame (see \textit{Local and Global Frames} in Chapter 10 of Lee \cite{LeeBook}) that is desired to be smooth is identified and the result is that the natural projection map is smooth (Proposition 10.24 of Lee \cite{LeeBook}). In the case of the tangent bundle a natural choice exists, but that is not so in our case. To make technical progress, researchers have found it convenient to identify the desired fibers via a family of projectors $P(\boldsymbol{k})$ that is associated with a particular set of energetically isolated bands since any such $P(\boldsymbol{k})$ is always smooth over the Brillouin zone torus.}.
Although this construction is intuitive, it is convenient to instead consider an isomorphic bundle over $\text{BZ}_{d}$ 
\footnote{In this paper, as is typical in condensed matter physics, we use the symbol $\text{BZ}_{d}$ to denote \textit{both} the Wigner-Seitz cell of $\Gamma^{*}$ and $\mathbb{R}^{d}/\sim_{\Gamma^{*}}\cong\mathbb{T}^{d}$ constructed using the equivalence relation for $\boldsymbol{\kappa},\boldsymbol{\kappa}'\in\mathbb{R}^{d}$ that $\boldsymbol{\kappa}\sim_{\Gamma^{*}} \boldsymbol{\kappa}' :\iff \boldsymbol{\kappa}'=\boldsymbol{\kappa}+\boldsymbol{G}$ for some $\boldsymbol{G}\in\Gamma^{*}$.} with total space defined by
\begin{align*}
	\mathcal{V}\equiv\big\{[\boldsymbol{\kappa},\ket{u}]_{\Gamma^{*}}\in (\mathbb{R}^{d}\times\mathcal{H})/\sim_{\Gamma^{*}} \, : \, \ket{u}\in\text{Ran}P_{\mathcal{V}}(\boldsymbol{\kappa})\big\},
\end{align*}
where
$\mathcal{H}$ is the Hilbert space of $\Gamma$-periodic functions,
$P_{\mathcal{V}}(\boldsymbol{\kappa})=\sum_{n=1}^{N}\ket{u_{n\boldsymbol{\kappa}}}\otimes\bra{u_{n\boldsymbol{\kappa}}}$ is the projector onto the space of $\Gamma$-periodic parts of occupied Bloch functions associated with $\boldsymbol{\kappa}\in\mathbb{R}^{d}$,
and the equivalence class $[\boldsymbol{\kappa},\ket{u}]_{\Gamma^{*}}\subset \mathbb{R}^{d}\times\mathcal{H}$ is defined by $(\boldsymbol{\kappa},\ket{u})\sim_{\Gamma^{*}} (\boldsymbol{\kappa}',\ket{u'}) :\iff \boldsymbol{\kappa}'=\boldsymbol{\kappa}+\boldsymbol{G} \text{ and } u'(\boldsymbol{r})=e^{-i\boldsymbol{G}\cdot\boldsymbol{r}}u(\boldsymbol{r})$ for some $\boldsymbol{G}\in\Gamma^{*}$.
The bundle projection map $\pi_{\mathcal{V}}:\mathcal{V}\rightarrow\text{BZ}_{d}$ is defined by $\pi_{\mathcal{V}}([\boldsymbol{\kappa},\ket{u}]_{\Gamma^{*}})\equiv[\boldsymbol{\kappa}]$.
It has been proved \cite{Panati2007,Monaco2014} that in any band insulator $\mathcal{V}\xrightarrow{\pi_{\mathcal{V}}}\text{BZ}_{d}$ is a vector bundle (in particular, a Hermitian bundle) and is therefore locally trivial: for every $\boldsymbol{k}\in\text{BZ}_{d}$ there exists an open neighborhood $U\subseteq\text{BZ}_{d}$ of $\boldsymbol{k}$ over which $(\pi_{\mathcal{V}}^{-1}(U)\xrightarrow{\pi_{\mathcal{V}}}U)\cong(U\times\mathbb{C}^{N}\xrightarrow{\text{pr}_{1}}U)$ holds.
In general, $\mathcal{V}$ need not be globally trivial; the above is not necessarily satisfied for $U=\text{BZ}_{d}$.
However, it has been shown \cite{Brouder,Panati2007,Monaco2014} that time-reversal symmetry implies $(\mathcal{V}\xrightarrow{\pi_{\mathcal{V}}}\text{BZ}_{d})\cong(\text{BZ}_{d}\times\mathbb{C}^{N}\xrightarrow{\text{pr}_{1}}\text{BZ}_{d})$ is satisfied \footnote{For example, when $d=2$ this is equivalent to the vanishing of the first Chern invariant associated with $\mathcal{V}\xrightarrow{\pi_{\mathcal{V}}}\text{BZ}_{\text{2D}}$.}.

In physics, the topology of $\mathcal V$ often manifests through the possible frames (also called gauge choices) thereof.
A local frame of $\mathcal{V}$ over an open subset $U\subseteq\text{BZ}_{d}$ can be obtained from a collection of maps $\tilde{u}_{i}$ in $\mathscr{U}\subset\mathbb{R}^{d}$ 
(the complete set of representatives of $U$ that is contained in the Wigner-Seitz cell of $\Gamma^{*}$)
defined by $\tilde{u}_{i}(\boldsymbol{\kappa})\equiv(\boldsymbol{\kappa},\ket{\tilde{u}_{i,\boldsymbol{\kappa}}})$ that satisfies $\forall\boldsymbol{\kappa}\in \mathscr{U}:(\ket{\tilde{u}_{i,\boldsymbol{\kappa}}})_{i\in\{1,\ldots,N\}}$ is a basis of $\text{Ran}P_{\mathcal{V}}(\boldsymbol{\kappa})$ (\textit{i.e.}~the related Bloch-type functions $|\tilde{\psi}_{i,\boldsymbol{k}}\rangle$ are a basis of $\mathcal{V}_{\boldsymbol{k}}$ for $\boldsymbol{k}\in[\boldsymbol{\kappa}]$); if the $\ket{\tilde{u}_{i,\boldsymbol{\kappa}}}$ are energy eigenvectors then the gauge is called Hamiltonian.
A general result \footnote{See, e.g., Proposition 10.19 and Corollary 10.20 of Lee \cite{LeeBook}.} is that a vector bundle $E\xrightarrow{\pi}M$ is locally trivial over an open subset $U\subseteq M$ if and only if there exists a smooth frame of $\pi^{-1}(U)\xrightarrow{\pi}U$.
Then, about every $\boldsymbol{k}\in\text{BZ}_{d}$ a smooth local frame of $\mathcal{V}$ always exists and if there is TRS then a smooth global frame of $\mathcal{V}$ exists.
For example, if a 2D insulator is characterized by a nonzero first Chern invariant $C$, then $\mathcal{V}$ is not globally trivial; nonzero $C$ acts as an obstruction to the existence of a smooth global gauge \cite{Panati2007}. 
In this case, that $\mathcal{V}$ is not globally trivial can be understood to manifest in calculations though the fact that the Berry connection cannot be made smooth over $\text{BZ}_{\text{2D}}$ and the integral expression for $C$ then returns a nonzero value.
The $\mathbb{Z}_{2}$ invariant relevant in this work is more subtle. TRS implies that $\mathcal{V}$ is globally trivial, but it has been shown \cite{Vanderbilt2011,Vanderbilt2012,Monaco2014,Monaco2017} that a non-trivial $\mathbb{Z}_2$ invariant acts as an obstruction to the existence of a global gauge that is both smooth and time-reversal-symmetric.
Thus, there need not exist a smooth global Hamiltonian frame of $\mathcal{V}$ \footnote{If the energy bands are not degenerate anywhere in $\boldsymbol{k}$ then the energy eigenvectors necessarily diagonalize the symmetry operators and are not be continuous over $\text{BZ}_{d}$ in a $\mathbb{Z}_2$-odd phase. And if there were degeneracies, but not over the entire $\text{BZ}_{d}$, then only linear combinations of the energy eigenvectors could be smooth in $\boldsymbol{k}$ \cite{Marzari2012}, regardless of the $\mathbb{Z}_2$ classification of the ground state.}.
And since $\alpha_{\text{CS}}$ must be calculated with respect to a smooth global gauge of $\mathcal{V}$, in a $\mathbb{Z}_2$-odd insulator the components of that gauge are topologically forbidden to be energy eigenvectors that simultaneously diagonalize the time-reversal operator.
Indeed, the generic nonexistence of a suitable Hamiltonian gauge means that to compute $\alpha_{\text{CS}}$ requires \cite{Vanderbilt2009} the employment of a Wannierization-like process \cite{Marzari2012} in order to construct Bloch-type functions $|\tilde{\psi}_{i,\boldsymbol{k}}\rangle$ that constitute a smooth global frame of $\mathcal{V}$. Within the model we employ, at half-filling and in the case of a band insulator, time-reversal and inversion symmetry together imply that the two occupied Bloch states are globally degenerate. Any frame associated with these bands is therefore Hamiltonian; every smooth global gauge of this $\mathcal{V}$ is Hamiltonian.
	
\subsubsection{Band Number Truncation in Tight-Binding Models and Crystal Momentum Dependence of Basis States }
\label{Sec:IIIa3}
	
Before we can employ this model to compute $\alpha_{\text{CS}}$, there is one more issue to address. Like most tight-binding models that appear in the literature, the Hamiltonian operator identified by Eq.~(\ref{HlayerTB}) and (\ref{H3d}) is not fully defined since the Bloch-type vectors $\ket{\bar{\psi}_{(\alpha,\sigma),\boldsymbol{k}}}$ that correspond with the operators $\hat{c}_{(\alpha,\sigma),\boldsymbol{k}}$, $\hat{c}^{\dagger}_{(\alpha,\sigma),\boldsymbol{k}}$ are not completely specified \footnote{For example, even in the case of graphene where it is said that the model is written with respect to $p_z$ orbitals, one centered at the position of each ion core, this is not sufficient because $p_z$ orbitals centered at different lattice sites are generally non-orthogonal if they have common support. So these at least need to be orthogonalized. In this sense there is an important distinction between mathematically consistent tight-binding Hamiltonians constructed via WFs and the physically motivated ones constructed from LCAOs.}. This can be particularly problematic when calculating, for example, band-diagonal components of the
Berry connection, in which 
case one must explicitly take $\boldsymbol{k}$-derivatives of the cell-periodic part of Bloch energy eigenvectors. (In $\boldsymbol{k}\cdot\boldsymbol{p}$ models this issue is avoided by construction, since these models are formulated with respect to a constant frame of the space of relevant Bloch states over the subset of $\text{BZ}_{d}$ for which the model is accurate \cite{Ando2005,Marconcini2011}.)
In analytical calculations using tight-binding models it is almost always \footnote{See, e.g., Sec.~II.~C.~1 of Xiao \textit{et al.}~\cite{Niu2009}.} assumed that the WFs (here the $\ket{W_{(\alpha,\sigma),\boldsymbol{R}}}$) with respect to which the model is specified are atomic-like, with negligible overlap between WFs that are associated with different unit cells. 
In this case, the variation of $\ket{\bar{u}_{(\alpha,\sigma),\boldsymbol{k}}}$ with $\boldsymbol{k}$ is negligible (within the above defined equivalence classification) allowing it to remain unspecified.
In this work we \textit{assume} that the Hamiltonian operator (\ref{H3d}) is written in a basis of Bloch-type functions $\ket{\bar{\psi}_{(\alpha,\sigma),\boldsymbol{k}}}$ with corresponding $\ket{\bar{u}_{(\alpha,\sigma),\boldsymbol{k}}}$ that are constant (modulo a phase) over $\text{BZ}_{\text{3D}}$, $\frac{\partial}{\partial k^{a}}\ket{\bar{u}_{(\alpha,\sigma),\boldsymbol{k}}}\equiv 0$ for $a=x$, $y$, or $z$ ($\equiv1$, $2$, or $3$) \footnote{In a related work Ref.~\cite{lei2023afm} in which we consider the TME in anti-ferromagnetic TIs, this approximation requires further consideration.}. While this may seem to be a drastic approximation, and in many cases it is indeed too drastic, we provide a physically motivated justification and argue that it is (at least) not mathematically inconsistent with the assumptions already inherent to our model.

A condition that one could potentially require of tight-binding models,
in which the full electronic Hilbert space of a crystal is truncated to include only the states associated with
a finite number of bands around the Fermi energy, is that the $\boldsymbol{k}$-dependence of the Bloch states be accurately
captured throughout the Brillouin zone. The local $\boldsymbol{k}$-dependence of the $\ket{u_{n,\boldsymbol{k}}}$ 
around some $\boldsymbol{k}_{0}$ can be calculated using $\bm{k}\cdot\bm{p}$ perturbation theory,
which yields
\iffalse
\begin{align*}
\ket{\partial_{a}u_{n,\boldsymbol{k}}}=\sum_{m\neq n}\ket{u_{m,\boldsymbol{k}}}\frac{\bra{u_{m,\boldsymbol{k}}}\partial_{a}H(\boldsymbol{k})\ket{u_{n,\boldsymbol{k}}}}{E_{n,\boldsymbol{k}}-E_{m,\boldsymbol{k}}}.
\end{align*}
\fi
\begin{align}
\frac{\partial}{\partial k^{a}}\ket{u_{n,\boldsymbol{k}}}=\frac{\hbar}{m_{e}}\sum_{m\neq n}\frac{\bra{u_{n,\boldsymbol{k}_{0}}}p^{a}\ket{u_{m,\boldsymbol{k}_{0}}}}{E_{n,\boldsymbol{k}_{0}}-E_{m,\boldsymbol{k}_{0}}}\ket{u_{m,\boldsymbol{k}_{0}}},
\label{eq:kdotp}
\end{align}
where $m_{e}$ is the electron mass.
It follows from Eq.~(\ref{eq:kdotp}) that a good criteria for the reliability of a 
tight-binding model is that all momentum matrix elements between the $\ket{u_{m,\boldsymbol{k}_{0}}}$ corresponding with included bands 
and neglected bands are small at all $\boldsymbol{k}_{0}$ of interest.
When such a set of bands that are dissociated in this sense across the entire 
Brillouin zone can be identified \footnote{If the states associated with all of the energy bands are employed, then in any material a $\bm{k}\cdot\bm{p}$ over the entire Brillouin zone exists (see, e.g., Chapter 2 of Winkler \cite{WinklerBook}) since the set $\{\ket{u_{m,\boldsymbol{k}_{0}}}\}_{n\in\mathbb{Z}}$ at any $\boldsymbol{k}_{0}$ span the space of $\Gamma$-periodic functions \cite{JonesKdotP}. In the application of $\boldsymbol{k}\cdot\boldsymbol{p}$ theory it is typically assumed (often accurately; see, e.g., Cardona \textit{et al.}~\cite{Cardona1966}) that the truncation of that basis is possible.}, 
we say that they satisfy a global isolation condition.
When this global isolation condition is satisfied,
we can always choose the finite-dimensional basis vectors of the included bands to be independent of $\bm{k}$, 
for example they could be the eigenstates at one particular $\boldsymbol{k}_{0}$.
Our expectation is that global isolation holds only when it is implied by chemistry,
{\it i.e.}~the set of energy bands derives from a linear combination of atomic or molecular orbitals.
	
This global isolation condition is probably difficult to satisfy in practice and probably unnecessary for many physically relevant 
calculations, but it greatly simplifies calculations of Berry connections and 
related quantities. When we start from a phenomenological tight-binding model, as in this calculation,
we know nothing about momentum matrix elements between included and neglected bands, so we 
have little choice but to assume the global isolation condition. It is probably common in {\it ab initio}
DFT-derived models that global isolation is not satisfied. In particular, for $\mathbb{Z}_2$ TIs, in which level inversion at one point in the Brillouin zone plays the essential 
physical role, it will never be possible for the set of occupied bands to satisfy the global isolation condition.

To consider the mathematical implications of taking $\frac{\partial}{\partial k^{a}}\ket{\bar{u}_{(\alpha,\sigma),\boldsymbol{k}}}\equiv0$ over $\text{BZ}_{\text{3D}}$, we return to the bundle-theoretic framework. 
In this paper we always consider the model at half-filling. 
Thus, at any $\boldsymbol{k}\in\text{BZ}_{\text{3D}}$, the $\ket{\bar{u}_{(\alpha,\sigma),\boldsymbol{k}}}$ ($\ket{\bar{\psi}_{(\alpha,\sigma),\boldsymbol{k}}}$)
need not be contained in $\text{Ran}P_{\mathcal{V}}(\boldsymbol{k})$ ($\mathcal{V}_{\boldsymbol{k}}$). 
Therefore, assuming the global isolation condition for (the space spanned by) the $\ket{\bar{u}_{(\alpha,\sigma),\boldsymbol{k}}}$ does
not constrain the topology of $\mathcal{V}\xrightarrow{\pi_{\mathcal{V}}}\text{BZ}_{\text{3D}}$.
However, this does have implications for the topology of the vector bundle of all relevant Bloch states, which we denote $\mathcal{B}\xrightarrow{\pi_{\mathcal{B}}}\text{BZ}_{\text{3D}}$. 
The total space $\mathcal{B}$ of that bundle is constructed in a manner similar to $\mathcal{V}$, but now with the fiber at each $\boldsymbol{k}\in\text{BZ}_{\text{3D}}$ isomorphic to $\mathcal{B}_{\boldsymbol{k}}\equiv\text{span}_{\mathbb{C}}(\{\ket{\bar{\psi}_{(\alpha,\sigma),\boldsymbol{k}'}}:\boldsymbol{k}'=\boldsymbol{k}\text{ and }\alpha\in\{0,1\},\sigma\in\{\uparrow,\downarrow\}\})$. 
(That $\mathcal{B}\xrightarrow{\pi_{\mathcal{B}}}\text{BZ}_{\text{3D}}$ is a vector bundle follows from an analogous argument \cite{Panati2007,Monaco2014} to $\mathcal{V}\xrightarrow{\pi_{\mathcal{V}}}\text{BZ}_{\text{3D}}$, which can indeed be applied to any set of isolated energy bands.)
In constructing the model, we have already explicitly
assumed, as one always does in employing tight-binding models,
that $\mathcal{B}\xrightarrow{\pi_{\mathcal{B}}}\text{BZ}_{\text{3D}}$ has a smooth global frame,
and is therefore globally trivial.
	
The assumption that we can label the WFs (\ref{WF}) by $I=(\alpha,\sigma)$ implies the existence of a smooth global frame of the form $(\ket{\bar{u}_{(\alpha,\sigma),\boldsymbol{k}}})_{\alpha\in\{0,1\},\sigma\in\{\uparrow,\downarrow\}}$. 
In particular, this means that $\mathcal{B}\xrightarrow{\pi_{\mathcal{B}}}\text{BZ}_{\text{3D}}$ can be decomposed as a product of vector bundles that correspond with spin-up ($\sigma=\uparrow$) and spin-down ($\sigma=\downarrow$) sectors \footnote{To construct these bundles, the prescription of Prodan \cite{Prodan2009} could be followed. Here, however, we are interested in spin sectors of both the total Bloch bundle and its valence subbundle. In the language of Prodan, the former corresponds with the projector being the identity map and the latter with the projector mapping to the occupied states of the electronic groundstate.}, and that each sector admits a smooth global frame $(\ket{\bar{u}_{(\alpha,\sigma),\boldsymbol{k}}})_{\alpha\in\{0,1\}}$ for a given $\sigma$. 
Thus the triple of Chern numbers (since $d=3$) that characterizes each sector vanishes and $\mathcal{B}\xrightarrow{\pi_{\mathcal{B}}}\text{BZ}_{\text{3D}}$ is characterized by a trivial $\mathbb{Z}_{2}$ invariant \cite{Roy2009,Roy2010}.
Therefore, the topology of $\mathcal{B}\xrightarrow{\pi_{\mathcal{B}}}\text{BZ}_{\text{3D}}$ does not forbid the existence of a smooth and symmetric global frame.
In addition, if we assume the global isolation condition then for any $\boldsymbol{k},\boldsymbol{k}'\in\text{BZ}_{\text{3D}}$ we have $\text{Ran}P_{\mathcal{B}}(\boldsymbol{k})=\text{Ran}P_{\mathcal{B}}(\boldsymbol{k}')$.
Then, while we do not make a general existence argument for a frame with components $\ket{\bar{u}_{(\alpha,\sigma),\boldsymbol{k}}}$ that are constant over $\text{BZ}_{\text{3D}}$, assuming one to exist is (at least) not generically forbidden by the topology of $\mathcal{B}\xrightarrow{\pi_{\mathcal{B}}}\text{BZ}_{\text{3D}}$. This argument applies to any tight-binding model that is specified with respect to WFs whose type labels are assumed to involve a spin index and satisfies the global isolation condition.

\subsection{Calculation of $\alpha_{\text{CS}}$}
\label{Sec:IIIb}
	
As mentioned above, in order to compute $\alpha_{\text{CS}}$ as the $\text{BZ}_{\text{3D}}$-integral of the Chern-Simons 3-form, which is valid strictly for band insulators with corresponding $\mathcal{V}\xrightarrow{\pi_{\mathcal{V}}}\text{BZ}_{\text{3D}}$ that is globally trivial, one is required to work in a smooth global gauge of $\mathcal{V}$.
Since Bloch energy eigenvectors are not generically smooth over $\text{BZ}_{\text{3D}}$, particularly in $\mathbb{Z}_{2}$ TIs, a Hamiltonian gauge choice is not typically viable.
In past works \cite{Vanderbilt2009,Malashevich2010,Essin2010,Mahon2020}, a gauge of that type is induced by the choice of WFs, where WFs are there assumed to be constructed from the (un)occupied energy eigenvectors alone. 
However, the WFs employed to construct tight-binding models are not typically of this type; the gauge $(\ket{\bar{u}_{(\alpha,\sigma),\boldsymbol{k}}})_{\alpha\in\{0,1\},\sigma\in\{\uparrow,\downarrow\}}$ is not valid for this calculation. 
	
As described in Sec.~\ref{Sec:IIIa2}, the model we employ has the positive feature that 
we are guaranteed a smooth global Hamiltonian gauge due to the combination of TRS and each set of isolated bands being completely degenerate over the entire $\text{BZ}_{\text{3D}}$. 
Using the $\ket{u_{n,\boldsymbol{k}}}$ related to (\ref{eigenvectorsNonMagneticTB}), 
we can define a suitable frame $\mathfrak{u}$ of $\mathcal{B}\xrightarrow{\pi_{\mathcal{B}}}\text{BZ}_{\text{3D}}$ pointwise by $\mathfrak{u}_{\boldsymbol{k}}\equiv(\ket{u_{n,\boldsymbol{k}}})_{n\in\{1,2,3,4\}}$; although $\mathfrak{u}$ is technically a frame of $\mathcal{B}$, since it is Hamiltonian it projects to a frame of $\mathcal{V}$.
This gauge choice corresponds to taking the bulk electronic polarization and 
orbital magnetization to be defined with respect to the set of WFs 
$\ket{W_{n,\boldsymbol{R}}}=\sqrt{\Omega_{uc}/(2\pi)^{3}}\int_{\text{BZ}_{\text{3D}}}e^{-i\boldsymbol{k}\cdot\boldsymbol{R}}\ket{\psi_{n,\boldsymbol{k}}}d^{3}k$. 
It has been shown \cite{Zhang2008,Vanderbilt2009,Malashevich2010,Essin2010,Mahon2020} that, with respect to any smooth global Hamiltonian gauge $\mathfrak{u}$ defined by $\mathfrak{u}_{\boldsymbol{k}}\equiv(\ket{u_{n,\boldsymbol{k}}})_{n\in\mathbb{N}}$,
\begin{widetext}
\begin{align}
    \alpha^{\mathfrak{u}}_{\text{CS}}
    &=-\frac{e^2}{2\hbar c}\epsilon^{abd}\int_{\text{BZ}_{\text{3D}}}\frac{d^{3}k}{(2\pi)^3}\left(\sum_{vv'}\xi^a_{vv'}(\boldsymbol{k})\frac{\partial}{\partial k^{b}}\xi^d_{v'v}(\boldsymbol{k})-\frac{2i}{3}\sum_{vv'v_1}\xi^a_{vv'}(\boldsymbol{k})\xi^b_{v'v_1}(\boldsymbol{k})\xi^d_{v_1v}(\boldsymbol{k})\right),
\label{alphaCS}
\end{align}
\end{widetext}
where the sums are over the initially occupied band indices (here $v,v',v_1\in\{1,2\}$) and $\xi^a_{nm}$ are the components of the non-Abelian Berry connection induced by $\mathfrak{u}$,
\begin{align}
    \xi^a_{nm}(\boldsymbol{k})\equiv i\braket{u_{n,\boldsymbol{k}}}{\frac{\partial}{\partial k^{a}}u_{m,\boldsymbol{k}}}.
    \label{connection1}
\end{align}
Although Eq.~(\ref{alphaCS}) is gauge dependent, transforming from one appropriate gauge of $\mathcal{V}$ to another, be it a Hamiltonian gauge or otherwise, can only change its value by an integer multiple of $e^2/hc$ \cite{Zhang2008,Vanderbilt2009}; that is, there is a quantum of indeterminacy associated with $\alpha_{\text{CS}}$.

 \begin{table*}
	\centering
	\begin{tabular}{c c}
		\begin{tabular}{||c c c||}
			\hline
			\text{\quad} $\Delta_{\text{D}}/\Delta_{\text{S}}$\text{\quad} & \text{\quad} $4B/\Delta_{\text{S}}$
			\text{\quad} & \text{\quad} $\alpha^{\mathfrak{u}}_{\text{CS}}$ $(\frac{e^2}{hc})$ \text{\quad} \\ [0.5ex] 
			\hline\hline
			\text{all} & 0 & 0 \\
			\hline
			$\pm 0.1$ & \begin{tabular}{@{}c@{}}0.5 \\ -0.5\end{tabular} & \begin{tabular}{@{}c@{}}1/2 \\ 0\end{tabular} \\
			\hline
			$\pm 0.1$ & \begin{tabular}{@{}c@{}}0.7 \\ -0.7\end{tabular} & \begin{tabular}{@{}c@{}}0 \\ 0\end{tabular} \\
			\hline
			$\pm 0.1$ & \begin{tabular}{@{}c@{}}1.0 \\ -1.0\end{tabular} & \begin{tabular}{@{}c@{}}-1 \\ 0\end{tabular} \\
			\hline
			$\pm 0.1$ & \begin{tabular}{@{}c@{}}1.2 \\ -1.2\end{tabular} & \begin{tabular}{@{}c@{}}0 \\ 0\end{tabular} \\
			\hline
			$\pm 1.1$ & \begin{tabular}{@{}c@{}}1.2 \\ -1.2\end{tabular} & \begin{tabular}{@{}c@{}}-1/2 \\ 1/2\end{tabular} \\
			\hline
		\end{tabular} \text{\qquad\qquad} &
		\begin{tabular}{||c c c||}
			\hline
			\text{\quad} $\Delta_{\text{D}}/\Delta_{\text{S}}$\text{\quad} & \text{\quad} $4B/\Delta_{\text{S}}$ \text{\quad} & \text{\quad} $\alpha^{\mathfrak{u}}_{\text{CS}}$ $(\frac{e^2}{hc})$ \text{\quad} \\ [0.5ex] 
			\hline\hline
			0 & \text{all} & 0 \\ 
			\hline
			$\pm 0.5$ & \begin{tabular}{@{}c@{}}0.3 \\ -0.3\end{tabular} & \begin{tabular}{@{}c@{}}1/2 \\ 0\end{tabular} \\
			\hline
			$\pm 0.8$ & \begin{tabular}{@{}c@{}}0.3 \\ -0.3\end{tabular} & \begin{tabular}{@{}c@{}}-1/2 \\ 0\end{tabular} \\
			\hline
			$\pm 1.1$ & \begin{tabular}{@{}c@{}}0.3 \\ -0.3\end{tabular} & \begin{tabular}{@{}c@{}}0 \\ 1/2\end{tabular} \\
			\hline
			$\pm 1.4$ & \begin{tabular}{@{}c@{}}0.3 \\ -0.3\end{tabular} & \begin{tabular}{@{}c@{}}0 \\ -1/2\end{tabular} \\ 
			\hline
			$\pm 1.7$ & \begin{tabular}{@{}c@{}}0.3 \\ -0.3\end{tabular} & \begin{tabular}{@{}c@{}}0 \\ 0\end{tabular} \\ 
			\hline
		\end{tabular}\\
	\end{tabular}
	\caption{Representative set of values of $\alpha^{\mathfrak{u}}_{\text{CS}}$ obtained from numerical integration of Eq.~(\ref{alphaCSintegralTB}) for (arbitrarily chosen) $\Delta_{\text{S}}=190$ meV. From these data the topological phase diagram (Fig.~\ref{Fig:phaseDiagram}) can be deduced. When the values of the crystal parameters are varied such that the band gap does not close, the value to which $\alpha^{\mathfrak{u}}_{\text{CS}}$ evaluates is unchanged.}
	\label{Table:alphaCSresults}
\end{table*}

\begin{figure*}
	\centering
	\includegraphics[width=0.65\textwidth]{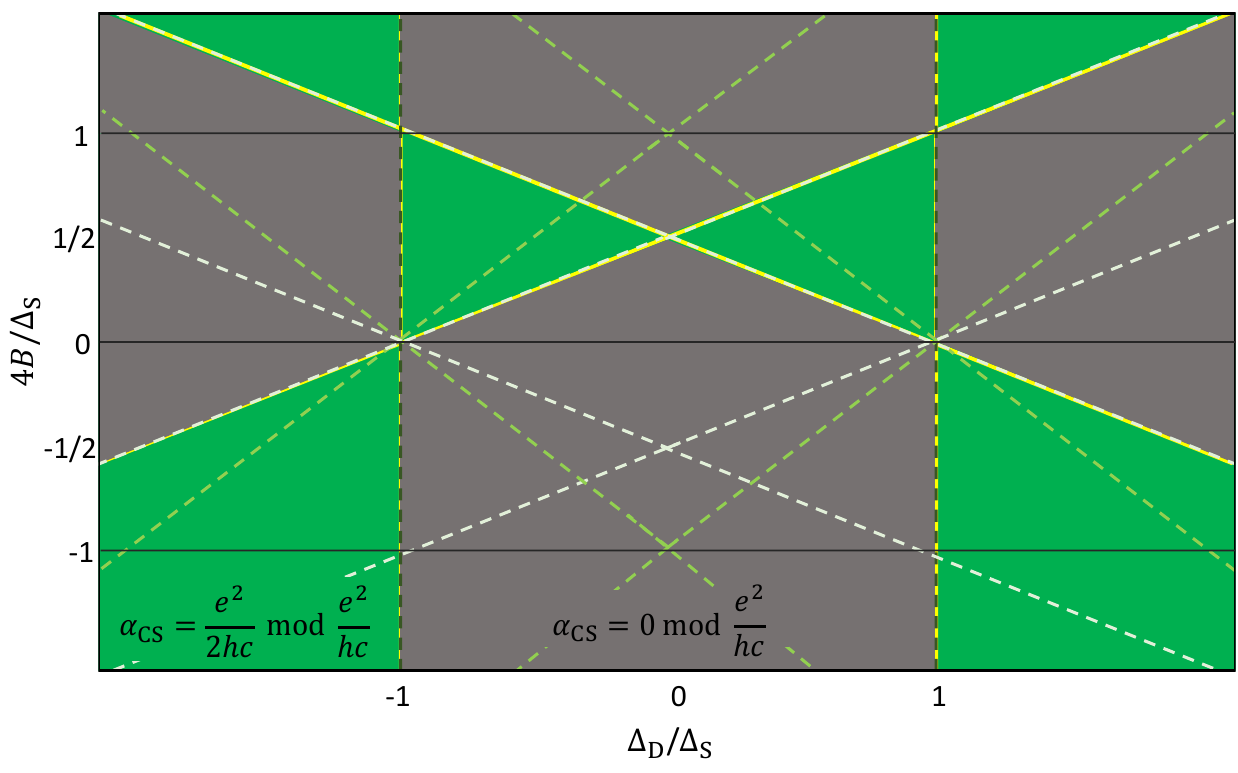}
	\caption{Phase diagram plotting the $\mathbb{Z}_{2}$ topological classification of the band insulating electronic ground state of $\hat{H}_{\text{reg}}^{(\text{3D})}$ (specified in Eq.~(\ref{H_nonmag}) and Sec.~\ref{Sec:IIIa3}) at half-filling. The darkest (vertical) dashed lines identify the points in parameter space at which a band gap closing occurs due to one of Eq.~(\ref{degenRelations1a},\ref{degenRelations2a}) evaluating to zero, the medium dashed lines identify points at which one of Eq.~(\ref{degenRelations1b},\ref{degenRelations2b}) evaluates to zero, and the lightest dashed lines identify points at which one of Eq.~(\ref{degenRelations1c},\ref{degenRelations2c}) evaluates to zero.}
	\label{Fig:phaseDiagram}
\end{figure*}

We now calculate $\alpha^{\mathfrak{u}}_{\text{CS}}$ via Eq.~(\ref{alphaCS}). Assuming that $\frac{\partial}{\partial k^{a}}\ket{\bar{u}_{(\alpha,\sigma),\boldsymbol{k}}}\equiv0$ (see Sec.~\ref{Sec:IIIa3}), Eq.~(\ref{connection1}) becomes
\begin{align}
    \xi^a_{nm}(\boldsymbol{k})=-\mathcal{T}^a_{nm}(\boldsymbol{k}),
    \label{connection}
\end{align}
where we have used $\ket{u_{n,\boldsymbol{k}}}=\sum_{\alpha,\sigma}\ket{\bar{u}_{(\alpha,\sigma),\boldsymbol{k}}}T^{\dagger}_{(\alpha,\sigma),n}(\boldsymbol{k})$ (obtained from (\ref{smoothFrame}) for $\alpha\in\{1,2\}$ and $\sigma\in\{\uparrow,\downarrow\}$) and defined the Hermitian matrix $\mathcal{T}^a$ populated by elements
\begin{align}
    \mathcal{T}^a_{nm}(\boldsymbol{k})\equiv i\sum_{\alpha,\sigma}\left(\frac{\partial}{\partial k^{a}}T_{n,(\alpha,\sigma)}(\boldsymbol{k})\right)T^{\dagger}_{(\alpha,\sigma),m}(\boldsymbol{k}).
\label{W}
\end{align}
% (The expression for $\mathcal{T}^a(\boldsymbol{k})$ in (\ref{W}) is sometimes referred to as the 
% termed the Maurer-Cartan form, and sometimes as a pure gauge connection.)
Using Eqs.~(\ref{eigenvectorsNonMagneticTB},\ref{smoothFrame}) in Eq.~(\ref{W}), 
we obtain
\begin{widetext}
\begin{gather}
    \mathcal{T}^{x}_{11}(\boldsymbol{k})=-\frac{A^2\text{c}k_{x}\text{s}k_{y}+2B\Delta_{\text{D}} \text{s}k_{x} \text{s}ak_{z}}{2\varepsilon(\boldsymbol{k})^2} \text{, } \mathcal{T}^{y}_{11}(\boldsymbol{k})=\frac{A^2\text{s}k_{x}\text{c}k_{y}-2B\Delta_{\text{D}}\text{s}k_{y}\text{s}ak_{z}}{2\varepsilon(\boldsymbol{k})^2} \text{, } \mathcal{T}^{z}_{11}(\boldsymbol{k})=-\frac{a\Delta_{\text{D}}\big(\Delta_{\text{D}}+\Delta_{\text{S}}(k_x,k_y)\text{c}ak_{z}\big)}{2\varepsilon(\boldsymbol{k})^2},\nonumber\\
    \mathcal{T}^{x}_{12}(\boldsymbol{k})=\frac{A\big(2B+\text{c}k_{x}(e^{iak_{z}}\Delta_{\text{D}}+\Delta_{\text{S}}-4B+2B\text{c}k_{y})-2iB\text{s}k_{x}\text{s}k_{y}\big)}{2\varepsilon(\boldsymbol{k})^2} \text{, }\nonumber\\	\mathcal{T}^{y}_{12}(\boldsymbol{k})=-\frac{iA\big(2B+\text{c}k_{y}(e^{iak_{z}}\Delta_{\text{D}}+\Delta_{\text{S}}-4B+2B\text{c}k_{x})+2iB\text{s}k_{x}\text{s}k_{y}\big)}{2\varepsilon(\boldsymbol{k})^2}\text{, } \mathcal{T}^{z}_{12}(\boldsymbol{k})=-\frac{a Ae^{iak_{z}}\Delta_{\text{D}}(i\text{s}k_{x}+\text{s}{k}_y)}{2\varepsilon(\boldsymbol{k})^2}.
    \label{connectionCompTB}
\end{gather}
All other components are related to those given above.
Note in particular that $\mathcal{T}^{a}$ is Hermitian, $\mathcal{T}^{a}_{nm}=(\mathcal{T}^{a}_{mn})^*$, and that by explicit calculation it can be shown
\begin{gather}
    \mathcal{T}^{a}_{11}=\mathcal{T}^{a}_{33} \text{, } \mathcal{T}^{a}_{22}=\mathcal{T}^{a}_{44} \text{, } \mathcal{T}^{a}_{11}=-\mathcal{T}^{a}_{22},\nonumber\\ 
    \mathcal{T}^{a}_{13}=-\mathcal{T}^{a}_{11} \text{, } \mathcal{T}^{a}_{24}=-\mathcal{T}^{a}_{22} \text{, } \mathcal{T}^{a}_{34}=\mathcal{T}^{a}_{12} \text{, } \mathcal{T}^{a}_{14}=\mathcal{T}^{a}_{32}=-\mathcal{T}^{a}_{12}.
\label{connectionCompTB2}
\end{gather}
The relations (\ref{connectionCompTB2}) need not be satisfied for a different choice of energy eigenvectors. Using Eqs.~(\ref{connection},\ref{connectionCompTB},\ref{connectionCompTB2}), the integrand of the first term of Eq.~(\ref{alphaCS}) is here found to be
		\iffalse
		\begin{align}
			\epsilon^{abc}\sum_{v=1}^{2}\xi^a_{vv}\partial_b\xi^c_{vv}=-\frac{aA^{2}\Delta_{\text{D}}(\Delta_{\text{D}}\text{c}k_{x}\text{c}k_{y}+(2B\text{c}k_{y}+\text{c}k_{x}(2B+(\Delta_{\text{S}}-4B)\text{c}k_{y}))\text{c}ak_{z})}{\varepsilon(\boldsymbol{k})^4},
		\end{align}
		plus
		\begin{align}
			\epsilon^{abc}\big(\xi^a_{12}\partial_b\xi^c_{21}+\xi^a_{21}\partial_b\xi^c_{12}\big)=-\frac{2aA^{2}\Delta_{\text{D}}(\Delta_{\text{D}}\text{c}k_{x}\text{c}k_{y}+(2B\text{c}k_{y}+\text{c}k_{x}(2B+(\Delta_{\text{S}}-4B)\text{c}k_{y}))\text{c}ak_{z})}{\varepsilon(\boldsymbol{k})^4},
		\end{align}
		and of the second term is
		\begin{align}
			-\frac{2i}{3}\epsilon^{abc}\sum_{v,v',v_1=1}^{2}\xi^a_{vv'}\xi^b_{v'v_1}\xi^c_{v_1v},
		\end{align}
		\fi
\begin{align}
    \epsilon^{abc}\sum_{v,v'=1}^{2}\xi^a_{vv'}\frac{\partial}{\partial k^{b}}\xi^c_{v'v}=-\frac{3aA^{2}\Delta_{\text{D}}\left(\Delta_{\text{D}}\text{c}k_{x}\text{c}k_{y}+\Big(2B(\text{c}k_{x}+\text{c}k_{y})+(\Delta_{\text{S}}-4B)\text{c}k_{x}\text{c}k_{y}\Big)\text{c}ak_{z}\right)}{\varepsilon(\boldsymbol{k})^4}.
\end{align}
The integrand of the second term of Eq.~(\ref{alphaCS}) is (in this case) $-1/3$ of the first.
Indeed, applying the general relation $\epsilon^{abc}\partial_{b}\xi^{c}_{nm}=i\epsilon^{abc}\sum_{r}\xi^b_{nr}\xi^c_{rm}$ (we now adopt the shorthand $\partial_{a}\equiv\partial/\partial k^{a}$) to this case (such that $r\in\{1,2,3,4\}$) in combination with the relations (\ref{connectionCompTB2}), one can explicitly show that $\epsilon^{abc}\sum_{v,v'=1}^{2}\xi^a_{vv'}\partial_{b}\xi^c_{v'v}=12i\epsilon^{abc}\xi^a_{11}\xi^b_{12}\xi^c_{21}=2i\big(\epsilon^{abc}\sum_{v,v',v_1=1}^{2}\xi^a_{vv'}\xi^b_{v'v_1}\xi^c_{v_1v}\big)$.
\iffalse
\begin{align}
    -\frac{2i}{3}\epsilon^{abc}\sum_{v,v',v_1=1}^{2}\xi^a_{vv'}\xi^b_{v'v_1}\xi^c_{v_1v}=\frac{aA^{2}\Delta_{\text{D}}(\Delta_{\text{D}}\text{c}k_{x}\text{c}k_{y}+(2B\text{c}k_{y}+\text{c}k_{x}(2B+(\Delta_{\text{S}}-4B)\text{c}k_{y}))\text{c}ak_{z})}{\varepsilon(\boldsymbol{k})^4}
\end{align}
\fi
With this, we obtain the final expression
\begin{align}
    \alpha^{\mathfrak{u}}_{\text{CS}}
    &=\frac{e^2}{\hbar c}\int_{-\pi/a}^{\pi/a}\frac{d(ak_z)}{2\pi}\int_{-\pi}^{\pi}\frac{dk_{x}dk_{y}}{(2\pi)^2}\frac{\Delta_{\text{D}}'\left(\Delta_{\text{D}}'\text{c}k_{x}\text{c}k_{y}+\Big(2B'(\text{c}k_{x}+\text{c}k_{y})+(\Delta_{\text{S}}'-4B')\text{c}k_{x}\text{c}k_{y}\Big)\text{c}ak_{z}\right)}{\varepsilon'(\boldsymbol{k})^4}.
\label{alphaCSintegralTB}
\end{align}
\end{widetext}
We have noted that $\alpha^{\mathfrak{u}}_{\text{CS}}$ is invariant under a change in energy scale and 
used this to remove the explicit dependence on $A$ in Eq.~(\ref{alphaCSintegralTB}) through the introduction of scaled parameters $B'\equiv B/A$, $\Delta_{\text{D}}'\equiv\Delta_{\text{D}}/A$ and $\Delta_{\text{S}}'\equiv\Delta_{\text{S}}/A$, and $\varepsilon'(\boldsymbol{k})\equiv\varepsilon(\boldsymbol{k})/A$.
In Table \ref{Table:alphaCSresults} we list values of $\alpha^{\mathfrak{u}}_{\text{CS}}$ obtained by 
extrapolating numerical estimates of the integrals over $\text{BZ}_{\text{3D}}$ in Eq.~(\ref{alphaCSintegralTB}) to
convergence for various combinations of model parameters; we were not able to perform an analytical integration. We are careful to avoid sets of parameters for which the band gap (at half-filling) vanishes since Eq.~(\ref{alphaCS}) applies only to band insulators.
%In fact, after obtaining Eq.~(\ref{alphaCSintegralTB}) we became aware of a previous work \cite{Franz2010} in which a formally similar expression to (\ref{alphaCSintegralTB}) was found and a similar issue encountered.

To obtain the topological phase diagram (Fig.~\ref{Fig:phaseDiagram}) from the data in Table \ref{Table:alphaCSresults}, we use that the numerical value of the $\alpha_{\text{CS}}^{\mathfrak{u}}$ integral is piecewise constant 
in regions of the model's parameter space over which the band gap is nonzero. That value can change when the parameters are varied in such a way that the band gap vanishes at some point in $\text{BZ}_{\text{3D}}$. From Eq.~(\ref{energyNonmagnetic})
we see that this can happen only when $\varepsilon(\boldsymbol{k})=0$,
that $\varepsilon(\boldsymbol{k})$ can vanish only at those $\bm{k}\in\text{BZ}_{\text{3D}}$ for which $k_{x},k_{y}\in\{0,\pm\pi\}$ and $k_{z}\in\{0,\pm\pi/a\}$, and that the condition for band gap closings
depends only on the ratio of the scaled parameters. 
The end result is that band gap closings occur along lines in $(\Delta'_{\text{D}}/\Delta'_{\text{S}})$--$(B'/\Delta'_{\text{S}})$ space (equivalently, $(\Delta_{\text{D}}/\Delta_{\text{S}})$--$(B/\Delta_{\text{S}})$ space).
Specifically, band gaps close only when at least one of the following energies vanishes: 
\begin{subequations}
    \begin{align}
    \varepsilon(0,0,0)&=|\Delta_{\text{S}}+\Delta_{\text{D}}|,\label{degenRelations1a}\\
    \varepsilon(\pi,0,0)&=\varepsilon(0,\pi,0)=|\Delta_{\text{S}}+\Delta_{\text{D}}-4B|,\label{degenRelations1b}\\
    \varepsilon(\pi,\pi,0)&=|\Delta_{\text{S}}+\Delta_{\text{D}}-8B|,\label{degenRelations1c}
    \end{align}
    \label{degenRelations1}
\end{subequations}
\begin{subequations}
    \begin{align}
    \varepsilon(0,0,\pi/a)&=|\Delta_{\text{S}}-\Delta_{\text{D}}|,\label{degenRelations2a}\\
    \varepsilon(\pi,0,\pi/a)&=\varepsilon(0,\pi,\pi/a)=|\Delta_{\text{S}}-\Delta_{\text{D}}-4B|,\label{degenRelations2b}\\
    \varepsilon(\pi,\pi,\pi/a)&=|\Delta_{\text{S}}-\Delta_{\text{D}}-8B|. \label{degenRelations2c}
    \end{align}
    \label{degenRelations2}
\end{subequations}
	
The topological phase diagram in $(\Delta_{\text{D}}/\Delta_{\text{S}})$--$(B/\Delta_{\text{S}})$ space is plotted in Fig.~\ref{Fig:phaseDiagram}. If $\Delta_{\text{D}}/\Delta_{\text{S}}=\mp1$ then from (\ref{degenRelations1a},{\ref{degenRelations2a}}) $\varepsilon(\boldsymbol{k})=0$ at either $(k_{x},k_{y},k_{z})=(0,0,0)$ or $(0,0,\pi/a)$. 
If $\Delta_{\text{D}}/\Delta_{\text{S}}=\mp(1-4B/\Delta_{\text{S}})$
then from (\ref{degenRelations1b},\ref{degenRelations2b}) $\varepsilon(\boldsymbol{k})=0$ at either $(\pi,0,0)$ and 
$(0,\pi,0)$ or $(\pi,0,\pi/a)$ and $(0,\pi,\pi/a)$. Finally, if $\Delta_{\text{D}}/\Delta_{\text{S}}=\mp(1-8B/\Delta_{\text{S}})$ then from (\ref{degenRelations1c},{\ref{degenRelations2c}}) $\varepsilon(\boldsymbol{k})=0$ at either $(\pi,\pi,0)$ or $(\pi,\pi,\pi/a)$.
In Fig.~\ref{Fig:phaseDiagram} we use dashed lines to identify the sets of points in parameter space where at least one of the energies in (\ref{degenRelations1}) or (\ref{degenRelations2}) vanishes. 
In the region of parameter space that we consider, we find that a topological phase transition can occur only if a band gap at $(0,0,0)$ or $(0,0,\pi/a)$ closes and re-opens, or one at $(\pi,\pi,0)$ or $(\pi,\pi,\pi/a)$ closes and re-opens (\textit{i.e.}~only along the lines defined by the vanishing of Eqs.~(\ref{degenRelations1a},{\ref{degenRelations2a}}) or (\ref{degenRelations1c},{\ref{degenRelations2c}})). When band gaps at $(\pi,0,0)$ and $(0,\pi,0)$ or $(\pi,0,\pi/a)$ and $(0,\pi,\pi/a)$ close and re-open, the $\mathbb{Z}_{2}$ topological classification of $\mathcal{V}\xrightarrow{\pi_{\mathcal{V}}}\text{BZ}_{\text{3D}}$ remains unchanged, but the value of $\alpha^{\mathfrak{u}}_{\text{CS}}$ can shift by an integer multiple of $e^2/hc$.
	
In Sec.~\ref{Sec:II}, the parameter $B$ did not appear. This can be understood by recalling that the tight-binding model we employ was developed as a lattice regularization of an effective model that assumes that the low-energy states of the materials of interest are only near $(k_x,k_y)=(0,0)$; this is a property of our lattice model only when $B/\Delta_{\text{S}}<0$ and $|\Delta_{\text{S}}| \approx |\Delta_{\text{D}}|$. 
In that case our expectation is that the largest contributions to the integrand of (\ref{alphaCSintegralTB}) 
come from near the line $(k_x,k_y)=(0,0)$, since $\varepsilon(\boldsymbol{k})$ is smallest there.
Expanding our expression (\ref{alphaCSintegralTB}) for $\alpha^{\mathfrak{u}}_{\text{CS}}$ to first order in $k_x$ and $k_y$ about $(0,0)$, we find that the terms involving $B$ cancel out and that 
\begin{align}
    \alpha^{\mathfrak{u}}_{\text{CS}}&\approx\frac{e^2}{\hbar c}\int_{-\pi/a}^{\pi/a}\frac{d(ak_z)}{2\pi}\int_{\mathbb{R}^2}\frac{dk_{x}dk_{y}}{(2\pi)^2}\frac{\Delta_{\text{D}}'\left(\Delta_{\text{D}}'+\Delta_{\text{S}}'\text{c}ak_{z}\right)}{\left(k_{x}^2+k_{y}^2+|\Delta_{k_{z}}'|^2\right)^2}\nonumber\\
    &=\begin{cases}
        0, & \text{if } |\Delta_{\text{S}}|>|\Delta_{\text{D}}|\\
        \frac{e^2}{2hc}, & \text{if } |\Delta_{\text{S}}|<|\Delta_{\text{D}}|
    \end{cases},
\label{eq:continuumapprox}
\end{align}
where we have artificially extended the domain of integration for $k_{x},k_{y}$ from the subset of $\text{BZ}_{\text{2D}}$ 
near the expansion line to $\mathbb{R}^{2}$.
For $B/\Delta_{\text{S}}<0$ and $|\Delta_{\text{D}}| \approx |\Delta_{\text{S}}|$, 
Eq.~(\ref{eq:continuumapprox}) is consistent with Fig.~\ref{Fig:phaseDiagram}. 
A similar approximate expression for $\alpha_{\text{CS}}$ was previously derived by Rosenberg and Franz \cite{Franz2010}
for models hosting Dirac-like low-energy states. We caution, however, that 
topological index calculations like this one, which focus only on contributions from regions near certain lines or 
points in $\boldsymbol{k}$-space, can fail. An explicit example is provided in Appendix \ref{Appendix:kdotp}. 

\section{Atomic-like and itinerant contributions to $\alpha_{\text{CS}}$}
\label{Sec:IV}
In Sec.~\ref{Sec:II} we employed a coupled-Dirac cone model of a multi-layer thin film -- which had a vanishing magnetic exchange mass (\textit{i.e.}~no magnetic dopants) in the interior layers, but allowed a finite exchange mass in outermost layers -- and found (see Fig.~\ref{fig:alphathickness}) that a perpendicular magnetic field can induce a charge 
density only near the surfaces, that this occurs only in the presence of 
local magnetic dopants, and that the corresponding magnetoelectric coefficient is quantized in sufficiently thick films only when the dopant configuration in the top and bottom layers is opposite.
In Sec.~\ref{Sec:GeneralLinearResponse} we argued from physical grounds that a meaningful bulk $\bm{P}^{(B)}$, particularly in non-magnetic $\mathbb{Z}_{2}$ TIs, must (at least) involve an implicit consideration of surfaces to account for these magnetic-dopant requirements.
We now ask whether there is a \textit{purely} bulk manifestation of those requirements.
	
To begin, we recall the equivalence of the susceptibilities in Eq.~(\ref{alpha}), which is at least valid
in bulk insulators for which $\mathcal{V}\xrightarrow{\pi_{\mathcal{V}}}\text{BZ}_{\text{3D}}$ is globally trivial.
We focus on the macroscopic electronic orbital magnetization $\boldsymbol{M}$ and recall from the modern theory \cite{Resta2005,Resta2006} (or from other formulations of the microscopic response theory \cite{Mahon2019}) that
\begin{align}
\boldsymbol{M} = \bar{\boldsymbol{M}}+\tilde{\boldsymbol{M}},
\label{Mdecomp}
\end{align}
where $\bar{\boldsymbol{M}}$ is called the atomic-like contribution and $\tilde{\boldsymbol{M}}$ the itinerant contribution. (In the modern theory these are typically denoted $\boldsymbol{M}_{\text{LC}}$ and $\boldsymbol{M}_{\text{IC}}$, respectively.)
In an unperturbed bulk insulator, or when the orbital magnetization is induced by a uniform electric and/or magnetic Maxwell field, $\boldsymbol{M}$ is uniform and static, as are $\bar{\boldsymbol{M}}$ and $\tilde{\boldsymbol{M}}$ \cite{Mahon2020a}. In the modern theory it has been argued \cite{Resta2005} that in ferromagnetic ($\boldsymbol{M}^{(0)}\neq\boldsymbol{0}$) bulk insulators with isolated energy bands, interior and surface currents 
in finite samples thereof can be used to uniquely determine $\bar{\boldsymbol{M}}^{(0)}$ and $\tilde{\boldsymbol{M}}^{(0)}$, respectively \footnote{In the case of a topologically trivial bulk insulator, the expressions for the electronic polarization and orbital magnetization that result from the different approaches \cite{Resta2005,Resta2006} and \cite{Mahon2019} agree. However, the proposed relation to interior and surface currents only arises in the former.}; we use the superscript $(0)$ to label spontaneous orbital magnetizations and later use $(E)$ to label magnetization that arises in linear response to the electric field. However, when there are degeneracies within the band structure, the decomposition of $\boldsymbol{M}^{(0)}$ as a sum of $\bar{\boldsymbol{M}}^{(0)}$ and $\tilde{\boldsymbol{M}}^{(0)}$ is gauge dependent \cite{Resta2006}. In this case, interior and surface currents do not uniquely determine $\bar{\boldsymbol{M}}^{(0)}$ and $\tilde{\boldsymbol{M}}^{(0)}$; additional data related to the complete set of WFs being employed is required. Still, the partition into contributions associated with
interior and surface currents may retain meaning. 
To investigate this possibility, we explicitly study the expressions for $\bar{\boldsymbol{M}}^{(E)}$ and $\tilde{\boldsymbol{M}}^{(E)}$, which, like their unperturbed counterparts, are individually gauge dependent, but their sum $\boldsymbol{M}^{(E)}$ is unique only modulo $e^2\bm{E}/hc$ \cite{Zhang2008,Vanderbilt2009}.

In the approach of the modern theory, the expressions for the atomic-like $\bar{\boldsymbol{M}}^{(E)}$ and itinerant $\tilde{\boldsymbol{M}}^{(E)}$ contributions to $\boldsymbol{M}^{(E)}$ (Eq.~(6) of Malashevich \textit{et al.}~\cite{Malashevich2010}) are given in terms of states that implicitly involve the electric field $\bm{E}$.
This is inconvenient for our purposes. 
As mentioned in Sec.~\ref{Sec:GeneralLinearResponse}, there exists a microscopic approach \cite{Mahon2019}, motivated by that of PZW, that identifies atomic-like and itinerant contributions to $\bm{M}$ by different means, but yields the same expressions in an unperturbed bulk insulator \cite{Mahon2020}. 
Indeed the microscopic approach reproduces the magnetoelectric susceptibility (\ref{alpha+cs}) derived via the modern theory.
At linear response to a uniform dc electric field the expressions resulting from those distinct approaches are similar in that $\bar{\boldsymbol{M}}^{(E)}$ originates from the effect of $\bm{E}$ on the electronic dynamics directly (called dynamical modifications; see Sec.~3.~C.~1 of Mahon and Sipe (MS) \cite{Mahon2020}), while $\tilde{\boldsymbol{M}}^{(E)}$ originates from a combination of dynamical modifications and changes in the form of the relevant matrix elements due to $\bm{E}$ (called compositional modifications).
A familiar example of this is paramagnetic and diamagnetic response of an atom to an electromagnetic field \footnote{See, e.g., Sec.~2.2 of Swiecicki \cite{Swiecicki2014}.}, 
which can be understood as arising from dynamical and compositional modifications, respectively.
Below we employ the expressions derived by MS \cite{Mahon2020}.

\iffalse
In $\mathbb{Z}_2$ TIs the expressions for $\bar{\boldsymbol{M}}^{(E)}$ and $\tilde{\boldsymbol{M}}^{(E)}$ derived via the modern theory \cite{Essin2010,Malashevich2010} and those from the microscopic approach \cite{Mahon2019,Mahon2020} agree. However, in the microscopic approach the linear response of a general quantity is naturally partitioned into so-called dynamical and compositional modifications (see Sec.~3 C 1 of Mahon and Sipe (MS) \cite{Mahon2020}). 
Dynamical modifications generally arise from the effect of an electromagnetic field on the electronic dynamics directly, while compositional modifications arise from a change in the form of the relevant matrix elements due to that field; a familiar example of this is paramagnetic and diamagnetic response of an atom to an electromagnetic field, 
which can be understood as arising from dynamical and compositional modifications, respectively.
Our comments below rely on the ability to distinguish these two types of modifications
and are therefore framed in terms of the microscopic approach to linear response. 
\fi
	
Working in the smooth global Hamiltonian gauge $\mathfrak{u}$ defined by the Bloch energy eigenvectors (\ref{eigenvectorsNonMagneticTB}) and using the degeneracy of the energy bands (see Eq.~(\ref{energyNonmagnetic})), Eq.~(70) of MS for the atomic-like contribution evaluates to 
\footnote{Working in a smooth and periodic Hamiltonian gauge, which, we reiterate, need not always exist, amounts to taking $U(\boldsymbol{k})=\text{id}_{4\times4}$ (where $v\in\{1,2\}$, $c\in\{3,4\}$) and therefore $\mathcal{W}_{nm}^{a}(\boldsymbol{k})\equiv0$ in the expressions derived in Ref.~\cite{Mahon2020,Mahon2020a}. 
In those works it was assumed the electronic ground state is such that there exists a $\Gamma^{*}$-periodic unitary matrix $U(\boldsymbol{k})$ that maps the (un)occupied Bloch energy eigenvectors $\ket{\psi_{n,\boldsymbol{k}}}$ to a set of smooth and orthogonal Bloch-type functions $\ket{\tilde{\psi}_{\alpha,\boldsymbol{k}}}$ 
that therefore live in the (un)occupied electronic Hilbert space. In the present work we find a set of smooth energy eigenvectors, so we can take that $U(\boldsymbol{k})$ as identity. That is, in this work the Bloch energy eigenvectors each map to a WF, with respect to which the bulk polarization and orbital magnetization are defined.}
\begin{widetext}
\begin{align}
    \bar{M}_{\mathfrak{u}}^{i(E)}
    &=\frac{e^2}{2\hbar c}\epsilon^{iab}E^l\sum_{cv}\int_{\text{BZ}_{\text{3D}}}\frac{d^{3}k}{(2\pi)^3}\left(-\sum_{v'}\text{Re}\big[i\xi^a_{vv'}\xi^b_{v'c}\xi^l_{cv}]+\sum_{c'}\text{Re}\big[i\xi^a_{cc'}\xi^b_{c'v}\xi^l_{vc}\big]\right)=0,
\label{nuBarE}
\end{align}
\end{widetext}
where $v,v'\in\{1,2\}$ and $c,c'\in\{3,4\}$. Here and elsewhere we often keep the $\boldsymbol{k}$-dependence of quantities implicit. In this gauge the atomic-like magnetization
vanishes as an immediate result of the relations (\ref{connectionCompTB2}) between the band components of $\xi^a$. The itinerant contribution $\tilde{M}^{i(E)}$ has both a dynamical modification $\tilde{M}^{i(E;\text{I})}$ and a compositional modification $\tilde{M}^{i(E;\text{II})}$ (given by Eqs.~(72) and (71) of MS, respectively).
In any smooth global Hamiltonian gauge $\mathfrak{u}'$ of our model, $\tilde{M}^{i(E;\text{I})}_{\mathfrak{u}'}=0$ since $E_{c,\boldsymbol{k}}+E_{v,\boldsymbol{k}}$ is constant over $\text{BZ}_{\text{3D}}$.
In the particular Hamiltonian gauge $\mathfrak{u}$ that we employ, Eq.~(71) of MS evaluates to
\begin{align}
    \tilde{M}_{\mathfrak{u}}^{i(E;\text{II})}&=\frac{e^2}{2\hbar c}\epsilon^{iab}E^l\sum_{vm}\int_{\text{BZ}_{\text{3D}}}\frac{d^{3}k}{(2\pi)^3} \text{Re}\big[\xi^l_{vm}\partial_{b}\xi^{a}_{mv}\big]\nonumber\\
    &=\alpha^{\mathfrak{u}}_{\text{CS}}\delta^{il}E^{l}.
\label{nuTildeEb}
\end{align}
To reach the final equality in Eq.~(\ref{nuTildeEb}) (in our case, $m\in\{1,2,3,4\}$) we have used (\ref{connectionCompTB},\ref{connectionCompTB2}) for $\xi^a$ and recognized the result as
the analytic expression (\ref{alphaCSintegralTB}) for $\alpha^{\mathfrak{u}}_{\text{CS}}$.
We find, therefore, that in the gauge $\mathfrak{u}$ the topological magnetoelectric response is entirely itinerant,
\textit{i.e.}~that 
\begin{equation}
    M_{\mathfrak{u}}^{(E)} = \tilde{M}_{\mathfrak{u}}^{i(E;\text{II})}.
\end{equation}
%While this result is gauge dependent, as we explicitly demonstrate below, it is at least true that in this model of 3D non-magnetic $\mathbb{Z}_{2}$ TIs there exists a gauge $\mathfrak{u}$ in which the Chern-Simons response is \textit{entirely} due to that of the itinerant contribution to the orbital magnetization.
	
Now consider working in some other smooth global Hamiltonian gauge $\mathfrak{u}'$. 
(Recall the discussion at the end of Sec.~\ref{Sec:IIIa2}.)
Then at each $\boldsymbol{k}\in\text{BZ}_{\text{3D}}$ there exists a unitary matrix $U(\boldsymbol{k})$ relating the components $|u'_{n,\boldsymbol{k}}\rangle$ and $\ket{u_{n,\boldsymbol{k}}}$ of the gauges $\mathfrak{u}'$ and $\mathfrak{u}$ \footnote{Since the components of these gauges are mutually orthogonal pointwise over BZ, $\braket{u'_{n,\boldsymbol{k}}}{u'_{m,\boldsymbol{k}}}=\delta_{nm}$ and $\braket{u_{n,\boldsymbol{k}}}{u_{m,\boldsymbol{k}}}=\delta_{nm}$, $T(\boldsymbol{k})$ is unitary.},
\begin{align}
\ket{u'_{n,\boldsymbol{k}}}=\sum_{m=1}^{4}\ket{u_{m,\boldsymbol{k}}}U_{m,n}(\boldsymbol{k}),
\end{align}
where $U_{c,v}(\boldsymbol{k})\equiv0$. 
The components of the non-Abelian Berry connection induced by $\mathfrak{u}'$ are
\begin{align*}
(\xi_{\mathfrak{u}'})_{nm}^{a}(\boldsymbol{k})=i\braket{u'_{n,\boldsymbol{k}}}{\partial_a u'_{m,\boldsymbol{k}}}
\end{align*}
and are related to those induced by $\mathfrak{u}$ (denoted in Eq.~(\ref{connection}) as $\xi^{a}_{nm}\equiv(\xi_{\mathfrak{u}})^{a}_{nm}$) by
\begin{align}
(\xi_{\mathfrak{u}'})_{nm}^{a}(\boldsymbol{k})=\sum_{l,s=1}^{4}U_{n,l}^{\dagger}(\boldsymbol{k})\Big(\xi^{a}_{ls}(\boldsymbol{k})+\mathcal{W}^a_{ls}(\boldsymbol{k})\Big)U_{s,m}(\boldsymbol{k}),
\label{gaugeTransf}
\end{align}
where
\begin{align}
\mathcal{W}^a_{nm}(\boldsymbol{k})\equiv i\sum_{r=1}^{4}\big(\partial_{a}U_{n,r}(\boldsymbol{k})\big)U^{\dagger}_{r,m}(\boldsymbol{k}).
\end{align}
Notably, $U_{c,v}(\boldsymbol{k})\equiv0$ implies $\mathcal{W}^a_{cv}(\boldsymbol{k})\equiv0$.
Employing Eq.~(\ref{gaugeTransf}), we are able to relate the atomic-like and itinerant contributions to the orbital magnetization that are identified with respect to the distinct gauges $\mathfrak{u}$ and $\mathfrak{u}'$. In particular, using Eq.~(\ref{nuBarE}) we find
\begin{widetext}
\begin{align}
\bar{M}_{\mathfrak{u}'}^{i(E)}
&=\bar{M}_{\mathfrak{u}}^{i(E)}+\frac{e^2}{2\hbar c}\epsilon^{iab}E^l\int_{\text{BZ}_{\text{3D}}}\frac{d^{3}k}{(2\pi)^3}\text{Re}\left[i\sum_{cvv'}\mathcal{W}^{b}_{vv'}\xi^{a}_{v'c}\xi^{l}_{cv}-i\sum_{vcc'}\mathcal{W}^{b}_{cc'}\xi^{a}_{c'v}\xi^{l}_{vc}\right]\equiv\bar{\alpha}^{li}_{\mathfrak{u}'}E^{l}.
\label{nuBarE2}
\end{align}
\end{widetext}
In general, the only restriction on $U(\boldsymbol{k})$ is that it be smooth and $\Gamma^{*}_{H}$-periodic, thus there is no reason to expect $\bar{\alpha}^{li}_{\mathfrak{u}'}$ to vanish. 
	
As an example, consider a gauge transformation $U(\boldsymbol{k})$ such that $U_{n,m}(\boldsymbol{k})\neq 0$ only if $n=m=1$. Since $U(\boldsymbol{k})$ is unitary, there exists a smooth function $\lambda(\boldsymbol{k})$ such that $\forall\boldsymbol{k}\in\text{BZ}_{\text{3D}}:\forall\boldsymbol{G}\in\Gamma^{*}_{H}:\lambda(\boldsymbol{k}+\boldsymbol{G})=\lambda(\boldsymbol{k})+2\pi j$ for $j\in\mathbb{Z}$, and $U_{1,1}(\boldsymbol{k})=e^{-i\lambda(\boldsymbol{k})}$. 
Taking for example $\lambda(\boldsymbol{k})\equiv\sin(k_{x})\cos(ak_{z})$ results in $\bar{\alpha}_{\mathfrak{u}'}^{zx}\neq0$ and taking $\lambda(\boldsymbol{k})\equiv\cos(k_{x})\sin(ak_{z})$ results in $\bar{\alpha}_{\mathfrak{u}'}^{zz}\neq0$. Then, since TRS implies that in any smooth global Hamiltonian gauge $\mathfrak{u}'$ we have $\boldsymbol{M}^{(E)}_{\mathfrak{u}'}=\bar{\boldsymbol{M}}^{(E)}_{\mathfrak{u}'}+\tilde{\boldsymbol{M}}^{(E)}_{\mathfrak{u}'}=\alpha^{\mathfrak{u}'}_{\text{CS}}\boldsymbol{E}$, it is generally \textit{not} the case that $\tilde{M}_{\mathfrak{u}'}^{i(E;\text{II})}$ equals $\alpha^{\mathfrak{u}'}_{\text{CS}}\delta^{il}E^{l}$.
	
The only feature that we \textit{might} think is special about the non-magnetic case considered here is that there exists a smooth global Hamiltonian gauge $\mathfrak{u}$ in which $\bar{\boldsymbol{M}}_{\mathfrak{u}}^{i(E)}=\boldsymbol{0}$ and $\tilde{\boldsymbol{M}}_{\mathfrak{u}}^{(E)}=\alpha^{\mathfrak{u}}_{\text{CS}}\boldsymbol{E}$.
In an related work \cite{lei2023afm} we investigate this further by considering the lattice regularization of the coupled-Dirac cone model for the MnV$_2$VI$_4$ family of anti-ferromagnetic $\mathbb{Z}_{2}$ TIs. In that work we show that there \textit{does not} generically exist a smooth global Hamiltonian gauge in which the topological magnetoelectric response is entirely itinerant.
Indeed whether such a gauge exists is related to the geometry of $\mathcal{V}\xrightarrow{\pi_{\mathcal{V}}}\text{BZ}_{\text{3D}}$ for the insulator under consideration -- a smooth global gauge of a vector bundle is equivalent to a global trivialization.  

Given that the topological magnetoelectric response in non-magnetic $\mathbb{Z}_{2}$ TIs is (with important qualifications) itinerant, from these purely bulk considerations alone we might expect it to originate entirely from surface currents. 
In cases for which this is true, there is good reason 
to regard the identification of the physically meaningful bulk magnetoelectric response as $\alpha_{\text{CS}}$ with suspicion, 
especially when we know that any finite size 
material sample will host gapless surface states whose role must also be considered. 
Surface magnetic dopants can generate gaps in the energy dispersion of the surface states and puts this 
identification on safer footing. To generate a gap everywhere on the surface,
it is also required that the outward normal of the magnetization characterizing the dopant configuration never change sign. 
For a thin film this leads to the requirement that this
magnetization retains a perpendicular component on the side walls and that it have opposite orientations on the top and bottom surfaces. Indeed these are the same magnetic-dopant requirements that were mentioned at the beginning of this section and it is only when they are satisfied that microscopic finite-size calculations agree with the simple interpretation of the
bulk magnetoelectric linear response calculation. 
When these requirement are not satisfied, there is no physically meaningful magnetoelectric response in non-magnetic $\mathbb{Z}_{2}$ TIs.

\iffalse
Given that the topological magnetoelectric response in non-magnetic TIs is (with important qualifications) itinerant, it may
originate entirely from surface currents. There is therefore good reason 
to regard the interpretation of the bulk calculation with suspicion, especially when we know that any finite size 
material sample will host gapless surface states. Surface magnetic dopants can generate surface-state gaps and puts the 
interpretation of the bulk calculation on safer footing. In order to generate a gap everywhere on the surface,
it is also required that the outward normal of the magnetization characterizing the dopant configuration never change sign. 
For a thin film this leads to the requirement that this
magnetization retains a perpendicular component on the side walls and that it have opposite orientations on the top and bottom surfaces. Indeed it is only in this case do microscopic finite-size calculations agree with the naive interpretation of the
bulk magnetoelectric linear response calculation. In non-magnetic $\mathbb{Z}_{2}$ TIs there is no physically meaningful magnetoelectric response without surface magnetization.
\fi
	
\section{Summary and Discussion}
In this paper we study the topological magnetoelectric effect in 3D non-magnetic $\mathbb{Z}_{2}$ topological insulators. We begin Sec.~\ref{Sec:GeneralLinearResponse} with a summary of the considerations that underlie notions of polarization and (orbital) magnetization in material media, and of magnetoelectric response. 
In doing so, we illustrate the precarious relationship between that response and 
time-reversal symmetry, in non-magnetic $\mathbb{Z}_{2}$ TIs in particular. 
Focused on materials of that type, we present a slightly modified interpretation of the argument \cite{Essin2010} used to
derive an expression for the bulk topological magnetoelectric coefficient, one that is manifestly consistent with time-reversal symmetry.
We conclude that the polarization adjacent to a particular surface position is 
activated locally (see below) by time-reversal symmetry breaking 
at that position.  This interpretation is supported by explicit calculations in finite sized (Sec.~\ref{Sec:II}) 
and bulk (Sec.~\ref{Sec:IV}) materials.
In the process of formulating this interpretation, we clarify a partially inconsistent conclusion related to
the adiabatic current expression from which the magnetoelectric response was previously \cite{Essin2010} deduced.
The central difference between our interpretation and the present one is that ours accounts for the fact that a quantized magnetoelectric effect occurs in non-magnetic $\mathbb{Z}_{2}$ TIs only if static magnetic surface dopants are present and are configured such that their out-of-plane magnetization
orientations at the top and bottom surfaces are nonzero and opposite. This resolves the tension between
the seemingly contradictory roles played by TRS in finite sized and bulk insulators.

The model that we employ in Sec.~\ref{Sec:II} is for an infinite quasi-2D thin film with 
two perfectly flat surfaces on top and bottom.  The explicit calculations on which our interpretation is 
based show that a magnetic field dependent charge density can arise only at those surfaces and depends on the surface normal of the 
magnetization on that surface.  This conclusion can be generalized to arbitrary surfaces by noting 
that only the sign of the surface normal magnetization component is important, and that the non-locality length along lines 
where the surface normal changes sign \cite{jackiw1976solitons}, $\lambda = \hbar v_{D}/m$, is finite. 
As long as the typical length scales of surface regions with a fixed sign of the surface normal 
of the magnetization is large compared with $\lambda$, the properties we have calculated for 
uniform surfaces apply to arbitrary surface magnetization profiles, and the term local here refers to 
averages over finite regions of area $\lambda^2$. 
One copy of the infinite flat thin film model in Sec.~\ref{Sec:II} can be associated with each locally flat spatial region of the sample. 
Within each such region, our interpretation of the adiabatic current calculation is that $\partial P^{i}/\partial B^{i}$ 
%magnetoelectric coefficient in bulk insulators that exhibit TRS
%change of bulk polarization with magnetic field 
%at every position along the boundary 
is constrained
by bulk topology to have a quantized value.  The value that is realized  
at a given position on a sample boundary is locally determined by the surface magnetization at the 
same position on the boundary \cite{Zhang2008}.  The magnitude of the polarization response to magnetic field tends to
be the smallest allowed quantized value, so changes from position to position tend to be changes in sign only.
Sign changes in the surface magnetization profile will produce locally ungapped regions on the surface that will support 
\cite{jackiw1976solitons} chiral edge states that separate 
polarization domains.

As we have explained, the topological magnetoelectric effect (TME) in our picture is a local surface 
charge density property.
If this is the meaning ascribed to the TME, then the 
quantum Hall effect that appears in the film as a whole when the magnetizations 
on top and bottom surfaces have the same orientation can be viewed to be a natural partner of the TME.    
This magnetic configuration implies opposite dependence of polarization on magnetic field on opposite surfaces of the sample, and 
via Eq.~(\ref{PM}) to a total charge density in the insulator that varies linearly with magnetic field.  A dependence of charge 
density on magnetic field in a quasi-2D insulator is equivalent to the quantum Hall effect via the Streda formula.  
(The relationship between the quantum Hall effect in TI thin films observed optically and 
the TME has been controversial \cite{TME_expt1,TME_expt2,TME_expt3,BeenakkerComment,Armitage2019}.)
When the magnetizations are anti-parallel on the two surfaces of a thin film, the same argument implies that in our 
interpretation the field-dependent polarization inside the material is spatially uniform, so that the total charge density is independent of magnetic field.
Thermodynamic identities nevertheless imply that the orbital magnetization must vary with the 
electric field applied across the system, implying that the orbital magnetization is sensitive to an energy 
shift between top and bottom surfaces in the local density-of-states even though no chiral edge states are present at the 
Fermi energy, as argued in Ref.~\cite{Pournaghavi2021}.

In Sec.~\ref{Sec:IV} we consider whether these surface magnetic-dopant qualifications 
on the magnetoelectric response of non-magnetic $\mathbb{Z}_{2}$ TIs manifest in the bulk expressions.
If they did, one might guess to use the equivalence of the susceptibility tensors (\ref{alpha}) and investigate the properties of the atomic-like $\bar{\bm{M}}$ and itinerant $\tilde{\bm{M}}$ contributions to the electric-field-induced magnetic dipole moment. 
Indeed $\tilde{\bm{M}}$ was originally associated with surface currents \cite{Resta2005}. 
The identification of $\bar{\bm{M}}$ and $\tilde{\bm{M}}$ is generally gauge dependent, even in unperturbed insulators \cite{Resta2006}, thus associating any physical meaning with that partitioning is suspect. 
Nevertheless, with respect to a particular smooth global Hamiltonian gauge we analytically demonstrate that the topological response $\delta^{il}\alpha_{\text{CS}}$ arises entirely through that of $\tilde{\bm{M}}$ (independent of a particular choice of parameter values, in all regions of parameter space for which there is always a band gap). 
However, this is not true in a generic Hamiltonian gauge. 
In a related work \cite{lei2023afm} we consider anti-ferromagnetic (AFM) $\mathbb{Z}_{2}$ TIs that exhibit a generalized TRS.
In that case we show that no gauge exists for 
which the response is entirely itinerant (in the above sense). 
In AFM materials, surface magnetic dopants are unnecessary for a magnetoelectric effect to manifest 
since the termination of the material already breaks the generalized TRS.   
Thus, it may be the case that the surface magnetic-dopant qualification in thin films is equivalent to the existence of a smooth global Hamiltonian gauge with respect to which the bulk response is entirely itinerant.
Indeed, this may be understood as a type of microscopic bulk-boundary correspondence relating the bulk geometry of $\mathcal{V}\xrightarrow{\pi_{\mathcal{V}}}\text{BZ}_{\text{3D}}$ to the
surface magnetic-dopant qualifications for the TME to manifest in a particular material.
Although the nonexistence of certain smooth global gauges has been related \cite{Monaco2017} to the topology of $\mathcal{V}\xrightarrow{\pi_{\mathcal{V}}}\text{BZ}_{\text{3D}}$,
we do not suspect this to be the case here;
we anticipate that in an AFM TI there could exist some smooth global gauges $\mathfrak{u}$ for which there exists a finely tuned point in the parameter space where $\tilde{\bm{M}}_{\mathfrak{u}}^{(E)}$ equals $\alpha^{\mathfrak{u}}_{\text{CS}}\bm{E}$.
Thus, whether such a gauge generically exists is taken to be a geometric property of $\mathcal{V}\xrightarrow{\pi_{\mathcal{V}}}\text{BZ}_{\text{3D}}$.

%While the above speculation may signify a connection between quantities defined in bulk $\mathbb{Z}_{2}$ TIs and those in finite samples thereof, one need display caution. A systematic approach that both makes manifest the central role of the bulk topology and also accurately predicts physical consequences is to first identify the bulk topological index of interest and how it manifests in susceptibility tensors (for example, in the response of $\bm{P}$ to $\bm{B}$ and $\bm{M}$ to $\bm{E}$, or $\bm{J}$ to $\bm{E}$). Next, develop an effective field-theoretic description of that phenomenon, which can then be used to identify any boundary modes that result as a consequence of the bulk topology. 
%The sum total response of both the bulk and the boundary degrees of freedom should be considered. For if such a prescription is not followed, then incorrect conclusions can be reached. We now illustrate this. 

That magnetic surface dopants activate the topological magnetoelectric effect in thin films of non-magnetic $\mathbb{Z}_2$ TIs can be understood on physical grounds, as we now describe.
Assume that the magnetoelectric coefficient in a bulk non-magnetic $\mathbb{Z}_{2}$ TI is given 
by $\alpha_{\text{CS}}=(n+1/2)e^{2}/hc$ for some $n\in\mathbb{Z}$. 
We seek the corresponding susceptibility tensor $\alpha_{\text{finite}}(\boldsymbol{r})$ in a finite sample thereof, which we assume can be partitioned into contributions from the sample's interior and surface regions, such that $\alpha_{\text{finite}}(\boldsymbol{r})=\alpha_{\text{CS}}f_{\text{int}}(\boldsymbol{r})+\alpha_{\text{surf}}(\boldsymbol{r})f_{\text{surf}}(\boldsymbol{r})$; $\{f_{\text{int}}(\boldsymbol{r}),f_{\text{surf}}(\boldsymbol{r}),f_{\text{ext}}(\boldsymbol{r})\}$ is a partition of unity over $\mathbb{R}^3$, where we take $f_{\text{int}}(\boldsymbol{r})$ to have support over the entire interior region of the sample and decay from $1$ to $0$ on some length scale near its boundary, $f_{\text{surf}}(\boldsymbol{r})$ to be nonvanishing only near the sample boundary, and $f_{\text{ext}}(\boldsymbol{r})$ to identify the region outside of the material. 

% The (\textit{incorrect}) assumption that the bulk physics determines that of its finite-sized counterpart amounts to neglecting $\alpha_{\text{surf}}(\boldsymbol{r})$, thereby assuming $\alpha_{\text{finite}}(\boldsymbol{r})=\alpha_{\text{CS}}(1-f_{\text{ext}}(\boldsymbol{r}))$. 

If we assume that bulk physics always determines the properties of
finite-sized samples such that $\alpha_{\text{surf}}(\boldsymbol{r})$ can be neglected, then $\alpha_{\text{finite}}(\boldsymbol{r})=\alpha_{\text{CS}}f_{\text{int}}(\boldsymbol{r})$.  We now show that this assumption leads to a contradiction.
Implementing this assumption in Eq.~(\ref{PM}), we find 
\footnote{Expressions that are similar to the following have been presented in previous works, including in Sec.~IV of Qi \textit{et al.}~\cite{Zhang2008} and in Chapter 6.4 of Vanderbilt \cite{VanderbiltBook}.}
	\begin{align*}
		J^{i(1)}_{\text{finite}}(\boldsymbol{r})&=c\alpha_{\text{CS}}\epsilon^{iab}\frac{\partial f_{\text{int}}(\boldsymbol{r})}{\partial r^a}E^{b}(\boldsymbol{r})+\ldots,\nonumber\\
		\varrho^{(1)}_{\text{finite}}(\boldsymbol{r})&=-\alpha_{\text{CS}}\frac{\partial f_{\text{int}}(\boldsymbol{r})}{\partial r^a}B^a(\boldsymbol{r})+\ldots.
	\end{align*}
That is, although in bulk insulators that exhibit TRS the linearly induced charge and current densities are insensitive to $\alpha_{\text{CS}}$, physical consequence of $\alpha_{\text{CS}}$ can manifest at the surface of a finite sample thereof. 
To reach this conclusion we assumed that $\boldsymbol{J}_{\text{finite}}^{(1)}(\boldsymbol{r})$ and $\varrho_{\text{finite}}^{(1)}(\boldsymbol{r})$ are found by restricting the bulk $\boldsymbol{P}^{(1)}(\boldsymbol{r})$ and $\boldsymbol{M}^{(1)}(\boldsymbol{r})$ to a finite region of space, and not the bulk $\boldsymbol{J}^{(1)}(\boldsymbol{r})$ and $\varrho^{(1)}(\boldsymbol{r})$ themselves. 
For if it were the latter, then $\alpha_{\text{CS}}$ would again not contribute. 
This is strange; who decides which quantity should be restricted?
It has also been pointed out \footnote{See, e.g., Sec.~IV of Qi \textit{et al.}~\cite{Zhang2008} or Chapters 6.2.2 and 6.3.2 of Vanderbilt \cite{VanderbiltBook}.} that the above expression for $\boldsymbol{J}^{(1)}(\boldsymbol{r})$ corresponds to a half-quantized quantum anomalous Hall current at each surface. 
But how can this be if the sample exhibits TRS?
%In the $\mathbb{Z}_2$-odd phase $\alpha_{\text{CS}}=\frac{e^2}{2hc}\text{ mod }\frac{e^2}{hc}$ and the Hall conductivity of each surface is half-quantized, while in the $\mathbb{Z}_2$-even phase $\alpha_{\text{CS}}=0\text{ mod }\frac{e^2}{hc}$ and the Hall conductivity of each surface is fully quantized relative to that in a Chern insulator. The usual interpretation of this is that there is an ambiguity in the number of Chern insulator wrappers at the surface \cite{Malashevich2010,Essin2010,VanderbiltBook}. 
Something has gone wrong.
Indeed, the result of Sec.~\ref{Sec:II} was that a magnetic-field-induced polarization can only occur in non-magnetic TI thin films when TRS is broken by magnetic dopants at the surface, which is moreover consistent with the usual symmetry analysis. In particular, if $m=0$ in Eq.~(\ref{alphaFinite}) prior to taking the bulk limit, then $\alpha_{\text{me}}=0$. 

Thus, the assumption that $\alpha_{\text{surf}}(\boldsymbol{r})$ can be neglected leads to contradictions. 
This illustrates that, while it may be true that the bulk $\alpha_{\text{CS}}$ can manifest at the surface of finite size samples, one must also consider the response of the surface states that exist as a consequence of the bulk topology. 
This conclusion is consistent with topological field theoretic considerations, where it has been demonstrated that the surface states implied by a non-trivial 3D bulk $\mathbb{Z}_{2}$ invariant (via the topological magnetoelectric coefficient $\alpha_{\text{CS}}$) are related to the restoration of the parity anomaly and their response identically cancels that of the bulk \cite{Burnell2013,Rosenow2013,Witten2016}. 
%That is, there is no magnetoelectric response in finite sized systems that exhibit TRS as one expects, even though the bulk magnetoelectric coefficient may be nonzero.
This contrasts the situation for the integer quantum Hall effect related to a non-trivial 2D bulk Chern invariant, in which case the bulk and surface responses add \cite{Yasuhiro1993,Hankiewicz2019,Hankiewicz2020}.

\iffalse
Further insight regarding these issues can be gleaned from topological field-theoretic considerations, in particular, from the nature of the anomalies that ``inflow'' from the bulk to the boundary. Roughly, the idea is that, given the microscopic physics (IQHE or TME) one writes down an appropriate effective field theory of the electromagnetic field, generally obeying modified Maxwell's equations with additional terms arising due to the Hall effect or the TME. 
To investigate a finite slab, one then restricts that theory to a manifold with a boundary. The presence of a boundary can result in the breaking of some fundamental features of the original model, e.g., gauge invariance, time-reversal or inversion symmetry, etc.; this is termed an anomaly. To restore these properties, it is argued that additional terms must be added to the theory that have support only on the boundary and are somehow anomalous in their own right. For example the presence of gapless chiral electronic excitations on the 1D boundary of a QH insulator restores gauge invariance \cite{WittenNotes2016}.
Similarly the odd number of gapless Dirac excitations on the 2D boundary of a non-magnetic TI
restores time-reversal symmetry. In finite-sized materials the boundary
excitations themselves contribute to response; in IQH insulators the bulk and surface responses add \cite{Yasuhiro1993} while in non-magnetic TI case they cancel \cite{Burnell2013,Rosenow2013,Witten2016} 
thereby restoring the time-reversal symmetry and correctly eliminating magnetoelectric response.
\fi

Ultimately, a ubiquitous aspect of theories considering polarization and orbital magnetization in crystalline solids has been the importance of interpretation \cite{VanderbiltBook}.
This work illustrates that in order to have a physically meaningful notion of bulk magnetoelectric response, in particular in insulators that exhibit TRS, the implicit involvement of a surface is unavoidable. 
Otherwise, as has been the case until now, the topological magnetoelectric response generally falls outside of the Peierls paradigm. Our hope is that this work will help to clarify misconceptions related to the topological magnetoelectric effect and 
thereby assist with its experimental demonstration.
We expect that a quantized TME will be observable in non-magnetic TI thin films only when the surface normal of the magnetization does not change direction on either top or bottom surfaces and
the magnetic gap induced at the surface is larger than the disorder potential on that surface.
Changes in sign of the surface normal and reductions in local gap size
on either surface will generically alter the global charge 
at which the density-of-states reaches a minimum and make it magnetic field dependent, swamping 
the topological magnetoelectric effect.  It follows that the robustness against disorder 
that is critical to accurate observation of the integer quantum Hall effect, does not apply to the 
topological magnetoelectric effect.

\iffalse
Finally, although the TME is often considered on the same footing as the IQHE, they have a profound difference. 
Although the corresponding susceptibilities are both sensitive to topological characterizations of the electronic ground state, symmetry play a central role in both characterizations.
the symmetry constraints under which the effect occurs is already accounted for in a nonzero Chern number, while TRS need explicitly be broken at/by the surface for TME.
\fi

%the bulk physics determines that of its finite-sized counterpart amounts to neglecting $\alpha_{\text{surf}}(\boldsymbol{r})$, thereby assuming $\alpha_{\text{finite}}(\boldsymbol{r})=\alpha_{\text{CS}}(1-f_{\text{ext}}(\boldsymbol{r}))$. 
	
\section{Acknowledgments}
Work at the University of Texas at Austin was supported by the
Simons Foundation: Targeted Grants in MPS No.~884934 and by the Robert A. Welch Foundation under Grant Welch F-2112.
The authors acknowledge the Texas Advanced Computing Center (TACC) at the University of Texas at Austin for providing HPC resources that have contributed to the research results reported within this paper. 
P.~T.~M.~gratefully acknowledges insightful conversations with John Sipe, Adarsh Patri, and Jason Kattan.
We also thank David Vanderbilt and Joel Moore for their useful comments and perspectives.
	
\appendix

\section{Flawed attempt to calculate $\alpha_{\text{CS}}$ via a low-energy effective Hamiltonian}
\label{Appendix:kdotp}
		
For illustrative purposes, in this section we perform a too-simplistic calculation of a low-energy approximation of $\alpha_{\text{CS}}$ using the bulk (\textit{i.e.}~periodic-in-$z$) version of the coupled-Dirac cone model introduced in Sec.~\ref{Sec:II}. 
In contrast to the quasi-2D models used in that section, this model is genuinely 3D.
Notably, in the smooth local gauge that motivates the eigenvectors (\ref{eigenvectorsNonMagneticTB}) of the tight-binding model obtained from a lattice regularization of this model, we will find the same result (\ref{eq:continuumapprox}). 
However, it is generally the case that one \textit{does not} produce the correct result for $\alpha_{\text{CS}}$ using a low-energy effective model alone. 
Indeed if one somehow knows that the local gauge choice on the $\text{BZ}_{\text{3D}}$ subset on which the effective model is valid can be extended to a smooth and periodic gauge over the entire $\text{BZ}_{\text{3D}}$ of some lattice regularization thereof, then the correct quantization can result; this was used implicitly in past work \cite{Franz2010}. 
Otherwise, incorrect half-quantization of the susceptibility tensor can arise, similar to the situation when too-simplistic attempts are made to calculate the 2D Hall conductivity in a Chern insulator.
		
The effective low-energy Hamiltonian of 3D bulk non-magnetic and magnetic multi-layer TIs that was previously studied by Lei \textit{et al.}~\cite{Lei2020} is essentially a $\boldsymbol{k}\cdot\boldsymbol{p}$ model about $\Gamma$ in the $\boldsymbol{b}_1$--$\boldsymbol{b}_2$ ($\boldsymbol{x}$--$\boldsymbol{y}$) plane and a lattice model in $\boldsymbol{b}_{3}$ ($\boldsymbol{z}$); by construction, the model is valid only on a subset of the 3D Brillouin zone. The model is written with respect to operators that generate basis vectors $\ket{v_{(\alpha,\sigma),\boldsymbol{k}}}$ that are products of 2D $\boldsymbol{k}\cdot\boldsymbol{p}$ basis vectors $\ket{\bar{u}_{(\alpha,\sigma),\boldsymbol{k}_{0}}}$, with the $\boldsymbol{k}\cdot\boldsymbol{p}$ expansion point $\boldsymbol{k}_{0}$ taken to be $(0,0)$, and 1D Bloch-type functions $\bar{\psi}_{(\alpha,\sigma),k_z}(z)$ that are associated with WFs $W_{(\alpha,\sigma),n\boldsymbol{a}_{3}}(z)$ that are localized in $\boldsymbol{a}_{3}\parallel\boldsymbol{z}$ and taken to coincide with a bottom (top) layer ``surface state'' for $\alpha=\text{even}$ (odd). Then, in this model the operators generate vectors $\ket{v_{(\alpha,\sigma),\boldsymbol{k}}}=\ket{v_{(\alpha,\sigma),(0,0,k_{z})}}\equiv\ket{v_{(\alpha,\sigma),k_{z}}}$, and, in fact, the possibility for variation in $k_z$ of the associated 1D cell-periodic functions $\bar{u}_{(\alpha,\sigma),k_z}(z)\propto e^{-ik_{z}z}\bar{\psi}_{(\alpha,\sigma),k_z}(z)$ is neglected. In the non-magnetic case, $\alpha\in\{0,1\}$ and the effective Hamiltonian is defined by its matrix representation
    \begin{align}
        \mathcal{H}^{(\text{3D})}_{\text{eff}}(\boldsymbol{k})=\left(\begin{array}{cccc}
		0 & \hbar v_{\text{D}}ik_{-} & \Delta_{k_z}^* & 0	\\
		-\hbar v_{\text{D}}ik_{+} & 0 & 0 & \Delta_{k_z}^*	\\
		\Delta_{k_z} & 0 & 0 & -\hbar v_{\text{D}}ik_{-}	\\
		0 & \Delta_{k_z} & \hbar v_{\text{D}}ik_{+} & 0
	\end{array}\right)
	\label{Heff_nonmag}
    \end{align}
in the basis $(\ket{v_{(0,\uparrow),k_{z}}},\ket{v_{(0,\downarrow),k_{z}}},\ket{v_{(1,\uparrow),k_{z}}},\ket{v_{(1,\downarrow),k_{z}}})$, where $k_{\pm}\equiv k_{x}\pm ik_{y}\equiv k_{\perp}e^{\pm i\theta_{k_{\perp}}}$, $\Delta_{k_z}\equiv\Delta_{\text{S}}+e^{iak_z}\Delta_{\text{D}}$; here $k_x$ and $k_y$ have units of $1/$length. The relevant velocity scale $v_{\text{D}}$ of the individual 2D layers is determined by the material parameters of one such layer. The eigenvalues of Eq.~(\ref{Heff_nonmag}) are $E_{1,2}(\boldsymbol{k})=-\epsilon(\boldsymbol{k})$ and $E_{3,4}(\boldsymbol{k})=\epsilon(\boldsymbol{k})$, where
    \begin{align}
	\epsilon(\boldsymbol{k})\equiv\sqrt{\hbar^2v_{\text{D}}^2(k_x^2+k_y^2)+|\Delta_{k_z}|^2}.
	\label{eigenvaluesNonMagnetic}
    \end{align}
In this model, the condition of a band insulator at half filling is $\forall k_{z}\in[-\pi/a,\pi/a]:\Delta_{k_z}\neq0$. Thus, as the ratio $|\Delta_{\text{D}}|/|\Delta_{\text{S}}|$ is varied, the band gap can vanish; in particular, if $\Delta_{\text{D}}=-\Delta_{\text{S}}$ then $\epsilon(0,0,0)=0$ and if $\Delta_{\text{D}}=\Delta_{\text{S}}$ then $\epsilon(0,0,\pi/a)=0$. These are the points in parameter space at which a topological phase transition may occur.
		
Due to the double degeneracy (\ref{eigenvaluesNonMagnetic}), at every $\boldsymbol{k}$ for which the model applies, the set of eigenvectors is highly non-unique. However, (\ref{Heff_nonmag}) has natural solutions in the $ak_{\perp}\rightarrow0$, $\infty$ limits ($k_{\perp}\equiv\sqrt{k_{x}^2+k_{y}^{2}}$), which motivates us to make a particular choice,
    \begin{widetext}
	\begin{align}
        \ket{\phi_{1,\boldsymbol{k}}}&=\frac{1}{\sqrt{2}}\left(-\frac{\Delta_{k_z}^*}{\epsilon(\boldsymbol{k})}\ket{v_{(0,\uparrow),k_{z}}}+\ket{v_{(1,\uparrow),k_{z}}}-\frac{i\hbar v_{\text{D}}k_{+}}{\epsilon(\boldsymbol{k})}\ket{v_{(1,\downarrow),k_{z}}}\right),\nonumber \\
        \ket{\phi_{2,\boldsymbol{k}}}&=\frac{1}{\sqrt{2}}\left(-\frac{i\hbar v_{\text{D}}k_{-}}{\epsilon(\boldsymbol{k})}\ket{v_{(0,\uparrow),k_{z}}}+\ket{v_{(0,\downarrow),k_{z}}}-\frac{\Delta_{k_z}}{\epsilon(\boldsymbol{k})}\ket{v_{(1,\downarrow),k_{z}}}\right), \nonumber\\
        \ket{\phi_{3,\boldsymbol{k}}}&=\frac{1}{\sqrt{2}}\left(\frac{\Delta_{k_z}^*}{\epsilon(\boldsymbol{k})}\ket{v_{(0,\uparrow),k_{z}}}+\ket{v_{(1,\uparrow),k_{z}}}+\frac{i\hbar v_{\text{D}}k_{+}}{\epsilon(\boldsymbol{k})}\ket{v_{(1,\downarrow),k_{z}}}\right), \nonumber\\
        \ket{\phi_{4,\boldsymbol{k}}}&=\frac{1}{\sqrt{2}}\left(\frac{i\hbar v_{\text{D}}k_{-}}{\epsilon(\boldsymbol{k})}\ket{v_{(0,\uparrow),k_{z}}}+\ket{v_{(0,\downarrow),k_{z}}}+\frac{\Delta_{k_z}}{\epsilon(\boldsymbol{k})}\ket{v_{(1,\downarrow),k_{z}}}\right).
	\label{eigenvectorsNonMagnetic}
    \end{align}
In the $ak_{\perp}\rightarrow0$ limit, the Hamiltonian (\ref{Heff_nonmag}) involves only inter-layer transitions and the spin index becomes preserved, while in the $ak_{\perp}\rightarrow\infty$ limit the spin-orbit interaction dominates and the layer index becomes preserved; these limits are evident in the chosen form (\ref{eigenvectorsNonMagnetic}). Notably, the set of vectors (\ref{eigenvectorsNonMagnetic}) forms an orthonormal basis of the relevant electronic Hilbert space, which are smooth in $\boldsymbol{k}$ for an insulator ($\epsilon(\boldsymbol{k})>0$). 

From (\ref{eigenvectorsNonMagnetic}) we can identify a unitary matrix $\breve{T}(\boldsymbol{k})$ analogous to that appearing in Eq.~(\ref{smoothFrame}), here with elements $\breve{T}_{n,(\alpha,\sigma)}(\boldsymbol{k})=\braket{\phi_{n,\boldsymbol{k}}}{v_{(\alpha,\sigma),k_{z}}}=\braket{v_{(\alpha,\sigma),k_{z}}}{\phi_{n,\boldsymbol{k}}}^*$. With this we explicitly find the matrices $\breve{\mathcal{T}}^{a}(\boldsymbol{k})\equiv i\big(\partial_{a}\breve{T}(\boldsymbol{k})\big)\breve{T}^{\dagger}(\boldsymbol{k})$ (for $a=x$, $y$, or $z$) analogous to those appearing in Eq.~(\ref{W}). Unsurprisingly, we find that these matrices satisfy relations analogous to (\ref{connectionCompTB2}), but now we find
    \begin{gather}
        \breve{\mathcal{T}}^{x}_{11}(\boldsymbol{k})=-\frac{\hbar^2v_{\text{D}}^2k_{y}}{2\epsilon(\boldsymbol{k})^2} \text{, } \breve{\mathcal{T}}^{y}_{11}(\boldsymbol{k})=\frac{\hbar^2v_{\text{D}}^2k_{x}}{2\epsilon(\boldsymbol{k})^2} \text{, } \breve{\mathcal{T}}^{z}_{11}(\boldsymbol{k})=-\frac{a\Delta_{\text{D}}(\Delta_{\text{D}}+\Delta_{\text{S}}\text{c}ak_{z})}{2\epsilon(\boldsymbol{k})^2},\nonumber\\
        \breve{\mathcal{T}}^{x}_{12}(\boldsymbol{k})=\frac{\hbar v_{\text{D}}\Delta_{k_{z}}}{2\epsilon(\boldsymbol{k})^2} \text{, } \breve{\mathcal{T}}^{y}_{12}(\boldsymbol{k})=-\frac{i\hbar v_{\text{D}}\Delta_{k_{z}}}{2\epsilon(\boldsymbol{k})^2}\text{, } \breve{\mathcal{T}}^{z}_{12}(\boldsymbol{k})=-\frac{\hbar v_{\text{D}}a\Delta_{\text{D}}e^{iak_{z}}(ik_{x}+k_y)}{2\epsilon(\boldsymbol{k})^2}.
        \label{Wnonmagnetic}
    \end{gather}
Indeed, the matrix elements (\ref{Wnonmagnetic}) are consistent with those (\ref{connectionCompTB}) that are obtained from the lattice regularized model; taking $B=0$ and $k_{x}$, $k_{y}$ to be small, Eq.~(\ref{connectionCompTB}) reduces to Eq.~(\ref{Wnonmagnetic}).

In principle, this model, like any $\boldsymbol{k}\cdot\boldsymbol{p}$ description, is insufficient to compute $\alpha_{\text{CS}}$. That is, while low-energy models may accurately describe the physics of the states associated with those crystal momentum $\boldsymbol{k}$ near the $\boldsymbol{k}\cdot\boldsymbol{p}$ expansion point, the topology of $\mathcal{V}$ imposes constraints on global properties of Bloch energy eigenvectors and of smooth gauge choices over all of $\text{BZ}_{d}$; those constraints are generally not relevant on subsets of $\text{BZ}_{d}$. Nevertheless, it is insightful to calculate a low-energy approximation of $\alpha_{\text{CS}}$ to see where issues arise. We take the initial electronic state of the insulator to be the zero-temperature ground state and there to be two electrons per unit cell; that is, we take $\ket{\phi_{1,\boldsymbol{k}}}$ and $\ket{\phi_{2,\boldsymbol{k}}}$ to be initially occupied and $\ket{\phi_{3,\boldsymbol{k}}}$ and $\ket{\phi_{4,\boldsymbol{k}}}$ to be initially unoccupied. The components of the Berry connection induced by (\ref{eigenvectorsNonMagnetic}) can be found via $\breve{\xi}^a_{nm}(\boldsymbol{k})=-\breve{\mathcal{T}}^a_{nm}(\boldsymbol{k})$ (recall the discussion above Eq.~(\ref{Heff_nonmag})). After some algebra we find
\begin{align}
  \epsilon^{abc}\sum_{vv'}\breve{\xi}^a_{vv'}\partial_b\breve{\xi}^c_{v'v}=-3\frac{a\hbar^2v_{\text{D}}^2\Delta_{\text{D}}(\Delta_{\text{D}}+\Delta_{\text{S}}\text{c}ak_z)}{\epsilon(\boldsymbol{k})^4}\text{,\space\space}
   -\frac{2i}{3}\epsilon^{abc}\sum_{vv'v_1}\breve{\xi}^a_{vv'}\breve{\xi}^b_{v'v_1}\breve{\xi}^c_{v_1v}=\frac{a\hbar^2v_{\text{D}}^2\Delta_{\text{D}}(\Delta_{\text{D}}+\Delta_{\text{S}}\text{c}ak_z)}{\epsilon(\boldsymbol{k})^4}.
\label{kpIntegrand}
\end{align}
To calculate quantities that are generally written as Brillouin zone integrals (such as $\alpha_{\text{CS}}$ (\ref{alphaCS})) via a $\boldsymbol{k}\cdot\boldsymbol{p}$ description, one usually restricts the domain of integration to the subset of $\text{BZ}_{\text{3D}}$ on which the model is valid (here to a limited range of $k_{x}$, $k_{y}$ about $(0,0)$) based on the assumption that outside of this region the integrand is negligible. In-line with this assumption, when $k_{x}$ and/or $k_{y}$ are large, the relevant integrand (\ref{kpIntegrand}) tends to 0; consequently, the domain of integration in the ``$\boldsymbol{k}\cdot\boldsymbol{p}$ dimensions'' is usually artificially extended to the entire plane and we do so here. With this, and moving to the cylindrical coordinates $(k_{\perp},\theta_{k_{\perp}},k_{z})$, we have
    \begin{align}
        \alpha_{\text{CS}}&\stackrel{\boldsymbol{.}}{=}\frac{e^2}{hc}\int_{-\pi/a}^{\pi/a}\frac{dk_{z}}{2\pi}\int_{0}^{\infty}dk_{\perp}k_{\perp}\frac{a\Delta_{\text{D}}'(\Delta_{\text{D}}'+\Delta_{\text{S}}'\text{c}ak_z)}{\epsilon'(\boldsymbol{k})^4},
    \label{CSnonmagnetic}
    \end{align}
where $\Delta_{\text{S}}'\equiv\Delta_{\text{S}}/\hbar v_{\text{D}}$, $\Delta_{\text{D}}'\equiv\Delta_{\text{D}}/\hbar v_{\text{D}}$, etc., and $\stackrel{\boldsymbol{.}}{=}$ denotes an equality under the above described assumptions of the $\boldsymbol{k}\cdot\boldsymbol{p}$ approach. Evaluating the integral,
    \begin{align}
        \int_{-\pi/a}^{\pi/a}\frac{dk_{z}}{2\pi}\int_{0}^{\infty}dk_{\perp}k_{\perp}\frac{a\Delta_{\text{D}}'(\Delta_{\text{D}}'+\Delta_{\text{S}}'\text{c}ak_z)}{(k_{\perp}^2+|\Delta'_{k_z}|^2)^2}=
        \begin{cases}
            0, & \text{if } |\Delta_{\text{S}}|>|\Delta_{\text{D}}|\\
            \frac{1}{2}, & \text{if } |\Delta_{\text{S}}|<|\Delta_{\text{D}}|
        \end{cases},
    \end{align}
yields the expected quantization of $\alpha_{\text{CS}}$. This is consistent with the result of the tight-binding calculation (see Eq.~(\ref{eq:continuumapprox}) and the surrounding text).
		
But what happens under a change of gauge? The result is that, in general, a change of gauge shifts the value of $\alpha_{\text{CS}}$ by an amount different than $n\frac{e^2}{hc}$ for $n\in\mathbb{Z}$, in contradiction to the known behavior for a model defined over the entire Brillouin zone \cite{Zhang2008,Vanderbilt2009}. This is an artifact of the use of a $\boldsymbol{k}\cdot\boldsymbol{p}$ description, which will not generally yield quantized topological invariants \footnote{See, e.g., Jotzu \cite{Jotzu2014}.}. In the case of $\alpha_{\text{CS}}$ in particular, the issue is that one does not know whether or not the eigenvectors employed in the gauge choice on the $\text{BZ}_{d}$ subset on which the $\boldsymbol{k}\cdot\boldsymbol{p}$ model is valid extend smoothly and periodically over the $\text{BZ}_{d}$ of a lattice regularization thereof. For example, in the model studied here, consider a gauge constructed from $\ket{v'_{n,\boldsymbol{k}}}=\sum_{m=1}^{4}\ket{v_{m,k_{z}}}U_{m,n}(\boldsymbol{k})$, where
    \begin{align}
        U(\boldsymbol{k})=U(k_z)=\left(\begin{array}{cccc} 
            0 & \Delta_{k_z}/|\Delta_{k_z}| & 0 & 0 \\
            -1 & 0 & 0 & 0	\\
            0 & 0 & 0 & \Delta_{k_z}/|\Delta_{k_z}|	\\
            0 & 0 & 1 & 0	
        \end{array}\right).
    \end{align}
This appears harmless within the context of the model (\ref{Heff_nonmag}) for an insulator ($\Delta_{k_z}\neq0$). But, in the lattice regularization presented in the main text, $\Delta_{k_z}\rightarrow\Delta_{\boldsymbol{k}}$ and there may exist $\boldsymbol{k}\in\text{BZ}_{\text{3D}}:|\Delta_{\boldsymbol{k}}|=0$ in a band insulator. Then attempting to define Bloch energy eigenvectors over the full Brillouin zone from the $\ket{v'_{n,\boldsymbol{k}}}$ (as was done in the main text (\ref{eigenvectorsNonMagneticTB}) for the $\ket{v_{n,\boldsymbol{k}}}$) is pathological at any $\boldsymbol{k}\in\text{BZ}_{\text{3D}}:|\Delta_{\boldsymbol{k}}|=0$; those vectors cannot be used to construct a smooth gauge over $\text{BZ}_{\text{3D}}$. Indeed a calculation of $\alpha_{\text{CS}}$ in that gauge would be invalid, since a smooth global gauge was assumed in deriving the $\text{BZ}_{\text{3D}}$ integral form of $\alpha_{\text{CS}}$.
		
\section{Anti-ferromagnetic 3D multi-layer tight-binding Hamiltonian}
\label{Appendix:AFM}
		
We now modify the considerations of Sec.~\ref{Sec:IIIa2} to the anti-ferromagnetic (AFM) case ($M_{l_z}=-M_{l_z\pm1}$). 
In this case the magnetic unit cell is twice the volume of the chemical unit cell. That is, the single-particle Bloch Hamiltonian $H(\boldsymbol{r},\boldsymbol{\mathfrak{p}}(\boldsymbol{r}))$ underlying a tight-binding description would not be invariant under translations by $a\boldsymbol{z}$, but would be invariant under translation by $2a\boldsymbol{z}$; it is useful to define $\Gamma_{\text{3D}}\equiv\Gamma_{\text{2D}}\times a\mathbb{Z}$ and $\Gamma_{H}\equiv\Gamma_{\text{2D}}\times 2a\mathbb{Z}$. Then $\boldsymbol{R}+a\boldsymbol{z}\notin\Gamma_{H}$ and strictly speaking Eq.~(\ref{H3d}) is not valid. Nevertheless we can modify that expression by relabelling the operators in such a way that the physical content is unchanged and the operators are identified in a technically correct manner. This achieved through the following: for $l_z\in\mathbb{Z}$ and $\boldsymbol{R}\in\Gamma_{\text{2D}}\times\{0\}$, take
		\begin{align}
			\ket{W_{(\alpha,\sigma),\boldsymbol{R}+l_{z}a\boldsymbol{z}}}\rightarrow\begin{cases}
				\ket{W_{(\alpha,\sigma),\boldsymbol{R}+l_{z}a\boldsymbol{z}}},& \text{for } l_z \text{ even} \\
				\ket{W_{(\alpha+2,\sigma),\boldsymbol{R}+(l_{z}-1)a\boldsymbol{z}}},& \text{for } l_z \text{ odd}
			\end{cases}.
			\label{WFrelabel}
		\end{align}
Under this transformation Eq.~(\ref{H3d}) is consistent; operators that are associated with WFs are labeled by elements of $\Gamma_{H}$ rather than by elements of $\Gamma_{\text{3D}}$. This relabeling results in eight types $(\alpha,\sigma)$ of WFs $W_{(\alpha,\sigma),\boldsymbol{R}}(\boldsymbol{r})$ associated with each Bravais lattice vector $\boldsymbol{R}\in\Gamma_{H}$. Separating out the $l_z=$ odd contributions and implementing the necessary relabeling (\ref{WFrelabel}), the tight-binding Hamiltonian (\ref{H3d}) is rewritten as
		\begin{align}
			\hat{H}^{(\text{3D})}=\sum_{l_z\in2\mathbb{Z}}\hat{H}'^{(\text{2D})}(l_z)+\Delta_{\text{D}}\sum_{\substack{\boldsymbol{R}\in\Gamma_{\text{2D}}\times\{0\}\\l_z\in2\mathbb{Z}\\\sigma\in\{\uparrow,\downarrow\}}}\left(\hat{c}^{\dagger}_{(2,\sigma),\boldsymbol{R}+l_za\boldsymbol{z}}\hat{c}_{(1,\sigma),\boldsymbol{R}+l_za\boldsymbol{z}}+\hat{c}^{\dagger}_{(0,\sigma),\boldsymbol{R}+(l_z+2)a\boldsymbol{z}}\hat{c}_{(3,\sigma),\boldsymbol{R}+l_za\boldsymbol{z}}+h.c.\right),
			\label{H3d_Afm}
		\end{align}
where $\hat{H}'^{(\text{2D})}(l_z)$ is obtained by implementing the relabeling (\ref{WFrelabel}) in $\hat{H}^{(\text{2D})}(l_z)+\hat{H}^{(\text{2D})}(l_z+1)$. 

There is another, more physical, complication in the case when $M_{l_{z}}\neq0$ in 3D, which is the possibility of exchange coupling between the dynamical degrees of freedom of a given layer with the static magnetization of all other layers. If we assume that the most relevant exchange interactions for a given WF $W_{(\alpha,\sigma),\boldsymbol{R}}(\boldsymbol{r})$ is that with the static magnetization of the layer for which it is associated (via $J_{\text{S}}$) and that with the static magnetization of the layer with which it is nearest (via $J_{\text{D}}$) -- for $\alpha=\text{odd}$ (even) this is the layer below (above) that for which $W_{(\alpha,\sigma),\boldsymbol{R}}(\boldsymbol{r})$ is associated -- and we focus on the AFM case such that both orbitals within a given layer feel the same $M_{l_{z}\pm1}=-M_{l_{z}}$, we can account for this by taking $J_{\text{S}}M_{l_{z}}\rightarrow(J_{\text{S}}-J_{\text{D}})M_{l_z}\equiv m_{l_{z}}$ in Eq.~(\ref{HlayerTB}). Taking $m_{l_{z}}=m$ for $l_{z}$ even and $=-m$ for $l_{z}$ odd, writing (\ref{H3d_Afm}) as a $\text{BZ}_{\text{3D}}$ integral via Eq.~(\ref{WF}), and using Eq.~(\ref{deltaIdentity}), we have
\begin{equation}
    \hat{H}^{(\text{3D})}_{\text{reg}}=\int_{\text{BZ}_{\text{3D}}}\frac{d^3k}{(2\pi)^3}\nonumber \hat{c}^{\dagger}(\boldsymbol{k}) \mathcal{H}^{(\text{3D})}_{\text{reg}}(\boldsymbol{k})\hat{c}(\boldsymbol{k}),
\end{equation}
which is of the same form as (\ref{Hoperator_general}), but in this case $\hat{c}(\boldsymbol{k})\equiv \left(\hat{c}_{(0,\uparrow),\boldsymbol{k}},\hat{c}_{(0,\downarrow),\boldsymbol{k}},\hat{c}_{(1,\uparrow),\boldsymbol{k}},\hat{c}_{(1,\downarrow),\boldsymbol{k}},\hat{c}_{(2,\uparrow),\boldsymbol{k}},\hat{c}_{(2,\downarrow),\boldsymbol{k}},\hat{c}_{(3,\uparrow),\boldsymbol{k}},\hat{c}_{(3,\downarrow),\boldsymbol{k}}\right)^{\text{T}}$ and $\hat{c}^{\dagger}(\boldsymbol{k})\equiv \left(\hat{c}^{\dagger}_{(0,\uparrow),\boldsymbol{k}},\hat{c}^{\dagger}_{(0,\downarrow),\boldsymbol{k}},\hat{c}^{\dagger}_{(1,\uparrow),\boldsymbol{k}},\hat{c}^{\dagger}_{(1,\downarrow),\boldsymbol{k}},\hat{c}^{\dagger}_{(2,\uparrow),\boldsymbol{k}},\hat{c}^{\dagger}_{(2,\downarrow),\boldsymbol{k}},\hat{c}^{\dagger}_{(3,\uparrow),\boldsymbol{k}},\hat{c}^{\dagger}_{(3,\downarrow),\boldsymbol{k}}\right)$, and the general $\mathcal{H}^{(d)}(\boldsymbol{k})$ is taken to be
\footnotesize
\begin{align}
&\mathcal{H}^{(\text{3D})}_{\text{reg}}(\boldsymbol{k}) = \nonumber\\
&\left(\begin{array}{cccccccc}
    m & iA(\text{s}k_x-i\text{s}k_y) & \Delta(k_x,k_y) & 0 & 0 & 0 & e^{-2iak_{z}}\Delta_{\text{D}} & 0	\\
    - iA(\text{s}k_x+i\text{s}k_y) & -m & 0 & \Delta(k_x,k_y) & 0 & 0 & 0 & e^{-2iak_{z}}\Delta_{\text{D}}	\\
    \Delta(k_x,k_y) & 0 & m & - iA(\text{s}k_x-i\text{s}k_y) & \Delta_{\text{D}} & 0 & 0 & 0	\\
    0 & \Delta(k_x,k_y) & iA(\text{s}k_x+i\text{s}k_y) & -m & 0 & \Delta_{\text{D}} & 0 & 0 \\
    0 & 0 & \Delta_{\text{D}} & 0 & -m & iA(\text{s}k_x-i\text{s}k_y) & \Delta(k_x,k_y) & 0	\\
    0 & 0 & 0 & \Delta_{\text{D}} & - iA(\text{s}k_x+i\text{s}k_y) & m & 0 & \Delta(k_x,k_y)	\\
    e^{2iak_{z}}\Delta_{\text{D}} & 0 & 0 & 0 & \Delta(k_x,k_y) & 0 & -m & - iA(\text{s}k_x-i\text{s}k_y)	\\
    0 & e^{2iak_{z}}\Delta_{\text{D}} & 0 & 0 & 0 & \Delta(k_x,k_y) & iA(\text{s}k_x+i\text{s}k_y) & m
\end{array}\right).
\label{H_AFM}
\end{align}
\normalsize

The eigenvalues of (\ref{H_AFM}) are 
\begin{align}
E_{1,2}(\boldsymbol{k})&=-	E_{7,8}(\boldsymbol{k})=-\sqrt{\frac{1}{2}A^2(2-\text{c}2k_{x}-\text{c}2k_{y})+\Delta_{\text{S}}(k_{x},k_{y})^2+\Delta_{\text{D}}^2+m^2+2\sqrt{\Delta_{\text{S}}(k_{x},k_{y})^2(\Delta_{\text{D}}^2\cos^2(ak_z)+m^2)}},\nonumber\\ E_{3,4}(\boldsymbol{k})&=-E_{5,6}(\boldsymbol{k})=-\sqrt{\frac{1}{2}A^2(2-\text{c}2k_{x}-\text{c}2k_{y})+\Delta_{\text{S}}(k_{x},k_{y})^2+\Delta_{\text{D}}^2+m^2-2\sqrt{\Delta_{\text{S}}(k_{x},k_{y})^2(\Delta_{\text{D}}^2\cos^2(ak_z)+m^2)}}.
\label{eigenvaluesAFM}
\end{align}
The double degeneracy of the energy bands at each $\boldsymbol{k}\in\text{BZ}$ follows from the combination of an inversion and a (fermionic) time-reversal symmetry (see Appendix \ref{Appendix:Symmetries}). Notice that taking $m=0$ in (\ref{eigenvaluesAFM}) does \textit{not} reproduce (\ref{eigenvaluesNonMagnetic}). This is not surprising because taking this limit implements an incorrect identification of the Bravais lattice for the non-magnetic Hamiltonian; in deriving (\ref{eigenvaluesAFM}) we explicitly use $\Gamma_{H}=\Gamma_{\text{2D}}\times 2a\mathbb{Z}$ rather than the correct $\Gamma_{H}=\Gamma_{\text{3D}}=\Gamma_{\text{2D}}\times a\mathbb{Z}$ implemented in deriving (\ref{H_nonmag}) in the non-magnetic case.

\section{Symmetry analysis of the non-magnetic and anti-ferromagnetic tight-binding models}
\label{Appendix:Symmetries}
In this appendix we demonstrate that both the non-magnetic and anti-ferromagnetic tight-binding Hamiltonian operators, specified by the matrix kernels (\ref{H_nonmag}) and (\ref{H_AFM}), respectively, have a center-of-inversion symmetry and a time-reversal symmetry.
We use the language of Refs.~\cite{Ryu2010,Ludwig2015} and we begin by reiterating (in a loose manner) some contents of those works to set notation.
Starting with a single-particle Hamiltonian $H:\mathcal{H}\rightarrow\mathcal{H}$, which is a linear and self-adjoint operator in the electronic Hilbert space $\mathcal{H}$, and an orthonormal set of basis vectors $\ket{\phi_{A}}$ of $\mathcal{H}$, for $A$ a general multi-index, we have $H=\sum_{A,B}H_{A,B}\ket{\phi_{A}}\otimes\bra{\phi_{B}}$, where $H_{A,B}\equiv\braket{\phi_{A}}{H\phi_{B}}\in\mathbb{C}$ for a given pair $(A,B)$. $H$ can be generalized to a single-body operator $\hat{\mathscr{H}}=\sum_{A,B}H_{A,B}\hat{c}^{\dagger}_{A}\hat{c}_{B}$ in the electronic Fock space $\mathcal{F}$, where $\hat{c}^{\dagger}_{A}$, $\hat{c}_{A}$ are fermionic operators that satisfy $\{\hat{c}^{\dagger}_{A},\hat{c}_{B}\}=\delta_{A,B}$ and $\hat{c}^{\dagger}_{A}\ket{\text{vac}}\equiv\ket{\phi_{A}}$, and $\mathcal{F}$ is constructed by antisymmeterizing sums of products of $\mathcal{H}$. $\hat{\mathscr{H}}$ agrees with $H$ when acting on one-particle states; this will be assumed of any generalization of an operator on $\mathcal{H}$ to one on $\mathcal{F}$. Products of operators on $\mathcal{H}$ or $\mathcal{F}$ is by composition, which we usually keep implicit. 
		
Wigner's theorem dictates that any symmetry transformation $S:\mathcal{H}\rightarrow\mathcal{H}$ is either linear and unitary or anti-linear and anti-unitary; we here neglect symmetry transformations that do not preserve particle number, for example, particle-hole symmetry. By definition, $\hat{\mathscr{S}}$ is a symmetry of $\hat{\mathscr{H}}$ if and only if $\hat{\mathscr{S}}\hat{\mathscr{H}}\hat{\mathscr{S}}^{-1}=\hat{\mathscr{H}}$ is satisfied, where $\hat{\mathscr{S}}$ is the generalization of $S$ to an operator on $\mathcal{F}$.
An insight is gained by noting that both $\hat{\mathscr{S}}\hat{c}_{A}\hat{\mathscr{S}}^{-1}$ and $\hat{c}_{A}$ map $\mathcal{H}^{(N)}\rightarrow\mathcal{H}^{(N-1)}$, and that both $\hat{\mathscr{S}}\hat{c}^{\dagger}_{A}\hat{\mathscr{S}}^{-1}$ and $\hat{c}^{\dagger}_{A}$ map $\mathcal{H}^{(N)}\rightarrow\mathcal{H}^{(N+1)}$ within $\mathcal{F}$. Thus, for every symmetry transformation $S$ there exists collections of numbers $(U_{S})_{A,B}\text{,\space}(V_{S})_{A,B}\in\mathbb{C}$ that satisfy
\begin{align}
\hat{\mathscr{S}}\hat{c}_{A}\hat{\mathscr{S}}^{-1}=\sum_{B}\big(U_{S}\big)_{A,B}\hat{c}_{B}\text{,\space\space}
\hat{\mathscr{S}}\hat{c}^{\dagger}_{A}\hat{\mathscr{S}}^{-1}=\sum_{B}\hat{c}^{\dagger}_{B}\big(V_{S}\big)_{B,A}.
\label{symmetryOnOperators}
\end{align}
It follows from $\hat{\mathscr{S}}\{\hat{c}_{A},\hat{c}_{B}^{\dagger}\}\hat{\mathscr{S}}^{-1}=\hat{\mathscr{S}}\{\hat{c}^{\dagger}_{A},\hat{c}_{B}\}\hat{\mathscr{S}}^{-1}=\delta_{A,B}$ that the matrices $V_{S}$ and $U_{S}$ are each others inverse; that is, the matrices $U_{S}$ and $V_{S}$ constructed from the coefficients in Eq.~(\ref{symmetryOnOperators}) satisfy $U_{S}V_{S}=V_{S}U_{S}=\mathbf{1}$, where matrix products are taken as the usual matrix multiplication. 
If $S$ is (anti-)unitary, then 
\begin{align}
\delta_{AB}&=\braket{\phi_{A}}{\phi_{B}}=\braket{S\phi_{A}}{S\phi_{B}}^{(*)}\equiv\bra{\text{vac}}(\hat{\mathscr{S}}\hat{c}^{\dagger}_{A}\hat{\mathscr{S}}^{-1})^{\dagger}\hat{\mathscr{S}}\hat{c}^{\dagger}_{B}\hat{\mathscr{S}}^{-1}\ket{\text{vac}}^{(*)}\nonumber\\
&=\Big(\sum_{C,D}\big(V_{S}\big)_{C,A}^{*}\big(V_{S}\big)_{D,B}\bra{\text{vac}}\hat{c}_{C}\hat{c}^{\dagger}_{D}\ket{\text{vac}}\Big)^{(*)}=\Big(\sum_{C}\big(V_{S}^{\dagger}\big)_{A,C}\big(V_{S}\big)_{C,B}\Big)^{(*)}.
\end{align}
Thus $V_{S}^{\dagger}=U_{S}=V_{S}^{-1}$ is unitary and $\big(V_{S}\big)_{B,A}=\big(U_{S}\big)_{A,B}^{*}$ in Eq.~(\ref{symmetryOnOperators}).
With this, if $\hat{\mathscr{S}}$ is a symmetry of $\hat{\mathscr{H}}$, then
\begin{gather}
\sum_{A,B}\big(\hat{\mathscr{S}}\hat{c}^{\dagger}_{A}\hat{\mathscr{S}}^{-1}\big)\big(\hat{\mathscr{S}}H_{A,B}\hat{\mathscr{S}}^{-1}\big)\big(\hat{\mathscr{S}}\hat{c}_{B}\hat{\mathscr{S}}^{-1}\big)=\sum_{I,J}\hat{c}^{\dagger}_{I}\Big(\sum_{A,B}\big(U_{S}\big)_{A,I}^{*}H^{(*)}_{A,B}\big(U_{S}\big)_{B,J}\Big)\hat{c}_{J}=\sum_{I,J}\hat{c}^{\dagger}_{I}H_{I,J}\hat{c}^{\dagger}_{J} \nonumber\\
\iff \sum_{A,B}\big(U^{\dagger}_{S}\big)_{I,A}H^{(*)}_{A,B}\big(U_{S}\big)_{B,J}=H_{I,J},
\label{symmRelation}
\end{gather}
where $H^{(*)}_{A,B}=H_{A,B}$ ($H_{A,B}^{*}$) for ${S}$ being (anti-)linear. 
\iffalse
In the case of unitary $S$, (\ref{symmRelation}) is the usual form of a symmetry relation on $\mathcal{H}$, and for anti-unitary $S$ it can be recast as 
\begin{align}
\sum_{A,B}\big(T^{-1}_{S}\big)_{I,A}H_{A,B}\big(T_{S}\big)_{B,J}=H_{I,J}\text{,\space\space} T_{S}\equiv K U_{S},
\end{align}
\textcolor{red}{where $K$ acts to complex conjugate and is defined such that $K^{-1} c K = c^{*}$ for $c\in\mathbb{C}$ and $K K=\text{id}_{\mathbb{C}}$}. 
\fi

Any anti-linear and anti-unitary symmetry $\hat{\mathscr{S}}$ of $\hat{\mathscr{H}}$ might be called a time-reversal symmetry because of its affect on the time-evolution operator $\hat{\mathscr{U}}(t)=e^{-i\hat{\mathscr{H}}t/\hbar}$; if $\hat{\mathscr{S}}$ is an anti-unitary symmetry, then $\hat{\mathscr{S}}\hat{\mathscr{U}}(t)\hat{\mathscr{S}}^{-1}=\hat{\mathscr{U}}(-t)$. See Footnote 13 of Ryu \textit{et al.}~\cite{Ryu2010} for a discussion on uniqueness. Nevertheless, there is typically one physically motivated operator named the time-reversal transformation.
When $S$ is anti-unitary we have
\begin{align}
\hat{\mathscr{S}}^2\hat{c}_{A}(\hat{\mathscr{S}}^{-1})^{2}&\equiv\hat{\mathscr{S}}(\hat{\mathscr{S}}\hat{c}_{A}\hat{\mathscr{S}}^{-1})\hat{\mathscr{S}}^{-1}=\sum_{B,D}\big(U_{S}\big)^{*}_{A,B}\big(U_{S}\big)_{B,D}\hat{c}_{D}=\sum_{D}\big(U_{S}^{*}U_{S}\big)_{A,D}\hat{c}_{D},\nonumber\\
\hat{\mathscr{S}}^2\hat{c}^{\dagger}_{A}(\hat{\mathscr{S}}^{-1})^{2}&\equiv\hat{\mathscr{S}}(\hat{\mathscr{S}}\hat{c}^{\dagger}_{A}\hat{\mathscr{S}}^{-1})\hat{\mathscr{S}}^{-1}=\sum_{B,D}\big(U_{S}\big)_{A,B}\big(U_{S}\big)_{B,D}^*\hat{c}^{\dagger}_{D}=\sum_{D}\big(U_{S}^{*}U_{S}\big)_{A,D}^*\hat{c}^{\dagger}_{D}.
\label{antiunitarySquared}
\end{align}
It has been shown \cite{Ryu2010} that in general $U_{S}^{*} U_{S}\in\{-\mathbf{1},\mathbf{1}\}$, thus either $\hat{\mathscr{S}}^2\hat{c}_{A}(\hat{\mathscr{S}}^{-1})^{2}=\hat{c}_{A}$ and $\hat{\mathscr{S}}^2\hat{c}^{\dagger}_{A}(\hat{\mathscr{S}}^{-1})^{2}=\hat{c}^{\dagger}_{A}$ or $\hat{\mathscr{S}}^2\hat{c}_{A}(\hat{\mathscr{S}}^{-1})^{2}=-\hat{c}_{A}$ and $\hat{\mathscr{S}}^2\hat{c}^{\dagger}_{A}(\hat{\mathscr{S}}^{-1})^{2}=-\hat{c}^{\dagger}_{A}$ are satisfied. 
This is familiar from the usual analysis on $\mathcal{H}$; \textit{i.e.}~in the case $U_{S}^{*} U_{S}=-\mathbf{1}$, Kramer's theorem results.
		
We now identify some symmetry operators $\hat{\mathscr{S}}$ (and the corresponding $U_{S}$) for which Eq.~(\ref{symmRelation}) is satisfied in the case of the non-magnetic (\ref{H_nonmag}) or the anti-ferromagnetic (\ref{H_AFM}) tight-binding Hamiltonian.
		
\subsection{Inversion symmetry}
An inversion operator $\mathcal{I}$ on the relevant electronic Hilbert space $\mathcal{H}$ is taken to be linear and unitary, and satisfy the physically motivated relation $\mathcal{I}^2=\text{id}_{\mathcal{H}}$. Any (anti-)linear operator can be specified by its action on a basis of the space in which it acts. For the non-magnetic Hamiltonian we define
\begin{gather}
\mathcal{I}\ket{W_{(0,\sigma),\boldsymbol{R}}}=\ket{W_{(1,\sigma),-\boldsymbol{R}}} \iff \mathcal{I}\ket{\bar{\psi}_{(0,\sigma),\boldsymbol{k}}}=\ket{\bar{\psi}_{(1,\sigma),-\boldsymbol{k}}}
\label{nonMag_inversion}
\end{gather}
for $\sigma\in\{\uparrow,\downarrow\}$ and $I=((\alpha,\sigma),\boldsymbol{k})$ is the relevant multi-index on the right hand side. This transformation corresponds to inversion about the (single) layer center within the unit cell. Eq.~(\ref{nonMag_inversion}) can be generalized to a relation involving Fock space operators $\hat{c}^{\dagger}_{(\alpha,\sigma),\boldsymbol{k}}$,
\begin{gather}
\hat{\mathscr{I}}\hat{c}^{\dagger}_{(0,\sigma),\boldsymbol{k}}\hat{\mathscr{I}}^{-1}=\hat{c}^{\dagger}_{(1,\sigma),-\boldsymbol{k}}\implies \big(U_{\mathcal{I}}\big)_{((\alpha,\sigma),\boldsymbol{k}),((\beta,\gamma),\boldsymbol{k}')}=\big(\tau_{x}\big)_{\alpha,\beta}\delta_{\sigma,\gamma}\delta(\boldsymbol{k}+\boldsymbol{k}'),
\label{inversion_nonMag}
\end{gather}
where the identification of $U_{\mathcal{I}}$ follows from comparison with (\ref{symmetryOnOperators}) and using $\big(V_{S}\big)_{B,A}=\big(U_{S}\big)_{A,B}^{*}$; equivalently one could consider relations for the $\hat{c}_{(\alpha,\sigma),\boldsymbol{k}}$. To ease notation, we do not include factors to make this unit-less, and take the unitarity to be understood as
\begin{align}
\sum_{\substack{\beta\in\{1,\ldots,N_{\text{orb}}\}\\\gamma\in\{\uparrow,\downarrow\}}}\int_{\text{BZ}_{d}}d^{d}k'\big(U_{S}\big)_{((\alpha,\sigma),\boldsymbol{k}),((\beta,\gamma),\boldsymbol{k}')}\big(U^{\dagger}_{S}\big)_{((\beta,\gamma),\boldsymbol{k}'),((\mu,\lambda),\boldsymbol{k}'')}=\delta_{\alpha,\mu}\delta_{\sigma,\gamma}\delta(\boldsymbol{k}-\boldsymbol{k}''),
\end{align}
where $N_{\text{orb}}=2 (4)$ in the non-magnetic (anti-ferromagnetic) case. For the AFM Hamiltonian we define
\begin{align}
\mathcal{I}\ket{W_{(0,\sigma),\boldsymbol{R}}}=\ket{W_{(1,\sigma),-\boldsymbol{R}}}&\impliedby\hat{\mathscr{I}}\hat{c}^{\dagger}_{(0,\sigma),\boldsymbol{k}}\hat{\mathscr{I}}^{-1}=\hat{c}^{\dagger}_{(1,\sigma),-\boldsymbol{k}},\nonumber\\
\mathcal{I}\ket{W_{(2,\sigma),\boldsymbol{R}}}=\ket{W_{(3,\sigma),-\boldsymbol{R}-\boldsymbol{a}_3}}&\impliedby\hat{\mathscr{I}}\hat{c}^{\dagger}_{(2,\sigma),\boldsymbol{k}}\hat{\mathscr{I}}^{-1}=e^{-i\boldsymbol{k}\cdot\boldsymbol{a}_3}\hat{c}^{\dagger}_{(3,\sigma),-\boldsymbol{k}},
\label{inversion_AFM}
\end{align}
which leads to
\begin{align}
\big(U_{\mathcal{I}}\big)_{((\alpha,\sigma),\boldsymbol{k}),((\beta,\gamma),\boldsymbol{k}')}=\left(\left(\begin{array}{cc}
1 & 0 \\
0 & e^{i2ak_z}
\end{array}\right)\otimes\tau_{x}\right)_{\alpha,\beta}\delta_{\sigma,\gamma}\delta(\boldsymbol{k}+\boldsymbol{k}'),
\label{inversion_U_AFM}
\end{align}
where $\boldsymbol{a}_{3}=2a\boldsymbol{z}$. This transformation corresponds to inversion about the center of the bottom layer in the unit cell; one could equally well define an inversion operation with respect to the center of the top layer. Physically, this corresponds to inversion about the position of a magnetic ion, which is indeed a symmetry of the crystal lattice \cite{Wang2019}. 

Then one can explicitly show that
\begin{align}
\hat{\mathscr{I}}\hat{H}_{\text{reg}}^{(\text{3D})}\hat{\mathscr{I}}^{-1}=\hat{H}_{\text{reg}}^{(\text{3D})}
\label{symm}
\end{align}
is satisfied for the $\hat{H}_{\text{reg}}^{(\text{3D})}$ the appropriate non-magnetic (\ref{H_nonmag}) or anti-ferromagnetic (\ref{H_AFM}) tight-binding Hamiltonian and $\hat{\mathscr{I}}$ the corresponding inversion operator (\ref{inversion_nonMag}) or (\ref{inversion_AFM},\ref{inversion_U_AFM}). In particular, in the non-magnetic case the inversion operator (\ref{inversion_nonMag}) is a symmetry of the Hamiltonian if and only if
\begin{align}
(\tau_{x}\otimes\text{id}_{2\times2})\mathcal{H}^{(\text{3D})}_{\text{reg}}(-\boldsymbol{k})(\tau_{x}\otimes\text{id}_{2\times2})=\mathcal{H}^{(\text{3D})}_{\text{reg}}(\boldsymbol{k})
\label{nonMag_invSym}
\end{align}
(which plays the role of (\ref{symmRelation}) in this case), where we have equated Brillouin zone integrands and taken $\boldsymbol{k}\rightarrow-\boldsymbol{k}$ in the $\text{BZ}_{\text{3D}}$-integral related to $\hat{\mathscr{I}}\hat{H}_{\text{reg}}^{(\text{3D})}\hat{\mathscr{I}}^{-1}$ on the LHS of Eq.~(\ref{symm}). Given the matrix kernel (\ref{H_nonmag}), one can explicitly show that Eq.~(\ref{nonMag_invSym}) is satisfied. Similarly, given the matrix kernel of the AFM Hamiltonian (\ref{H_AFM}), the inversion operator (\ref{inversion_AFM}) is a symmetry since one can explicitly show that
\begin{align}
\left(\left(\begin{array}{cc}
1 & 0 \\
0 & e^{-i2ak_z}
\end{array}\right)\otimes\tau_{x}\otimes\text{id}_{2\times2}\right)\mathcal{H}^{(\text{3D})}_{\text{reg}}(\boldsymbol{k})\left(\left(\begin{array}{cc}
1 & 0 \\
0 & e^{i2ak_z}
\end{array}\right)\otimes\tau_{x}\otimes\text{id}_{2\times2}\right)=\mathcal{H}^{(\text{3D})}_{\text{reg}}(-\boldsymbol{k})
\label{AFM_invSym}
\end{align}
is satisfied. To arrive at Eq.~(\ref{AFM_invSym}) we have again equated $\text{BZ}_{\text{3D}}$-integrands, but have here taken $\boldsymbol{k}\rightarrow-\boldsymbol{k}$ in the $\text{BZ}_{\text{3D}}$-integral related to $\hat{H}_{\text{reg}}^{(\text{3D})}$ on the RHS of Eq.~(\ref{symm}). Note that the AFM Hamiltonian (\ref{H_AFM}) is not symmetric with respect to inversion about the center point between the two layers within a unit cell.
		
\subsection{Time-reversal symmetry}
\subsubsection{One option}
A time-reversal operator $\mathcal{T}$ is always taken to be an anti-linear and anti-unitary operator on the electronic Hilbert space $\mathcal{H}$. We begin by seeking a traditional time-reversal transformation $\mathcal{T}$, which takes a state with spin components $(s_x,s_y,s_z)$ to one with $(-s_x,-s_y,-s_z)$. One option is \footnote{See, e.g., pg.~277-280 of Sakurai \cite{Sakurai}.}
\begin{align}
\mathcal{T}\ket{W_{(\alpha,\uparrow),\boldsymbol{R}}}=-\ket{W_{(\alpha,\downarrow),\boldsymbol{R}}} \text{,\space\space} \mathcal{T}\ket{W_{(\alpha,\downarrow),\boldsymbol{R}}}=\ket{W_{(\alpha,\uparrow),\boldsymbol{R}}},
\label{TRS1}
\end{align}
for $\alpha\in\{0,1\}$ ($\alpha\in\{0,1,2,3\}$) in the non-magnetic (AFM) case. A generalization to a Fock space relation is
\begin{gather}
\hat{\mathscr{T}}\hat{c}^{\dagger}_{(\alpha,\uparrow),\boldsymbol{k}}\hat{\mathscr{T}}^{-1}=-\hat{c}^{\dagger}_{(\alpha,\downarrow),-\boldsymbol{k}} \text{,\space\space} \hat{\mathscr{T}}\hat{c}^{\dagger}_{(\alpha,\downarrow),\boldsymbol{k}}\hat{\mathscr{T}}^{-1}=\hat{c}^{\dagger}_{(\alpha,\uparrow),-\boldsymbol{k}} \nonumber\\
\implies	\hat{\mathscr{T}}^{2}\hat{c}^{\dagger}_{(\alpha,\sigma),\boldsymbol{k}}(\hat{\mathscr{T}}^{-1})^{2}=-\hat{c}^{\dagger}_{(\alpha,\sigma),\boldsymbol{k}},
\label{TRSrelation}
\end{gather}
which is consistent with (\ref{antiunitarySquared}) for $U^*_{\mathcal{T}}U_{\mathcal{T}}=-\mathbf{1}$. In particular, comparing the second line of (\ref{antiunitarySquared}) and the second line of (\ref{TRSrelation}) gives $(U^*_{\mathcal{T}}U_{\mathcal{T}})_{((\alpha,\sigma),\boldsymbol{k}),((\beta,\gamma),\boldsymbol{k}')}=-\delta_{\alpha,\beta}\delta_{\sigma,\gamma}\delta(\boldsymbol{k}-\boldsymbol{k}')$.
Then from (\ref{symmetryOnOperators}) and the first line of (\ref{TRSrelation}) we identify 
\begin{align}
\big(U_{\mathcal{T}}\big)_{((\alpha,\sigma),\boldsymbol{k}),((\beta,\gamma),\boldsymbol{k}')}=\delta_{\alpha,\beta}(-i\sigma_{y})_{\sigma,\gamma}\delta(\boldsymbol{k}+\boldsymbol{k}').
\label{TRSop}
\end{align} 
With this one can show
\begin{align}
\hat{\mathscr{T}}\hat{H}_{\text{reg}}^{(\text{3D})}\hat{\mathscr{T}}^{-1}=\hat{H}_{\text{reg}}^{(\text{3D})}
\label{symm1}
\end{align}
is satisfied for the non-magnetic Hamiltonian (\ref{H_nonmag}), but not for the anti-ferromagnetic (AFM) Hamiltonian (\ref{H_AFM}). 

In particular, given the matrix kernel of the non-magnetic Hamiltonian (\ref{H_nonmag}), one can explicitly show that
\begin{align}
(\text{id}_{2\times2}\otimes(i\sigma_{y}))\mathcal{H}^{(\text{3D})}_{\text{reg}}(-\boldsymbol{k})^*(\text{id}_{2\times2}\otimes(-i\sigma_{y}))=\mathcal{H}^{(\text{3D})}_{\text{reg}}(\boldsymbol{k}),
\end{align}
which is (\ref{symmRelation}) in this case, and again we have equated Brillouin zone integrands in Eq.~(\ref{symm1}). 

In contrast, given the matrix kernel of the AFM Hamiltonian (\ref{H_AFM}), the time-reversal transformation (\ref{TRS1}) is not a symmetry due to the terms related to the exchange interaction. Indeed, upon explicit evaluation one finds
\begin{align}
\Big((\text{id}_{4\times4}\otimes(i\sigma_{y}))\mathcal{H}^{(\text{3D})}_{\text{reg}}(-\boldsymbol{k})^*(\text{id}_{4\times4}\otimes(-i\sigma_{y}))\Big)- \mathcal{H}^{(\text{3D})}_{\text{reg}}(\boldsymbol{k})=-2m\big(\text{id}_{4\times4}\otimes\sigma_{z}\big).
\end{align}
Nevertheless, there is a different time-reversal-like transformation that is a symmetry of the AFM Hamiltonian (\ref{H_AFM}). 

\subsubsection{A second option}
In the AFM case we are physically motivated to consider an operation that involves a spin flip, as in the preceding subsection, followed by a translation by $a\boldsymbol{z}$. This can be obtained by a product of operators, $T_{a\boldsymbol{z}}$ and $\mathcal{T}$, where $\mathcal{T}$ is as before (Eq.~(\ref{TRSop})) and we take $T_{a\boldsymbol{z}}:\mathcal{H}\rightarrow\mathcal{H}$ to be linear and unitary, and such that
\begin{align}
T_{a\boldsymbol{z}}\ket{W_{(\alpha,\sigma),\boldsymbol{R}}}=\begin{cases}
    \ket{W_{(\alpha+2,\sigma),\boldsymbol{R}}}, & \text{if } \alpha\in\{0,1\} \\
    \ket{W_{(\alpha-2,\sigma),\boldsymbol{R}+\boldsymbol{a}_3}}, & \text{if } \alpha\in\{2,3\}
    \end{cases}
\impliedby \hat{T}_{a\boldsymbol{z}}\hat{c}^{\dagger}_{(\alpha,\sigma),\boldsymbol{k}}\hat{T}_{a\boldsymbol{z}}^{-1}=\begin{cases}
\hat{c}^{\dagger}_{(\alpha+2,\sigma),\boldsymbol{k}}, & \text{if } \alpha\in\{0,1\} \\
e^{-i\boldsymbol{k}\cdot\boldsymbol{a}_3}\hat{c}^{\dagger}_{(\alpha-2,\sigma),\boldsymbol{k}}, & \text{if } \alpha\in\{2,3\}
\end{cases}.
\end{align}
Then,
\begin{align}
\big(\hat{T}_{a\boldsymbol{z}}\hat{\mathscr{T}}\big)\hat{c}^{\dagger}_{(\alpha,\uparrow),\boldsymbol{k}}\big(\hat{T}_{a\boldsymbol{z}}\hat{\mathscr{T}}\big)^{-1}&=\begin{cases}
-\hat{c}^{\dagger}_{(\alpha+2,\downarrow),-\boldsymbol{k}}, & \text{if } \alpha\in\{0,1\} \\
-e^{i\boldsymbol{k}\cdot\boldsymbol{a}_3}\hat{c}^{\dagger}_{(\alpha-2,\downarrow),-\boldsymbol{k}}, & \text{if } \alpha\in\{2,3\}
\end{cases},\nonumber\\
\big(\hat{T}_{a\boldsymbol{z}}\hat{\mathscr{T}}\big)\hat{c}^{\dagger}_{(\alpha,\downarrow),\boldsymbol{k}}\big(\hat{T}_{a\boldsymbol{z}}\hat{\mathscr{T}}\big)^{-1}&=\begin{cases}
\hat{c}^{\dagger}_{(\alpha+2,\uparrow),-\boldsymbol{k}}, & \text{if } \alpha\in\{0,1\} \\
e^{i\boldsymbol{k}\cdot\boldsymbol{a}_3}\hat{c}^{\dagger}_{(\alpha-2,\uparrow),-\boldsymbol{k}}, & \text{if } \alpha\in\{2,3\}
\end{cases},
\end{align}
and thus
\begin{align}
\big(U_{T_{a\boldsymbol{z}}\mathcal{T}}\big)_{((\alpha,\sigma),\boldsymbol{k}),((\beta,\gamma),\boldsymbol{k}')}=\left(\left(\begin{array}{cc}
	0 & 1 \\
	e^{-i2ak_z} & 0
\end{array}\right)\otimes\text{id}_{2\times2}\right)_{\alpha,\beta}(-i\sigma_{y})_{\sigma,\gamma}\delta(\boldsymbol{k}+\boldsymbol{k}').
\end{align}
Then, given the matrix kernel of the AFM Hamiltonian (\ref{H_AFM}), this time-reversal-like transformation is a symmetry, 
\begin{align}
\big(\hat{T}_{a\boldsymbol{z}}\hat{\mathscr{T}}\big)\hat{H}_{\text{reg}}^{(\text{3D})}\big(\hat{T}_{a\boldsymbol{z}}\hat{\mathscr{T}}\big)^{-1}=\hat{H}_{\text{reg}}^{(\text{3D})},
\label{symm2}
\end{align}
since one can explicitly show
\begin{align}
\left(\left(\begin{array}{cc}
	0 & e^{i2ak_z} \\
	1 & 0
\end{array}\right)\otimes\text{id}_{2\times2}\otimes(i\sigma_{y})\right)\mathcal{H}^{(\text{3D})}_{\text{reg}}(\boldsymbol{k})^{*}\left(\left(\begin{array}{cc}
	0 & 1 \\
	e^{-i2ak_z} & 0
\end{array}\right)\otimes\text{id}_{2\times2}\otimes(-i\sigma_{y})\right)=\mathcal{H}^{(\text{3D})}_{\text{reg}}(-\boldsymbol{k}).
\end{align}
We have again equated Brillouin zone integrands and here taken $\boldsymbol{k}\rightarrow-\boldsymbol{k}$ in the $\text{BZ}_{\text{3D}}$-integral related to $\hat{H}_{\text{reg}}^{(\text{3D})}$ on the RHS of Eq.~(\ref{symm2}). Indeed, in the AFM case it is this symmetry that leads to a $\mathbb{Z}_{2}$ topological classification of bulk insulating electronic ground states \cite{Moore2010,Wang2019} similar to the non-magnetic case \cite{Kane2005,Fu2006}.
		
In summary, both the non-magnetic and anti-ferromagnetic Hamiltonian operators, specified by their matrix kernels (\ref{H_nonmag}) and (\ref{H_AFM}), respectively, have a center-of-inversion symmetry and a time-reversal symmetry.
\end{widetext}

\bibliography{nonMagnetic_TME.bib}
	
\end{document}